\documentclass[11pt,showpacs]{revtex4}
\input epsf.sty

\usepackage{ulem}
\usepackage{cancel}
\usepackage{color}

\usepackage{graphicx}
\usepackage{amsmath}
\usepackage{amsfonts}
\usepackage{amssymb}
\usepackage{rotating}
\usepackage{booktabs}
\usepackage{xcolor}
\usepackage{soul}

\def\red#1{\textcolor{red}{#1}}

\def\comment#1{}

\def\E{{\mathcal E}}

\usepackage{graphicx}
\begin{document}
\title{Hierarchy spectrum of SM fermions: from top quark to electron neutrino}
\author{She-Sheng Xue}
\email{xue@icra.it}
\affiliation{ICRANeT, Piazzale della Repubblica, 10-65122, Pescara,\\
Physics Department, University of Rome ``La Sapienza'', Rome,
Italy} 
%\affiliation{$^{(b)}$Department of Physics, Isfahan University of Technology, Isfahan 84156-83111, Iran}

%\date{Received version \today}

\begin{abstract}
In the SM gauge symmetries and fermion content of neutrinos, charged leptons and quarks, we study the effective four-fermion operators 
of Einstein-Cartan type and their contributions to the Schwinger-Dyson equations of fermion self-energy functions. 
The study is motivated by the speculation that these four-fermion operators are probably originated due to 
the quantum gravity that provides the natural regularization for 
chiral-symmetric gauge field theories.
In the chiral-gauge symmetry breaking phase, 
as to achieve the energetically favorable ground state, 
only the top-quark mass is generated via the spontaneous symmetry breaking, 
and other fermion masses are generated via the explicit symmetry breaking 
induced by the top-quark mass, four-fermion interactions and 
fermion-flavor mixing matrices. A phase transition from the symmetry breaking phase
to the chiral-gauge symmetric phase
at TeV scale occurs and the drastically fine-tuning problem can be resolved.  
In the infrared fixed-point domain of the four-fermion coupling 
for the SM at low energies, we qualitatively obtain the hierarchy patterns 
of the SM fermion Dirac masses, Yukawa couplings and 
family-flavor mixing matrices with 
three additional right-handed neutrinos $\nu^f_R$. 
Large Majorana masses and lepton-symmetry breaking 
are originated by the four-fermion interactions 
among $\nu^f_R$ and their left-handed conjugated fields $\nu^{fc}_R$. 
Light masses of gauged Majorana neutrinos in the normal hierarchy 
($10^{-5}-10^{-2}$ eV) are obtained consistently with neutrino oscillations. 
We present some discussions on the composite Higgs phenomenology 
and forward-backward asymmetry of $t\bar t$-production, as well as remarks on the candidates of 
light and heavy dark matter particles 
(fermions, scalar and pseudoscalar bosons). 
\end{abstract}


\maketitle

\newpage

\setcounter{footnote}{0}

\tableofcontents
\newpage

%\vskip0.1cm
\section
{\bf Introduction}
\hskip0.1cm
The parity-violating (chiral) gauge symmetries and spontaneous/explicit 
breaking of these symmetries for the hierarchy pattern of fermion masses have 
been at the center of a conceptual elaboration that has played a major 
role in donating to mankind the beauty of the Standard Model (SM) for fundamental
particle physics. On the one hand the {\it composite} 
Higgs-boson model or the Nambu-Jona-Lasinio (NJL) \cite{njl} with
effective four-fermion operators, and on the other  %its effective counterpart, 
the phenomenological model \cite{higgs}
of the {\it elementary} Higgs boson, %and its Yukawa-coupling to fermions 
they are effectively equivalent for the SM at low energies and provide an elegant 
and simple description for the chiral electroweak symmetry 
breaking and intermediate gauge boson masses. 
The experimental measurements of Higgs-boson mass 126 GeV \cite{ATLASCMS} and top-quark mass 173 GeV \cite{CDFD0}, as well as the other SM fermion masses and family-mixing angles, 
in particular neutrino oscillations, begin to shed light 
on this most elusive and fascinating arena of fundamental particle physics.

The patterns of the SM fermion masses and family-mixing matrices are equally fundamental, and closely related. Since Gatto {\it et. al}
\cite{1} tried to find the relation between the Cabibbo mixing angle and light-quark masses, the tremendous effort and many models 
have been made to study the relation of the SM fermion masses and family-mixing 
matrices from the phenomenological and/or theoretical view points \cite{1}-\cite{so10_5}, 
where the references are too many to be completely listed. 
In literature the most of effort based on phenomenological models assuming 
a particular texture in the original fermion-mass matrices in 
quark and/or lepton sectors to find the
fermion-family mixing matrices as functions of observed fermion masses, i.e., the eigenvalues of the original fermion-mass matrices. Whereas some other 
models try to find the relations of fermion masses and family-mixing matrices on the 
basis of theoretically model-building approaches, for example, the 
left-right symmetric scenario \cite{1}-\cite{5} and \cite{12,37}, string theory phenomenology \cite{faraggi2014,Dine2015} or the scenario 
of effective vector-like $W^\pm$-coupling at high energies \cite{xue1997mx,xue1999nu}. 
In the model-independent approach, the fermion-mass matrices with 
different null matrix elements (texture zeros) are considered to find the relations of
fermion mass and mixing patterns \cite{38}-\cite{43}. The gauge 
symmetries of grand unification theories, 
like $SO(10)$-theory, and/or the fermion-flavor symmetries, like horizontal or family discrete 
symmetry, are adopted to find non-trivial relations of fermion mass and mixing 
patterns \cite{11}-\cite{14}, \cite{43}-\cite{48} and \cite{so10_0}-\cite{so10_5}. 
As the precision measurements for neutrino oscillations are progressing \cite{35,49,50}, the study of neutrino mass pattern and lepton-flavor 
mixing becomes vigorously crucial \cite{51,52,qlmixing2015}.

%In Ref.~\cite{xue2015}, we 
%distinguished physically relevant four-fermion operators from 
%irrelevant operators in the both domains of IR- and UV-fixed points, and 
%focused on the discussion of relevant operators and their 
%resonant and nonresonant new phenomena for experimental searches.

In this article, we approach to this long-standing problem by 
considering effective four-fermion operators in the framework of the
SM gauge symmetries and fermion content: neutrinos, charged leptons and quarks. 
In order to accommodate high-dimensional operators of fermion 
fields in the SM-framework of a well-defined quantum field theory at the 
high-energy scale $\Lambda$, it is essential and necessary to study: 
(i) what physics beyond the SM at the scale $\Lambda$ explains the 
origin of these operators; (ii) which dynamics of these operators 
undergoes in terms of their dimensional couplings (e.g., $G$) 
and energy scale $\mu$; (iii) associating to these dynamics, 
where infrared (IR) and ultraviolet (UV) 
stable fixed points of these couplings locate and what characteristic energy 
scales are; (iv) in the IR-domain and 
UV-domain (scaling regions) of these stable IR and UV fixed points, 
which operators become physically relevant (effectively dimension-4) 
and renormalizable following renormalization group (RG) equations (scaling laws), 
and other irrelevant operators are suppressed by the cutoff at least 
$\mathcal O(\Lambda^{-2})$. 
      
We briefly recall that the strong technicolor dynamics 
of extended gauge theories at the 
TeV scale was invoked \cite{hill1994,bhl1990a}
to have a natural scheme incorporating the 
four-fermion operator
\begin{eqnarray}
L = L_{\rm kinetic} + G(\bar\psi^{ia}_Lt_{Ra})(\bar t^b_{R}\psi_{Lib}),%\quad G\sim 1/\Lambda^2
\label{bhl}
\end{eqnarray}
of Bardeen, Hill and Lindner (BHL) $\langle \bar t t \rangle$-condensate model \cite{bhl1990} 
in the context of a well-defined quantum field theory at the 
high-energy scale $\Lambda$.
%where notations will be given later in Eq.~(\ref{bhlx}) 
The four-fermion operator (\ref{bhl}) undergoes 
the spontaneous symmetry breaking (SSB) dynamics 
responsible for the generation of top-quark and Higgs-boson masses 
in the domain %(small $G\gtrsim G_c$) 
of IR-stable fixed point $G_c$ (critical value associated with the SSB) 
and characteristic energy scale (vev) $v\approx 239.5$ GeV. 
The analysis of this composite Higgs boson model shows \cite{bhl1990} 
that Eq.~(\ref{bhl}) effectively becomes a bilinear 
and renormalizable Lagrangian following RG equations, together
with the composite Goldstone modes %(pseudo scalars $\bar\psi\gamma_5\psi$) 
for the longitudinal components of massive $W^\pm$ and $Z^0$ gauge bosons, 
and the composite scalar %($H\sim\bar\psi\psi$) 
for the Higgs boson.  
The low-energy SM physics, including the values of top-quark and Higgs-boson masses,
was supposed to be achieved by the RG-equations in the domain 
of the IR-stable fixed point \cite{bhl1990a,Marciano1989,bhl1990}.
On the other hand, the relevant operator (\ref{bhl}) can be constructed 
on the basis of the SM phenomenology at low-energies.  
It was suggested \cite{bhl1990,nambu1989,Marciano1989} 
that the symmetry breakdown of the SM could be 
a dynamical mechanism of the NJL type that 
intimately involves the top quark at the high-energy scale $\Lambda$, 
since then, many models based on this idea have been studied \cite{DSB_review}.  

Nowadays, the known top-quark and Higgs boson masses completely determine the boundary 
conditions of the RG equations for the top-quark Yukawa coupling $\bar g_t(\mu)$ 
and Higgs-boson quartic coupling $\tilde \lambda(\mu)$
in the composite Higgs boson model (\ref{bhl}).
Using the experimental values of top-quark and Higgs boson masses, 
we obtained \cite{xue2013,xue2014} the unique solutions $\bar g_t(\mu)$ 
and $\tilde \lambda(\mu)$to these RG equations, 
provided the appropriate non-vanishing 
form-factor $\tilde Z_H(\mu)=1/\bar g^2_t(\mu)$ of the composite Higgs boson 
at the energy-scale $\E\sim$ TeV, 
where the effective quartic coupling $\tilde \lambda(\mu)$ 
of composite Higgs bosons vanishes.

The form-factor of composite Higgs boson $H\sim(\bar\psi\psi)$ is finite and  
does not vanish in the SSB phase (composite Higgs phase for small $G\gtrsim G_c$), 
indicating that the tightly bound composite Higges particle behaves 
as if an elementary particle. On the other hand, 
due to large four-fermion coupling $G$, massive composite fermions $\Psi \sim (H\psi)$ 
are formed by combining a composite Higgs boson $H$ with an elementary fermion $\psi$ 
in the symmetric phase where the SM gauge symmetries are exactly preserved \cite{xue1997}. 
This indicates that a second-order phase transition from the SSB phase to the 
SM gauge symmetric phase takes place at the critical point $G_{\rm crit}> G_c$. 
In addition the effective quartic coupling of composite Higgs bosons vanishing 
at $\E\sim$ TeV scales indicates the characteristic energy scale of such phase transition. 
The energy scale $\E$ is much lower than the cutoff scale 
$\Lambda$ ($\E \ll \Lambda$) so that the drastically fine-tuning (hierarchy) problem that
fermion masses $m_f\ll \Lambda$ or the pseudoscalar 
decay constant $f_\pi\ll \Lambda$ can be possibly avoided by the replacements 
$m_f< \E$ or the pseudoscalar 
decay constant $f_\pi< \E$ \cite{xue2013}. 

In Ref. \cite{xue2016}, after a short review that recalls and 
explains the quantum-gravity origin 
of four-fermion operators at the cutoff $\Lambda$, 
the BHL $\langle \bar t t\rangle$-condensate model and the SSB, 
we show that due to four-fermion operators
(i) there are the SM gauge symmetric vertexes 
of quark-lepton interactions; 
(ii) the one-particle-irreducible (1PI) 
vertex-function of $W^\pm$-boson coupling becomes 
approximately vector-like at TeV scale. 
Both interacting vertexes contribute the explicit symmetry breaking (ESB) 
terms to the Schwinger-Dyson (SD) equations of fermion self-energy functions. 
As a result, once the top-quark mass is generated via the SSB, 
the masses of third fermion family $(\nu_\tau, \tau, b)$ are generated 
by the ESB via quark-lepton interactions and $W^\pm$-boson vector-like coupling.  
Within the third fermion family, we qualitatively study the hierarchy 
of fermion masses and effective Yukawa couplings %$\bar g_{t,b,\tau}$ 
in terms of the top-quark mass and Yukawa coupling \cite{xue2016}. 
\comment{
In the concluding section, a summary of basic points of the scenario and 
its extension to three fermion families is given. In addition, we present some 
discussions on the possible experimental relevance of running Yukawa couplings obtained and 
parity-conservation feature of the $W^\pm$-boson coupling at TeV scale.}
   
In this article, we generalize this study into three fermion families of the SM
by taking into account the flavor mixing of three fermion families. 
Such flavor mixing inevitably introduces 
the 1PI vertex-functions of quark-lepton interactions 
and approximately vector-like $W^\pm$-boson coupling
among three fermion families at TeV scale. 
As a consequence, these 1PI vertex-functions 
introduce the ESB terms into the SD-equations of
the fermion self-energy functions for all SM fermions in three families.
Once the top-quark mass is generated via the SSB, 
all other SM fermions acquire their masses via the ESB terms 
by (i) four-fermion interactions among 
fermion flavors via family mixing matrices; 
(ii) the $W^\pm$-boson coupling among fermion flavors
via the CKM or PMNS mixing matrix. 
The latter is dominate particularly for light quarks and leptons. 
As a result, we quatitatively obtain the hierarchy patterns 
of the SM fermion masses and family-mixing matrices, and all fermion masses and 
Yukawa couplings are functions of the top-quark mass and Yukawa coupling.
Neutrino masses will separately be studied in the last part of the article, for its
peculiarity.

This lengthy article is organized as follow. 
In Sec.~\ref{four-fermionS}, 
we give an argument why four-fermion operators 
should be present in an effective Lagrangian at the high-energy cutoff at which the 
quantum gravity introduces a natural regulator for chiral gauge theories.
In the framework of the SM gauge symmetries and fermion content, we discuss 
four-fermion operators, including quark-lepton interactions. 
In Sec.~\ref{mixing-matrixS}, we describe fermion-flavor mixing matrices in lepton and 
quark sectors, as well as quark-lepton interaction sector. In Sec.~\ref{SSBS},
we give a brief recall that the SSB is responsible only for 
the top-quark and Higgs boson masses, whose values determine the unique solution to 
the RG equations for the top-quark Yukawa and composite Higgs quartic couplings.
In Secs.~\ref{esbS}and \ref{sdS}, we discuss the ESB terms and massive solutions of 
SD equations of other SM fermions. 
In Sec.~\ref{hiS}, we qualitatively present the hierarchy patterns of the SM fermions 
and fermion-flavor mixing matrices. 
In the last section \ref{neutrinoS}, we focus on the discussions
of gauged and sterile neutrinos of Dirac or Majorana type, 
and their masses, mixing and oscillation. A brief summary and some remarks are given at 
the end of the article \footnote{More discussions on the experimental aspects of this scenario can be found in the Refs.~S.-S.~Xue,
arXiv1601.06845 and references therein.}.

%\vskip0.1cm
\noindent
\section {Four-fermion operators beyond the SM 
%from quantum gravity
}\label{four-fermionS}
%\hskip0.1cm
\subsection{Regularization and quantum gravity}
Up to now the theoretical and experimental studies tell us 
the chiral gauge-field interactions to fermions 
in the lepton-quark family that is replicated three times and mixed. 
The spontaneous breaking of these chiral gauge symmetries and generating of fermion masses are made by the Higgs field sector. 
In the IR-fixed-point domain of weak four-fermion coupling 
or equivalently weak Yukawa coupling, the SM Lagrangian with all relevant operators (parametrizations) is realized and behaves an effective and renormalizable field theory in low energies. 
To achieve these SM relevant operators, a finite field theory of chiral-gauge interactions should be well-defined by including the quantum gravity that naturally provides a space-time regularization (UV cutoff).
%For these purposes, 
As an example, the finite superstring theory is proposed by postulating that instead of a simple space-time point, the fundamental space-time ``constituents" is a space-time ``string". 
The Planck scale is a plausible cut-off, at which 
%the gravitational coupling is of comparable strength
%to the gauge couplings Grand Unified Theories (GUT).
all principle and symmetries are fully respect by gauge fields and 
particle spectra, fermions and bosons. \comment{NPB The superstring compactification
on the Calabi-Yau manifold or orbifold results in the pointlike supergravity theories, inducing 
high-dimensional operators below the
Planck scale that provides a perturbatively consistent scheme to understand the SM parameters, e.g.~particle masses, and their logarithmic running with energy, see review Ref.~\cite{faraggi2014}}
%, which  show the superstring theory to be the most promising candidate 
%for the fundamental theory.

In this article, we do not discuss how a fundamental theory
at the Planck scale %, such as the superstring theory, 
induces high-dimensional operators. Instead, as a postulation or motivation, we argue the presence of at least four-fermion operators beyond the SM from the following point view. A well-defined quantum field theory for the SM Lagrangian requires a natural regularization (UV cutoff $\Lambda$) fully preserving the SM 
chiral-gauge symmetry. The quantum gravity naturally provides a such regularization of discrete 
space-time with the minimal length $\tilde a\approx 1.2\,a_{\rm pl}$ \cite{xue2010}, 
where the Planck length 
$a_{\rm pl}\sim 10^{-33}\,$cm and scale $\Lambda_{\rm pl}=\pi/a_{\rm pl}\sim 10^{19}\,$GeV. 
However, the no-go theorem \cite{nn1981} tells us 
that there is no any consistent way to regularize 
the SM bilinear fermion Lagrangian to exactly preserve the SM chiral-gauge symmetries, which 
must be explicitly broken at the scale of fundamental space-time cutoff $\tilde a$. 
This implies that the natural quantum-gravity regularization for the SM should
lead us to consider at least dimension-6 four-fermion operators originated from quantum gravity effects at short distances \footnote{In the regularized and quantized EC theory \cite{xue2010}
with a basic space-time cutoff, in addition to dimension-6 four-fermion operators,
%the leading term $J^dJ_d$ in Eq.~(\ref{ec0}) 
there are high-dimensional fermion operators ($d>6$), 
e.g., $\partial_\sigma J^\mu\partial^\sigma J{_\mu}$, which are suppressed
at least by ${\mathcal O}(\tilde a^4)$.
}. As a model, we adopt the four-fermion operators of the torsion-free Einstein-Cartan Lagrangian within the framework of the SM fermion content and gauge symmetries. We stress that a fundamental theory at the UV cutoff is still unknown.
\comment{ 
The
quantum-gravity origin of four-fermion operators is just an argumentation or a speculation. This
however motivates us to adopt the EC Lagrangian as an effective Lagrangian with the SM chiral
gauge symmetries and fermion content to study the SM fermion masses and their hierarchy.
This article does not focus on the origin of the four-fermion operators from the underlying
theory, nevertheless, it is expected that the most promising candidate for the fundamental theory
of quantum gravity should be the superstring theory, and there have been many detailed studies of
the origin of fermion generations and masses in the context of superstring theory and the resultant
pointlike supergravity theories, see for example the recent monograph [77]on the issue. 
}

\subsection{Einstein-Cartan theory with the SM gauge symmetries and fermion content}

The Lagrangian of torsion-free Einstein-Cartan (EC) theory reads,
\comment{The four-fermion operators of the classical and 
torsion-free Einstein-Cartan (EC) theory are naturally
obtained by integrating over ``static'' torsion fields at the Planck length.}
\begin{eqnarray}
{\mathcal L}_{EC}(e,\omega,\psi)&=&  {\mathcal L}_{EC}(e,\omega) + 
\bar\psi e^\mu {\mathcal D}_\mu\psi +GJ^dJ_d,
\label{ec0}
\end{eqnarray}
where the gravitational Lagrangian ${\mathcal L}_{EC}={\mathcal L}_{EC}(e,\omega)$, 
tetrad field $e_\mu (x)= e_\mu^{\,\,\,a}(x)\gamma_a$,
spin-connection field $\omega_\mu(x) = \omega^{ab}_\mu(x)\sigma_{ab}$,  
the covariant derivative ${\mathcal D}_\mu =\partial_\mu - ig\omega_\mu$ and 
the axial current $J^d=\bar\psi\gamma^d\gamma^5\psi$ of massless fermion fields. 
The four-fermion coupling $G$ relates to the gravitation-fermion gauge 
coupling $g$ and fundamental space-time cutoff $\tilde a$. 

Within the SM fermion content, we consider massless left- and 
right-handed Weyl fermions $\psi^f_{_L}$ and $\psi^f_{_R}$ carrying 
quantum numbers of the SM symmetries, as well as three
right-handed Weyl sterile neutrinos 
$\nu^f_{_R}$ and their left-handed conjugated fields 
$\nu^{f\, c}_{_R}=i\gamma_2(\nu_{_R})^*$, where ``$f$'' is the fermion-family index.  
Analogously to the EC theory (\ref{ec0}), we obtain a torsion-free, 
diffeomorphism and {\it local} gauge-invariant 
Lagrangian % (see for example Refs.~\cite{xue2010,art1989,contact}), 
\begin{eqnarray}
{\mathcal L}
&=&{\mathcal L}_{EC}(e,\omega)+\sum_f\bar\psi^f_{_{L,R}} e^\mu {\mathcal D}_\mu\psi^f_{_{L,R}} 
+ \sum_f\bar\nu^{f c}_{_{R}} e^\mu {\mathcal D}_\mu\nu^{f c}_{_{R}}\nonumber\\
&+&G\left(J^{\mu}_{_{L}}J_{_{L,\mu}} + J^{\mu}_{_{R}}J_{_{R,\mu}} 
+ 2 J^{\mu}_{_{L}}J_{_{R,\mu}}\right)\nonumber\\
&+&G\left(j^{\mu}_{_{L}}j_{_{L,\mu}} + 2J^{\mu}_{_L}j_{_{L,\mu}} 
+ 2 J^{\mu}_{_R}j_{_{L,\mu}}\right),
\label{art}
\end{eqnarray}
where we omit the gauge interactions in ${\mathcal D}_\mu$ 
and axial currents read
\begin{eqnarray}
J^{\mu}_{_{L,R}}\equiv \sum_f\bar\psi^f_{_{L,R}}\gamma^\mu\gamma^5\psi^f_{_{L,R}}, \quad
j^{\mu}_{_L}\equiv \sum_f\bar\nu^{fc}_{_R}\gamma^\mu\gamma^5\nu^{fc}_{_R}. 
\label{acur}
\end{eqnarray}
The four-fermion coupling $G$ is unique for all four-fermion operators and 
high-dimensional fermion operators ($d>6$) are neglected. 
\comment{If torsion fields that couple to fermion fields 
are not exactly static, propagating a short distance 
$\tilde \ell \gtrsim \tilde a$, 
characterized by their large masses 
$\Lambda\propto \tilde \ell^{-1}$, this implies the four-fermion 
coupling $G\propto \Lambda^{-2}$.
We will in future 
address the issue how the space-time cutoff $\tilde a$ due to quantum 
gravity relates to the cutoff scale $\Lambda(\tilde a)$ 
possibly by intermediate torsion fields 
or the Wilson-Kadanoff renormalization group approach.}

By using the Fierz theorem \cite{itzykson}, the dimension-6 four-fermion operators in Eq.~(\ref{art}) can be written as \cite{xue2015} 
\begin{eqnarray}
&+&(G/2)\left(J^{\mu}_{_{L}}J_{_{L,\mu}} + J^{\mu}_{_{R}}J_{_{R,\mu}} 
+ j^{\mu}_{_{L}}j_{_{L,\mu}} + 2J^{\mu}_{_L}j_{_{L,\mu}}\right)\label{art0}\\
&-&G\sum_{ff'}\left(\, \bar\psi^f_{_L}\psi^{f'}_{_R}\bar\psi^{f'}_{_R} \psi^f_{_L}
+\, \bar\nu^{fc}_{_R}\psi^{f'}_{_R}\bar\psi^{f'}_{_R} \nu^{fc}_{_R}\right),
\label{art0'}
\end{eqnarray}
which preserve the SM gauge symmetries. Equations (\ref{art0}) and 
(\ref{art0'}) represent repulsive and attractive operators respectively.
The former (\ref{art0}) 
are suppressed by the cutoff ${\mathcal O}(\Lambda^{-2})$, and cannot become relevant and renormalizable operators of effective dimension-4. 
\comment{and in the UV-domain where the 
formation of composite fermions occurs. Therefore they are irrelevant operators. 
Instead, four-fermion operators (\ref{art0'}) are
relevant operators in the both domains of IR- and UV-stable fixed points, respectively 
associated with the SSB dynamics and formation of composite fermions. 
We presented some discussions of these relevant operators and their 
resonant and nonresonant new phenomena for experimental searches. 
We will proceed further discussions in the last section 
of this article. However, main attention will be given to the mass generation of the 
third fermion family in the IR-domain where the SSB dynamics occurs.
}
Thus the torsion-free EC theory with the attractive
four-fermion operators read,
\begin{eqnarray}
{\mathcal L}
&=&{\mathcal L}_{EC}+\sum_{f}\bar\psi^f_{_{L,R}} e^\mu {\mathcal D}_\mu\psi^f_{_{L,R}} 
%+ \bar\nu^{ f}_{_{R}} e^\mu {\mathcal D}_\mu\nu^{ f}_{_{R}}
+ \sum_{f}\bar\nu^{ fc}_{_{R}} e^\mu {\mathcal D}_\mu\nu^{ fc}_{_{R}}\nonumber\\
&-&G\sum_{ff'}\left(\, \bar\psi^{f}_{_L}\psi^{f'}_{_R}\bar\psi^{f'}_{_R} \psi^{f}_{_L}
+\, \bar\nu^{fc}_{_R}\psi^{f'}_{_R}\bar\psi^{f'}_{_R} \nu^{fc}_{_R}\right)+{\rm h.c.},
\label{art1}
\end{eqnarray}
where the two component Weyl fermions $\psi^{f}_{_L}$ and $\psi^{f}_{_R}$  
respectively are the $SU_L(2)\times U_Y(1)$ gauged doublets and singlets of the SM. 
For the sake of compact notations, $\psi^{f}_{_R}$ are also used to represent 
$\nu^f_R$, %and $\nu_R^{f c}=i\gamma_2(\nu_R^{f})^*$ are their the conjugate fields.
which have no any SM quantum numbers. 
All fermions are massless, 
they are four-component Dirac fermions 
$\psi^f=(\psi_L^f+\psi_R^f)$, two-component right-handed Weyl neutrinos $\nu^f_L$ 
and four-component sterile Majorana neutrinos $\nu_M^f=(\nu_R^{fc}+\nu_R^f)$ whose 
kinetic terms read
\begin{eqnarray}
\bar\nu^f_{_{L}} e^\mu {\mathcal D}_\mu\nu^f_{_{L}}, \quad
\bar\nu_{_{M}}^f e^\mu {\mathcal D}_\mu\nu_{_{M}}^f =
\bar\nu^f_{_{R}} e^\mu {\mathcal D}_\mu\nu^f_{_{R}} 
+ \bar\nu^{ fc}_{_{R}} e^\mu {\mathcal D}_\mu\nu^{ fc}_{_{R}}.
\label{mnu}
\end{eqnarray} 
In Eq.~(\ref{art1}), $f$ and $f'$ ($f,f'=1,2,3$) are 
fermion-family indexes summed over respectively for three 
lepton families (charge $q=0,-1$) and three quark families ($q=2/3,-1/3$). 
Eq.~(\ref{art1}) preserves not only the SM gauge symmetries 
and global fermion-family symmetries, but also the global symmetries for 
fermion-numbers conservations. We adopt the effective four-fermion operators (\ref{art1}) 
in the context of a well-defined quantum field theory 
at the high-energy scale $\Lambda$.
\comment{Relating to the gravitation-fermion gauge 
coupling $g$, the effective four-fermion coupling $G$ is unique 
for all four-fermion operators, and its strength depends on energy scale and 
characterizes: (i) the domain of IR fixed point where the 
spontaneous breaking of SM gauge-symmetries occurs (see for example 
\cite{bhl1990}) 
and (ii) the domain of UV fixed point where the SM gauge-symmetries 
are restored and massive (TeV) composite Dirac fermions are formed \cite{xue2014}. 
}

%\vskip0.1cm
\noindent
\subsection{SM gauge-symmetric four-fermion operators}\label{sm4f}
\hskip0.1cm
Neglecting the flavor-mixing
of three fermion families ($f=f'$) to simply notations, 
we explicitly show SM gauge symmetric four-fermion operators in Eq.~(\ref{art1}).
In the quark sector, the four-fermion operators are  
\begin{eqnarray}
G\left[(\bar\psi^{ia}_Lt_{Ra})(\bar t^b_{R}\psi_{Lib})
+ (\bar\psi^{ia}_Lb_{Ra})(\bar b^b_{R}\psi_{Lib})\right]+{\rm ``terms"},
\label{bhlx}
\end{eqnarray}
where $a,b$ and $i,j$ are the color and flavor indexes 
of the top and bottom quarks, the quark $SU_L(2)$ doublet 
$\psi^{ia}_L=(t^{a}_L,b^{a}_L)$ 
and singlet $\psi^{a}_R=t^{a}_R,b^{a}_R$ are the eigenstates 
of electroweak interaction. 
The first and second terms in Eq.~(\ref{bhlx}) are respectively 
the four-fermion operators of top-quark channel \cite{bhl1990} 
and bottom-quark channel,
whereas ``terms" stands for 
the first and second quark families that can be obtained 
by substituting $t\rightarrow u,c$ 
and $b\rightarrow d,s$ \cite{xue2013_1,xue2013,xue2014}. 

In the lepton sector with three right-handed sterile neutrinos $\nu^\ell_R$ 
($\ell=e,\mu,\tau$), the four-fermion operators 
in terms of gauge eigenstates are,
\begin{eqnarray}
G\left[(\bar\ell^{i}_L\ell_{R})(\bar \ell_{R}\ell_{Li})
+ (\bar\ell^{i}_L\nu^\ell_{R})(\bar \nu^\ell_{R}\ell_{Li}) 
+ (\bar\nu^{\ell\, c}_R\nu^{\ell\,}_{R})(\bar \nu^{\ell}_{R}\nu^{\ell c}_{R})\right],
\label{bhlxl}
\end{eqnarray}
preserving all SM gauge symmetries, 
where the lepton $SU_L(2)$ doublets $\ell^i_L=(\nu^\ell_L,\ell_L)$, singlets 
$\ell_{R}$ and the conjugate fields of sterile neutrinos 
$\nu_R^{\ell c}=i\gamma_2(\nu_R^{\ell})^*$.  
Coming from the second term in Eq.~(\ref{art1}), the last term in Eq.~(\ref{bhlxl}) 
preserves the symmetry 
$U_{\rm lepton}(1)$ for the lepton-number 
conservation, although $(\bar \nu^{\ell}_{R}\nu^{\ell c}_{R})$ 
violates the lepton number of family ``$\ell$'' by two units. 

Similarly, from the second term in Eq.~(\ref{art1}) there are following four-fermion operators  
\begin{eqnarray}
G\left[(\bar\nu^{\ell\, c}_R\ell_{R})(\bar \ell_{R}\nu^{\ell\, c}_{R})
+(\bar\nu^{\ell\, c}_R u^{\ell}_{a,R})(\bar u^{\ell}_{a,R}\nu^{\ell c}_{R})
+(\bar\nu^{\ell\,c}_R d^\ell_{a,R})(\bar d^{\ell}_{a,R}\nu^{\ell c}_{R})\right],
\label{bhlbv}
\end{eqnarray}
where quark fields $u^{\ell}_{a,R}=(u,c,t)_{a,R}$ and $d^{\ell}_{a,R}=(d,s,b)_{a,R}$.

\subsection{Four-fermion operators of quark-lepton interactions}

Although the four-fermion operators in Eq.~(\ref{art1}) do not have 
quark-lepton interactions, we consider the following
SM gauge-symmetric four-fermion 
operators that contain quark-lepton interactions \cite{xue1999nu}, 
\begin{eqnarray}
G\left[(\bar\ell^{i}_Le_{R})(\bar d^a_{R}\psi_{Lia})
+(\bar\ell^{i}_L\nu^e_{R})(\bar u^a_{R}\psi_{Lia})\right] +(\cdot\cdot\cdot),
\label{bhlql}
\end{eqnarray}
where $\ell^i_L=(\nu^e_L,e_L)$ and $\psi_{Lia}=(u_{La},d_{La})$
for the first family. The ($\cdot\cdot\cdot$) represents   
for the second and third families with substitutions: 
$e\rightarrow \mu, \tau$, $\nu^e\rightarrow \nu^\mu, \nu^\tau$, and 
$u\rightarrow c, t$ and $d\rightarrow s, b$. 
%Here we do not consider baryon-number violating operators. 
The four-fermion operators (\ref{bhlql}) of quark-lepton interactions 
are not included in Eq.~(\ref{art1}), since leptons and quarks are in separated
representations of SM gauge groups. They should be expected
in the framework of Einstein-Cartan theory and $SO(10)$ unification theory \cite{lq1997}.

In order to study the mass generation of three fermion families by the mixing of three fermion families we  generalize the quark-lepton interacting operators (\ref{bhlql}) to 
\begin{eqnarray}
G\sum_{ff'}\Big\{(\bar\ell^{if}_Le^{f'}_{R})(\bar d^{af'}_{R}\psi^f_{Lia})
+(\bar\ell^{if}_L\nu^{f'}_{eR})(\bar u^{af'}_{R}\psi^f_{Lia})\Big\},
\label{bhlqlm}
\end{eqnarray}
analogously to the four-fermion operators in Eq.~(\ref{art1}).

%\vskip0.1cm
\section
{\bf Gauge vs mass eigenstates in fermion-family space}\label{mixing-matrixS}
\hskip0.1cm
Due to the unique four-fermion coupling $G$ 
and the global fermion-family $U_L(3)\times U_R(3)$ symmetry 
of Eq.~(\ref{art1}), 
one is allowed to perform chiral transformations ${\mathcal U}_L\in U_L(3)$ and 
${\mathcal U}_R\in U_R(3)$ so that $f=f'$, the four-fermion operators (\ref{art1})
are only for each fermion family without the family-flavor-mixing 
and all fermion fields are Dirac mass eigenstates. 
In this section, neglecting gauge interactions 
we discuss the unitary chiral transformations from gauge eigenstates to mass eigenstates in 
quark and lepton sectors, so as to diagonalize in the fermion-family space 
the four-fermion operators (\ref{art1}) and two-fermion operators ($\psi\bar\psi$), the latter 
is relating to fermion mass matrices. 

\subsection{Quark sector}

For the quark sector, the four-fermion operators (\ref{art1}) are  
\begin{eqnarray}
-G\, \sum_{ff'}\Big[\bar\psi^{f}_{_L}\psi^{f'}_{_R}\bar\psi^{f'}_{_R} \psi^{f}_{_L}\Big]_{2/3}
-G\, \sum_{ff'}\Big[\bar\psi^{f}_{_L}\psi^{f'}_{_R}\bar\psi^{f'}_{_R} \psi^{f}_{_L}\Big]_{-1/3},
\label{q}
\end{eqnarray}
where the $SU_L(2)\times U_Y(1)$ doublets $\psi^{f}_{_L}$ and singlets $\psi^{f}_{_R}$ are 
the SM gauge eigenstates, $SU(3)$-color index ``$a$''is summed over 
$\bar\psi^{af}_{_L}\psi^{f'}_{_{aR}}
\rightarrow \bar\psi^{f}_{_L}\psi^{f'}_{_{R}}$, $f$ and $f'$ are 
family indexes of three fermion families. 
The first term is for the $(2/3)$-charged sector, $\psi^{f'}_{_{R}}\Rightarrow u^{f'}_{_{R}}$
represented by the $u$-quark sector $u^{f'}\Rightarrow(u,c,t)$, 
the second term is for the $(-1/3)$-charged sector, $\psi^{f'}_{_{R}}\Rightarrow d^{f'}_{_{R}}$ 
represented by the $d$-quark sector $d^{f'}\Rightarrow(d,s,b)$.  

Due to the unique four-fermion coupling $G$ 
and the global fermion-family $U^u_L(3)\times U^u_R(3)$ symmetry for 
the $u$-quark sector and $U^d_L(3)\times U^d_R(3)$ symmetry for 
the $d$-quark sector in Eq.~(\ref{q}), 
we perform four unitary chiral transformations from gauge-eigenstates to mass-eigenstates: 
\begin{eqnarray}
\psi^u_L&\rightarrow & {\mathcal U}^{u}_L~\psi^u_L,\quad 
\psi^u_R\rightarrow  {\mathcal U}^{u}_R~\psi^u_R;\quad 
{\mathcal U}^{u}_{L,R}\in U^{u}_{L,R}(3),
\label{cqu}
\end{eqnarray}
and
\begin{eqnarray}
\psi^d_L&\rightarrow & {\mathcal U}^{d}_L~\psi^d_L,\quad 
\psi^d_R\rightarrow  {\mathcal U}^{d}_R~\psi^d_R;\quad 
{\mathcal U}^{d}_{L,R}\in U^{d}_{L,R}(3),
\label{cqd}
\end{eqnarray}
so that in Eq.~(\ref{q}) the fermion-family indexes $f=f'$, i.e., $\delta_{ff'}$ 
respectively for the $u$-quark sector and 
the $d$-quark sector. As a result, all quark fields are mass eigenstates, 
the four-fermion operators (\ref{q}) are ``diagonal'' only for each quark family 
without family-mixing, 
\begin{eqnarray}
-G\, \sum_{f=1,2,3}\Big[\bar\psi^{f}_{_L}\psi^{f}_{_R}\bar\psi^{f}_{_R} \psi^{f}_{_L}\Big]_{2/3}
-G\, \sum_{f=1,2,3}\Big[\bar\psi^{f}_{_L}\psi^{f}_{_R}\bar\psi^{f}_{_R} \psi^{f}_{_L}\Big]_{-1/3}.
\label{q1}
\end{eqnarray}
In this representation, the vacuum expectation values of two-fermion operators 
$\langle\psi^{f}_{_R} \bar \psi^{f}_{_L}\rangle+{\rm h.c.}$, i.e., quark-mass matrices are diagonalized in the fermion-family space by the biunitary transformations
\begin{eqnarray}
M^u\Rightarrow M^u_{\rm diag}= (m^u_1, m^c_2,m^t_3)={\mathcal U}^{u\dagger}_L M^u {\mathcal U}^u_R,\label{mqu}\\
M^d\Rightarrow M^d_{\rm diag}=(m^d_1, m^s_2,m^b_3)= {\mathcal U}^{d\dagger}_L M^d {\mathcal U}^d_R,\label{mqd}
\end{eqnarray}
where all quark masses (eigenvalues) are positive, ${\mathcal U}_L$ 
and ${\mathcal U}_R$ are related by
\begin{eqnarray}
{\mathcal U}^{u,d}_L={\mathcal V}^{u,d} {\mathcal U}^{u,d}_R,
\label{ulur}
\end{eqnarray}
and ${\mathcal V}^{u,d}$ is an unitary matrix, see for example \cite{zli,mohapatra}.
 
%In usual notations, $M^{u}_{\rm diag}=(m^u_1, m^c_2,m^t_3)$
%and $M^{d}_{\rm diag}=(m^d_1, m^s_2,m^b_3)$ 
%are the diagonal mass-matrices. 

Using unitary matrices ${\mathcal U}^u_{L,R}$ (\ref{cqu}) 
and ${\mathcal U}^d_{L,R}$ (\ref{cqd}), up to a diagonal phase matrix we define the unitary quark-family mixing matrices,
\begin{eqnarray}
{\mathcal U}^{u\dagger}_L{\mathcal U}^{d}_L, &\quad&  
%e^{i\varphi_1}
{\mathcal U}^{u\dagger}_L{\mathcal U}^d_R,\nonumber\\
%e^{i\varphi_2}
{\mathcal U}^{u\dagger}_R{\mathcal U}^d_L, &\quad&%~~~~~
{\mathcal U}^{u\dagger}_R{\mathcal U}^{d}_R.
\label{mmud}
\end{eqnarray}
where the first element is the CKM matrix $U=U^q_L\equiv 
{\mathcal U}_L^{u\dagger}{\mathcal U}_L^d$. 
The experimental values \cite{pdg2012}
of CKM matrix are adopted to calculate the fermion spectrum in this article. 
   
\comment{$e^{i\varphi_1}$ is a relative phase between ${\mathcal U}^{u\dagger}_L$ and 
${\mathcal U}^d_R$, and $e^{i\varphi_2}$ is another relative phase between ${\mathcal U}^{u\dagger}_R$ and ${\mathcal U}^d_L$. Actually, Eq.~(\ref{mmud}) is a $6\times 6$ mixing matrix,
which is unitary, if the relative phases $\varphi_2-\varphi_1=n\pi, ~n=1,2,3$.} 

\subsection{Lepton sector}\label{lnmass}

For the lepton sector, the four-fermion operators (\ref{art1}) are 
\begin{eqnarray}
-G\, \sum_{ff'}\left[\bar\ell^{f}_{_L}\ell^{f'}_{_R}\bar\ell^{f'}_{_R} \ell^{f}_{_L}+
(\bar\ell^{f}_L\nu^{f'}_{R})(\bar \nu^{f'}_{R}\ell^f_{L})+ 
(\bar\nu^{f\, c}_R\nu^{f'\,}_{R})(\bar \nu^{f'}_{R}\nu^{f c}_{R})
\right],
\label{l}
\end{eqnarray}
where Dirac lepton fields $\ell^{f}_{_L}$ and 
$\ell^{f}_{_R}$ are the SM $SU_L(2)$-doublets and singlets
respectively, $\nu^{f}_{R}$ are three sterile (Dirac) neutrinos and 
$\nu_R^{f c}=i\gamma_2(\nu_R^{f})^*$ are their the conjugate fields.
Analogously to the quark sector (\ref{q}), 
we perform four unitary chiral transformations from gauge eigenstates
to mass eigenstates 
\begin{eqnarray}
{\nu}_L&\rightarrow & {\mathcal U}^{\nu}_L~{\nu}_L,\quad 
{\nu}_R\rightarrow  {\mathcal U}^{\nu}_R~{\nu}_R;\quad 
{\mathcal U}^{\nu}_{L,R}\in U^{\nu}_{L,R}(3),
\label{clnu}
\end{eqnarray}
and
\begin{eqnarray}
\ell_L &\rightarrow& {\mathcal U}^\ell_L~\ell_L,\quad 
\ell_R\rightarrow {\mathcal U}^\ell_R~\ell_R;\quad {\mathcal U}^\ell_{L,R}\in U^\ell_{L,R}(3)
\label{cl}
\end{eqnarray}
so that in Eq.~(\ref{l}) the fermion-family indexes $f=f'$, i.e., $\delta_{ff'}$ 
respectively for the Dirac $\nu$-neutrino sector $f\mapsto\nu\Rightarrow(\nu_e,\nu_\mu,\nu_\tau)$ 
and the charged $\ell$-lepton sector  
$f\mapsto\ell\Rightarrow(e,\mu,\tau)$.
As a result, all lepton fields are mass eigenstates, 
the four-fermion operators (\ref{l}) are ``diagonal'' only for each lepton family 
without family-mixing, 
\begin{eqnarray}
-G\, \sum_{f=1,2,3}\left[\bar\ell^{f}_{_L}\ell^{f}_{_R}\bar\ell^{f}_{_R} \ell^{f}_{_L}+
(\bar\ell^{f}_L\nu^{f}_{R})(\bar \nu^{f}_{R}\ell^f_{L})+ 
(\bar\nu^{f\, c}_R\nu^{f\,}_{R})(\bar \nu^{f}_{R}\nu^{f c}_{R})
\right],
\label{l1}
\end{eqnarray}
and the vacuum expectation values of two-lepton operators 
$\langle\ell^{f}_{_R} \bar\ell^{f}_{_L}\rangle+{\rm h.c.}$, 
$\langle\nu^{f}_L\bar\nu^{f}_{R}\rangle+{\rm h.c.}$ 
and $\langle \nu^{f\, c}_R\bar\nu^{f\,}_{R}\rangle+{\rm h.c.}$, i.e., 
lepton-mass matrices 
are diagonalized in the fermion-family space by the biunitary transformations
\begin{eqnarray}
M^\ell\Rightarrow M^\ell_{\rm diag} &=&(m^e_1, m^\mu_2,m^\tau_3)= {\mathcal U}^{\ell\dagger}_L M^\ell {\mathcal U}^\ell_R,
\label{ml}
\end{eqnarray}
and
\begin{eqnarray}
M^\nu\Rightarrow M^\nu_{\rm diag}&=&(m^{\nu_e}_1, m^{\nu_\mu}_2,
m^{\nu_\tau}_3)= {\mathcal U}^{\nu\dagger}_L M^\nu {\mathcal U}^\nu_R,
\label{mlnu}\\
M\Rightarrow M_{\rm diag} &=&(m^M_1, m^M_2,m^M_3)=({\mathcal U}^{\nu\dagger}_R)^* M {\mathcal U}^\nu_R= 
\tilde {\mathcal U}^{\nu}_R M {\mathcal U}^\nu_R,
\label{mm}
\end{eqnarray}
where all lepton masses (eigenvalues) are positive. 
The Dirac neutrino mass matrix can be expressed as 
$
M^\nu={\mathcal H}^\nu{\mathcal V}^{\nu},
\label{nmmatrix}
$ 
and ${\mathcal H}^{\nu}$ is a 
hermitian matrix. The ${\mathcal U}^{\nu}_L$ and ${\mathcal U}^{\nu}_R$ are related by
\begin{eqnarray}
{\mathcal U}^{\nu}_L={\mathcal V}^{\nu} {\mathcal U}^{\nu}_R,
\label{rlr}  
\end{eqnarray}
and ${\mathcal V}^{\nu}$ is an unitary matrix. This also applies for charged lepton sector 
$(\nu\rightarrow \ell)$, see \cite{zli,mohapatra}.
In the following sections, we adopt the bases of mass-eigenstates %($f,f'=1,2,3$)
and drop the subscriptions $1,2,3$ for simplifying the notations in $M^{u,d}_{\rm diag}$ 
(\ref{mqu},\ref{mqd}), $M^{\nu,\ell}_{\rm diag}$ (\ref{mlnu},\ref{ml}) and 
$M_{\rm diag}$ (\ref{mm}).  

Using unitary matrices ${\mathcal U}^\nu_{L,R}$ (\ref{clnu}) 
and ${\mathcal U}^\ell_{L,R}$ (\ref{cl}), up to a phase we define the unitary lepton-family mixing matrices,
\begin{eqnarray}
{\mathcal U}^{\nu\dagger}_L{\mathcal U}^{\ell}_L, &\quad&  %e^{i\varphi_1}
{\mathcal U}^{\nu\dagger}_L{\mathcal U}^\ell_R,\nonumber\\
%e^{i\varphi_2}
{\mathcal U}^{\nu\dagger}_R{\mathcal U}^\ell_L, &\quad&%~~~~~
{\mathcal U}^{\nu\dagger}_R{\mathcal U}^{\ell}_R.
\label{mmnul}
\end{eqnarray}
where the first element is the PMNS matrix $U^\ell=U^\ell_L\equiv 
{\mathcal U}_L^{\nu\dagger}{\mathcal U}_L^\ell$. We adopt the most recent updated 
range \cite{PMNS} of PMNS matrix elements to calculate the fermion spectrum in this article.
We can also define the notation for 
the last element
\begin{eqnarray}
U^\ell_R\equiv {\mathcal U}_R^{\nu\dagger}{\mathcal U}_R^\ell,
\label{rtan}
\end{eqnarray}
that will be used later.
\comment{$e^{i\varphi_1}$ is a relative phase between ${\mathcal U}^{\nu\dagger}_L$ and 
${\mathcal U}^\ell_R$, and $e^{i\varphi_2}$ is another relative 
phase between ${\mathcal U}^{\nu\dagger}_R$ and ${\mathcal U}^\ell_L$.
Analogously to Eq.~(\ref{mmud}), 
Eq.~(\ref{mmnul}) is a $6\times 6$ mixing matrix,
which is unitary, if the relative phases $\varphi_2-\varphi_1=n\pi, ~n=1,2,3$.}
Note that each of the unitary matrices ${\mathcal U}^{\nu,\ell,u, d}_L$ 
in Eqs.~(\ref{cqu},\ref{cqd}) 
and (\ref{clnu},\ref{cl}) 
is unique up to a diagonal phase 
matrix $P^{\nu,\ell,u, d}_{\rm diag}=(e^{i\phi_1},e^{i\phi_2},e^{i\phi_3})$ \cite{zli}. These 
phase degrees of freedom are used here to ensure all mass eigenvalues are positive, and we do not
consider the question of CP-violation at the moment.
% and mixing angles.

The Majorana mass matrix $M$ (\ref{mm}) is a symmetric matrix, 
relating to the vacuum expectation value of 
two fermion operator $(\bar\nu^{f\, c}_R\nu^{f\,}_{R})$. 
%and ${\mathcal U}^\nu_R$ is a real and orthogonal matrix. 
Using (\ref{rlr}), Eq.~(\ref{mm}) can be rewritted as,
\begin{eqnarray}
M_{\rm diag} &=&({\mathcal U}^{\nu\dagger}_L {\mathcal V}^{\nu})^* M {\mathcal U}^\nu_R=({\mathcal U}^{\nu\dagger}_L {\mathcal V}^{\nu} M {\mathcal U}^\nu_R)^*={\mathcal U}^{\nu\dagger}_L ({\mathcal V}^{\nu} M) {\mathcal U}^\nu_R,
\label{mm1}
\end{eqnarray}
where in the last equality we assume the CP-conservation 
for Majorana fields $\nu^{fc}_R$ and $\nu^{f}_R$ so that their matrix $M=M^*$ 
and transformation $({\mathcal U}^\nu_R)^*={\mathcal U}^\nu_R$ are real.  
Comparing Eq.~(\ref{mlnu}) 
to Eq.~(\ref{mm1}), we find that the Dirac neutrino mass matrix 
$M^\nu=H^\nu{\mathcal V}^{\nu}$ (\ref{mlnu})
and the matrix ${\mathcal V}^{\nu} M$ (\ref{mm1}) are diagonalized by the same biunitary transformation, and they have common eigen-vectors. In fact,
both mass matrices are related to the 
$\bar\nu^f_R$-field condensation, i.e., the Dirac mass matrix $M^\nu \sim 
\langle\nu^{f}_L\bar\nu^{f}_{R}\rangle$ 
and the Majorana  mass matrix $M \sim \langle \nu^{f\, c}_R\bar\nu^{f\,}_{R}\rangle$. Therefore, 
we expect that they should have a similar structure of eigenvalues, for example 
the normal hierarchy structure, $m^{\nu_e}_1< m^{\nu_\mu}_2 < m^{\nu_\tau}_3$ in Eq.~(\ref{mlnu}) 
and $m^{M}_1< m^{M}_2 < m^{M}_3$ in Eq.~(\ref{mm}). We will present detail discussions
on the first, second and third four-fermion operators involving $\nu_R^f$ 
in Eq.~(\ref{l1}), as well as the Dirac mass matrix $M^\nu$ (\ref{mlnu}),
the Majorana mass matrix $M$ (\ref{mm}) and mixing matrix (\ref{mmnul}) in the last section 
specially for neutrinos. 

\subsection{Quark-lepton interaction sector}

Using the same chiral transformations (\ref{cqu}), (\ref{cqd}),
(\ref{clnu}) and (\ref{cl}) in quark and lepton sectors, 
we obtain that in the fermion-family space 
the four-fermion operators (\ref{bhlqlm}) are 
``diagonal'' ($f=f'$), and we rewrite these operators 
in terms of Dirac mass eigenstates  
\begin{eqnarray}
{\rm Eq}.~(\ref{bhlqlm})&=&G\sum_{ff'}\Big\{[(\bar\ell^{i}_L{\mathcal U}^{e\dagger}_L)^f({\mathcal U}^{e}_Re_{R})^{f'}]
[(\bar d^{a}_{R}{\mathcal U}^{d\dagger}_R)^{f'}({\mathcal U}^{d}_L\psi_{Lia})^f]\nonumber\\
&+&[(\bar\ell^{i}_L{\mathcal U}^{\nu\dagger}_L)^f({\mathcal U}^{\nu}_R\nu_{eR})^{f'}]
[(\bar u^{a}_{R}{\mathcal U}^{u\dagger}_R)^{f'}({\mathcal U}^{u}_L\psi_{Lia})^f]
\Big\}\label{bhlqlm1}\\
&=&G\sum_{ff'}\Big\{[\bar\ell^{if}_L(U_R^{de}e_{R})^{f'}]
[\bar d^{af'}_{R}(U^{ed}_L\psi_{Lia})^f]\nonumber\\
&+&[\bar\ell^{if}_L(U_R^{u\nu}\nu_{eR})^{f'}]
[\bar u^{af'}_{R}(U_L^{\nu u}\psi_{Lia})^f]\Big\} .
\label{bhlqlm2}
\end{eqnarray}
where $f=f'$ and four unitary mixing matrices between lepton and quark families 
are defined by
\begin{eqnarray}
U_R^{de}&=&{\mathcal U}_R^{d\dagger}{\mathcal U}_R^e,\quad
U_L^{ed}={\mathcal U}_L^{e\dagger}{\mathcal U}_L^d,
\nonumber\\
U_R^{u\nu}&=&{\mathcal U}_R^{u\dagger}{\mathcal U}_R^{\nu},\quad
U_L^{\nu u}={\mathcal U}_L^{\nu\dagger}{\mathcal U}_L^u,
\label{mql1}
\end{eqnarray}
analogously to the mixing matrices (\ref{mmud}) in the quark sector and (\ref{mmnul}) in the 
lepton sector. Relating to the $U_L^{d}$ ($U_L^{u}$) 
in the CKM matrix $U_L^{u\dagger}U_L^{d}$ , the matrix $U_L^{ed}$ 
($U_L^{\nu u}$) is expected to have a hierarchy structure, namely, in the fermion-family space 
the diagonal elements are the order of unit, while the off-diagonal elements are much smaller than  the order of unit.   

Equations (\ref{mmud}), (\ref{mmnul}), and (\ref{mql1}) give the mixing matrices 
of mass and gauge eigenstates of three fermion families, due to the 
$W^\pm$-boson interaction and four-fermion interactions (\ref{art}).  
The elements of these unitary matrices are not completely independent each other, 
as we have already known from the CKM and PMNS matrices. As will be shown, 
these mixing matrices and mass spectra of the SM fermions are fundamental, and 
closely related. 

Henceforth, all fermion fields are mass eigenstates, two-fermion mass operators and 
four-fermion operators are ``diagonal'' in the fermion-family space.  

\comment{NPB To end this Section, it is worthwhile to point out that the 
unitary mixing matrices (\ref{mmud}), (\ref{mmnul}) and (\ref{mql1}) are necessarily relevant to the two-fermion mass operators, 
due to the presence of
three-family four-fermion operators of quark sector (\ref{q}), 
lepton sector (\ref{l}) and quark-lepton sector (\ref{bhlqlm1}).  
Recall that only the CKM matrix in Eq.~(\ref{mmud}) 
and the PMNS matrix in Eq.~(\ref{mmnul}) relating 
to the $W^\pm$-interacting vertex are discussed in the literature. 
As the main results, we will show below that
the unitary mixing matrices (\ref{mmud}), (\ref{mmnul}) 
and (\ref{mql1}) play essential roles in the twelve SD equations for 
fermion masses and the resultant hierarchy spectrum, i.e.,
play much more important roles than the CKM matrix or the PMNS matrix. 
The latter is relevant only for the light-quark 
sector or the light-lepton sector.}

%\vskip0.1cm
\section
{\bf Spontaneous symmetry breaking}\label{SSBS}
\hskip0.1cm In this section, we briefly recall and discuss that in the IR-domain of 
the IR-stable fixed point $G_c$, the relevant four-fermion operator (\ref{bhlx})
undergoes the SSB and 
becomes an effectively bilinear and renormalizable Lagrangian that follows 
the RG-equations to approach the SM physics in the low-energy. This is necessary and fundamental for studying the origin of SM fermion masses in this article.   

\subsection{The IR fixed-point domain and only top-quark mass generated via the SSB %relevant operators
}\label{onlyt}

Apart from what is possible new physics at the scale $\Lambda$ explaining the 
origin of these effective four-fermion operators (\ref{art1}), 
it is essential and necessary to study: (i) which dynamics of these operators 
undergo in terms of their couplings as functions of running energy 
scale $\mu$; (ii) associating to these dynamics where the infrared (IR) 
or ultraviolet (UV) stable fixed point of physical couplings locates; 
(iii) in the domains (scaling regions) of these stable fixed points, 
which physically relevant operators that 
become effectively dimensional-4 renormalizable 
operators following RG equations (scaling laws), 
while other irrelevant operators are suppressed by the cutoff at least 
$\mathcal O(\Lambda^{-2})$.    

%\vskip0.1cm
%\noindent
%{\bf Relevant and irrelevant four-fermion operators in the IR- and UV-domain.}
%\hskip0.1cm 
In the IR-domain of the IR-stable fixed point $G_c$,
the four-fermion operator (\ref{bhl})
%$G(\bar\psi^{ia}_Lt_{Ra})(\bar t^b_{R}\psi_{Lib})$   
was shown \cite{bhl1990} to become physically relevant and 
renormalizable operators of effective dimension-4, due to 
the SSB dynamics of NJL-type. Namely, the Lagrangian (\ref{bhl}) becomes 
the effective SM Lagrangian with {\it bilinear} top-quark mass term and 
Yukawa-coupling to the composite Higgs boson $H$, 
which obeys the RG-equations approaching to
the low-energy SM physics characterized by the energy 
scale $v\approx 239.5$ GeV. In addition, the top-quark and composite 
Higgs-boson masses are correctly 
obtained by solving RG-equations with the 
appropriate non-vanishing form-factor of the Higgs 
boson in TeV scales \cite{xue2014,xue2013}. 
%The composite Higgs boson behaves 
%as an elementary particle, as long as its form-factor 
%(wave-function renormalization) is finite.

It seems that via the SSB dynamics the four-fermion operator 
%(\ref{bhlx}) or 
(\ref{q1}) leads to 
the quark-condensation $M^q_{ff'} =-G\langle \bar \psi^f \psi^{f'}\rangle/2N_c=m\delta_{ff'}\not=0$ (the color number $N_c$), 
and two diagonal mass matrices $M^{u}_{\rm diag}=(m^u_1, m^c_2,m^t_3)$
and $M^{d}_{\rm diag}=(m^d_1, m^s_2,m^b_3)$ of quark sectors $q=2/3$ and $q=-1/3$ satisfying $3+3$ mass-gap equations of NJL type.  
%It seems that nontrivial 
%quark matrices $M^{u}_{\rm diag}=(m^u_1, m^c_2,m^t_3)$
%and $M^{d}_{\rm diag}=(m^d_1, m^s_2,m^b_3)$ are resulted. 
%These mass-gap equations, contributed by the tadpole diagram Fig.~\ref{figt}, 
%are actually the simplest SD equation for fermion self-energy 
%functions $\Sigma_f(p)$, neglecting gauge interactions. 
It was demonstrated \cite{xue2013_1} that as an energetically favorable 
solution of the SSB ground state of the SM, 
only top quark is massive ($m^{\rm sb}_t=- G\langle \bar\psi_t \psi_t\rangle\not=0$), 
otherwise there would be more Goldstone modes 
in addition to those become the longitudinal modes of massive gauge bosons $W^\pm$ and $Z^0$. 
Extra Goldstone modes have positive contributions to the ground-state energy, and thus make the ground-state energy increase. As a result, $M^{u}_{\rm diag}=(0, 0,m^t_3)$
and $M^{d}_{\rm diag}=(0, 0,0)$, only the top-quark channel (\ref{bhl}) undergoes the SSB dynamics and  
becomes relevant operator following the RG equations 
in the IR domain.

We turn to the lepton sector. The first and second   
four-fermion operators in Eq.~(\ref{bhlxl}) or (\ref{l1}) relate to the 
lepton Dirac mass matrix.
At first glance, it seems that the four-lepton operators
undergo the SSB leading to
the lepton-condensation 
$M^\ell_{ff'} =-G\langle \bar \ell^f \ell^{f'}\rangle/2=m_\ell\delta_{ff'}$, and
two diagonal mass matrices 
$M^{\nu}_{\rm diag}=(m^{\nu_e}_1, m^{\nu_\mu}_2,m^{\nu_\tau}_3)$
and $M^{\ell}_{\rm diag}=(m^e_1, m^\mu_2,m^\tau_3)$ 
of the lepton sector ($q=0$ and $q=-1$) 
satisfying $3+3$ mass-gap equations of NJL type. 
Actually, the first and second   
four-fermion operators in Eq.~(\ref{bhlxl}) or (\ref{l1}) do not 
undergo the SSB and two lepton Dirac mass matrices ($q=0$ and $q=-1$) are 
zero matrices, i.e., $M^{\nu}_{\rm diag}=(0, 0,0)$
and $M^{\ell}_{\rm diag}=(0, 0,0)$. The reason is that
the effective four-lepton coupling $(GN_c)/N_c$ is $N_c$-smaller 
than the critical value $(GN_c)$ of the effective four-quark 
coupling for the SSB in the quark sector, 
in addition to the reason of energetically favorable 
solution for the SSB ground state discussed above. 
 
Therefore, in the IR-domain where the SSB occurs, except the top quark, 
all quarks and leptons are massless and their four-fermion operators 
(\ref{q1}) and (\ref{l1}), as well as repulsive 
four-fermion operators (\ref{art0}), 
are irrelevant dimension-6 operators. 
Their tree-level amplitudes of four-fermion scatterings are 
suppressed ${\mathcal O}(\Lambda^{-2})$, thus such deviations 
from the SM are experimentally inaccessible today \cite{xue2015}.  

The heaviest quark which acquires its mass via the SSB is identified and named as the top
quark. The heaviest fermion family is named as the third fermion family
of fermions $\nu_\tau, \tau, t, b$, where the top quark is. We study their mass spectra
in Ref.~\cite{xue2016}. As will be discussed, these third-family quarks 
and leptons are grouped together for their heavy masses, due to the fermions
$\nu_\tau, \tau, b$ have the largest mixing with the top quark.  

\subsection
{The $\langle\bar t t\rangle$-condensate model}\label{SSBSt}
\hskip0.1cm 
We briefly recall the BHL $\langle\bar t t\rangle$-condensate model 
 \cite{bhl1990} for the full effective Lagrangian of the low-energy SM 
in the IR-domain, and the analysis \cite{xue2013,xue2014} of RG equations based on experimental boundary conditions, as well as experimental indications of the composite Higgs boson.  
\comment{NPB In this section, briefly recalling the BHL $\langle\bar t t\rangle$-condensate model 
 \cite{bhl1990} for the full effective Lagrangian of the low-energy SM 
in the scaling region (IR-domain) of the IR-fixed point, 
we explain why our solution is radically different from the BHL one though the 
same renormalization procedure and RG-equations are adopted 
in the IR-domain. It is important to compare our solution with the BHL one, as well as 
discuss its difference from the elementary Higgs model. 
}
  
%\vskip0.1cm
\subsubsection
{The scaling region of the IR-stable fixed point and BHL analysis}
\hskip0.1cm 
Using the approach of large $N_c$-expansion with a fixed value 
$GN_c$, %NPB to involve the most of fermion loops (since each loop provides a factor of $N_c$),
%dominate Feynman diagrams are those involving the most of fermion loops, since each loop provides a factor of $N_c$,   
it is shown \cite{bhl1990} that the top-quark channel of 
operators (\ref{bhlx}) undergoes the SSB dynamics in the IR-domain 
of IR-stable fixed point $G_c$, leading to the generation of top-quark mass  
\begin{eqnarray}
m_t&=&-(1/2N_c)G\sum_a\langle \bar t^a t_a\rangle
=-(G/N_c)\sum_a\langle \bar t^a_L t_{aR}\rangle
\label{tqmassd}
\end{eqnarray}
by the $\langle \bar t t \rangle$-condensate.
As a result, the $\Lambda^2$-divergence 
(tadpole-diagram) is removed by the mass gap-equation, the 
top-quark channel of four-fermion operator (\ref{bhl}) 
becomes physically relevant and 
renormalizable operators of effective dimension-4. Namely, 
the effective SM Lagrangian with the {\it bilinear} top-quark mass term and 
Yukawa coupling to the composite Higgs boson $H$ at the low-energy 
scale $\mu$ is given by \cite{bhl1990}
\begin{eqnarray}
L &=& L_{\rm kinetic} + g_{t0}(\bar \Psi_L t_RH+ {\rm h.c.})
+\Delta L_{\rm gauge}\nonumber\\ 
&+& Z_H|D_\mu H|^2-m_{_H}^2H^\dagger H
-\frac{\lambda_0}{2}(H^\dagger H)^2,
\label{eff}
\end{eqnarray}
%\begin{eqnarray}
%L_{\rm eff} &=& L_{\rm kinetic} + L_{\rm gauge} + 
%m_t \bar t t+ \bar g_{t}\bar t t H \nonumber\\ 
%&+& \tilde Z_H|D_\mu H|^2-m_{_H}^2H^\dagger H
%-\frac{\tilde \lambda}{2}(H^\dagger H)^2,
%\label{eff}
%\end{eqnarray} 
all renormalized quantities received fermion-loop contributions are 
defined with respect to the low-energy scale $\mu$. 
%At high-energy scale $\E$ we have a finite coupling $G=g^2_{t0}/M^2_H$ and $M_H=\E$. 
The conventional renormalization $Z_\psi=1$ for fundamental 
fermions and the unconventional wave-function renormalization (form factor)
$\tilde Z_H$ for the composite Higgs boson are 
adopted
\begin{equation}
\tilde Z_{H}(\mu)=\frac{1}{\bar g^2_t(\mu)},\, \bar g_t(\mu)=\frac{Z_{HY}}{Z_H^{1/2}}g_{t0}; \quad \tilde \lambda(\mu)=\frac{\bar\lambda(\mu)}{\bar g^4_t(\mu)},\,\bar\lambda(\mu)=\frac{Z_{4H}}{Z_H^2}\lambda_0,
\label{boun0}
\end{equation}
where $Z_{HY}$ and $Z_{4H}$ are proper renormalization constants of 
the Yukawa coupling and quartic coupling in Eq.~(\ref{eff}). 
The SSB-generated top-quark mass $m_t(\mu)=\bar g_t^2(\mu)v/\sqrt{2}$. 
The composite Higgs-boson is described by its pole-mass 
$m^2_H(\mu)=2\tilde \lambda(\mu) v^2$, form-factor $\tilde Z_H(\mu)=1/\bar g_t^2(\mu)$, 
and effective quartic coupling $\tilde\lambda(\mu)$, provided that 
$\tilde Z_H(\mu)>0$ and $\tilde\lambda(\mu)>0$ are obeyed. After the proper wave-function 
renormalization $\tilde Z_H(\mu)$, the Higgs boson behaves 
as an elementary particle, as long as $\tilde Z_H(\mu)\not=0$ is finite.

In the IR-domain 
where the SM 
of particle physics is realized, %we utilize 
the full one-loop RG equations for running couplings $\bar g_t(\mu^2)$ and $\bar \lambda(\mu^2)$
read
\begin{eqnarray}
16\pi^2\frac{d\bar g_t}{dt} &=&\left(\frac{9}{2}\bar g_t^2-8 \bar g^2_3 - \frac{9}{4}\bar g^2_2 -\frac{17}{12}\bar g^2_1 \right)\bar g_t,
\label{reg1}\\
16\pi^2\frac{d\bar \lambda}{dt} &=&12\left[\bar\lambda^2+(\bar g_t^2-A)\bar\lambda + B -\bar g^4_t \right],\quad t=\ln\mu \label{reg2}
\end{eqnarray}
where one can find $A$, $B$ and RG equations for 
running gauge couplings $\bar g^2_{1,2,3}$ in Eqs.~(4.7), (4.8) of 
Ref.~\cite{bhl1990}. The solutions to these ordinary differential equations are uniquely 
determined, once the boundary conditions are fixed.
\comment{NPB In 1990, when the top-quark and Higgs masses were unknown, {\it using 
the composite conditions $\tilde Z_H=0$ and $\tilde\lambda=0$ as the 
boundary conditions at the cutoff} $\Lambda$, 
the analysis of the RG equations (\ref{reg1}) and (\ref{reg2}) for $\tilde Z_H(\mu)$ 
and $\tilde\lambda(\mu)$ %approaching to the low-energy SM physics 
was made to calculate the top-quark and Higgs-boson 
masses by varying the values of cutoff $\Lambda$. It was found that the cutoff $\Lambda$
varies from $10^{4}$ to $10^{19}$ GeV, the obtained top-quark and Higgs-boson masses are
larger than $200$ GeV.  
}

\begin{figure}%[!h]
\begin{center}
\includegraphics[height=1.40in]{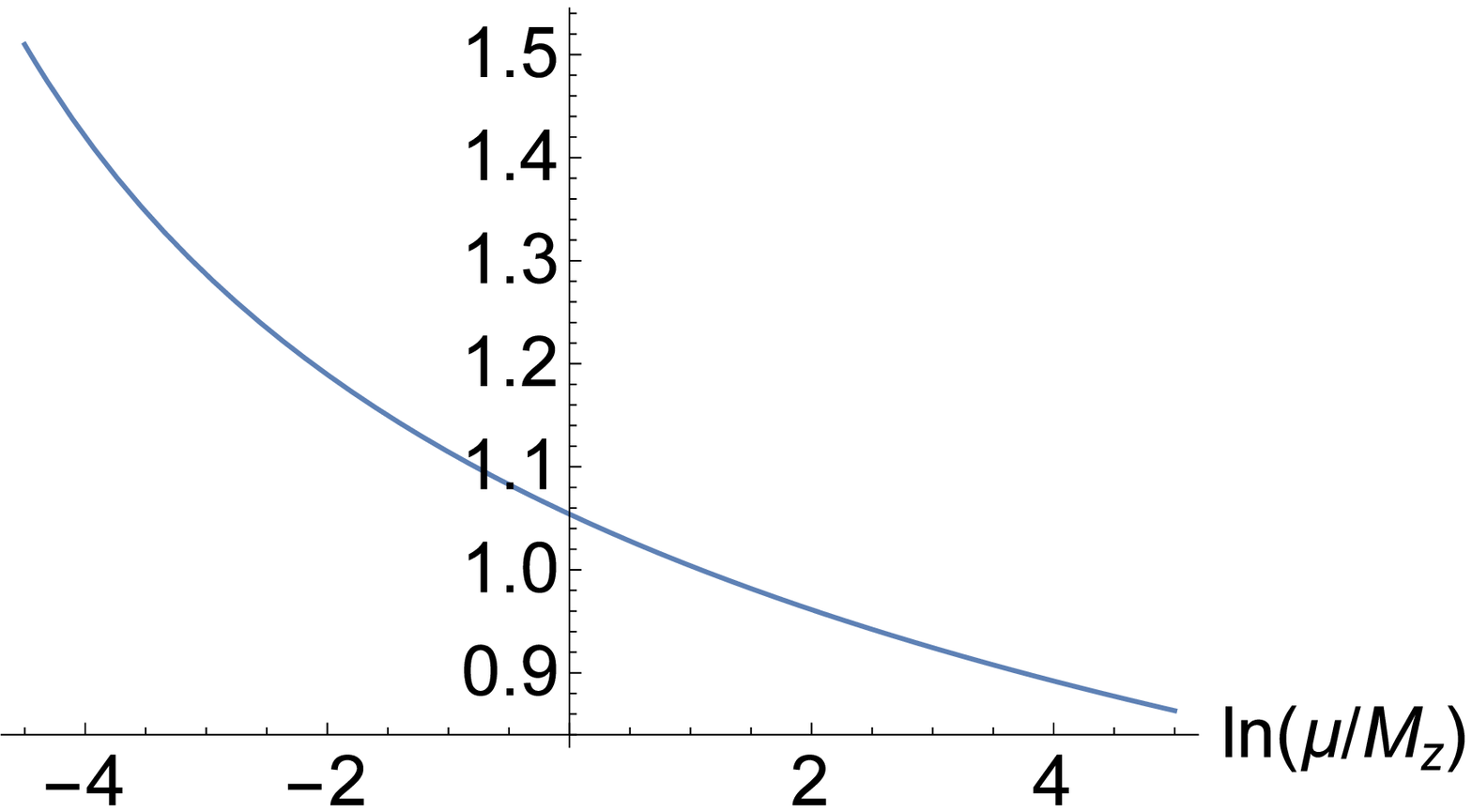}
\includegraphics[height=1.40in]{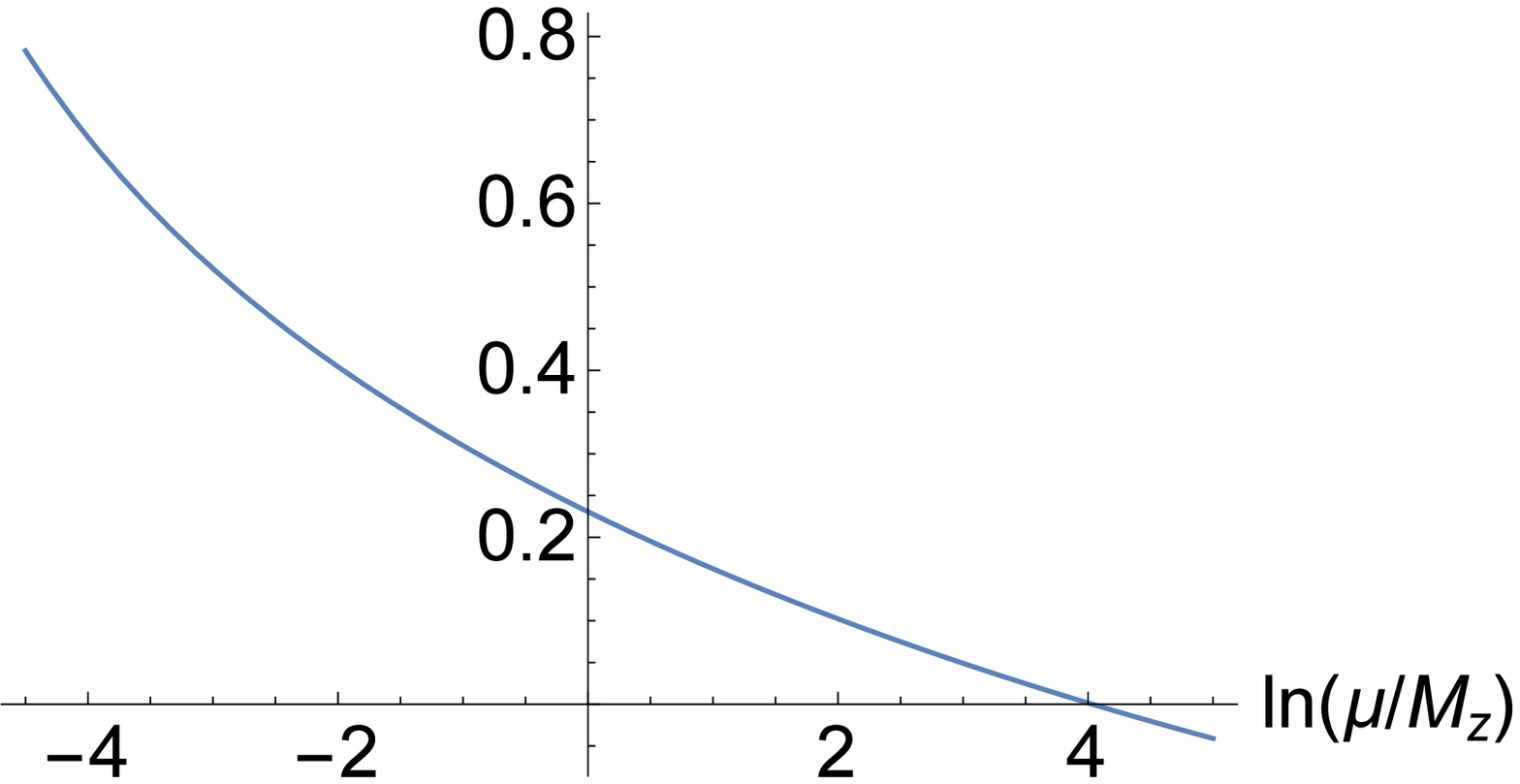}
\put(-120,105){\footnotesize $\tilde\lambda(\mu)$}
\put(-320,105){\footnotesize $\bar g_t(\mu)$}
%\put(-180,95){\footnotesize $\bar\lambda(\mu)$}
\caption{Using
experimentally measured SM quantities (including $m_t$ and $m_{_H}$) 
as boundary values, we uniquely solve the RG equations
for the composite Higgs-boson model \cite{bhl1990}, 
we find \cite{xue2013,xue2014} the effective 
top-quark Yukawa coupling $\bar g_t(\mu)$ (left) 
and effective Higgs quartic coupling $\tilde \lambda(\mu)$ (right) in the range
$1.0 ~{\rm GeV}\lesssim \mu \lesssim 13.5~ {\rm TeV}$.
%$1.01 ~{\rm GeV}\lesssim \mu \lesssim 13.53~ {\rm TeV}$.  
Note that $\tilde\lambda(\E)=0$ at $\E\approx 5.14~{\rm TeV}$ and $\tilde\lambda(\mu)<0$ 
for $\mu >\E$. 
%indicates an instability or the quadratic Higgs-boson term  
%$m^2_H/2=\tilde\lambda(m_H) v^2< 0$ indicates 
%the phase transition to a symmetric phase. 
} \label{figyt}
\end{center}
\end{figure}

In Ref.~\cite{xue2013,xue2014}, we analyzed the RG equations 
(\ref{reg1}) and (\ref{reg2}) by using the boundary conditions 
based on the experimental values of top-quark and Higgs-boson masses, $m_t\approx 173$ GeV and $m_H\approx 126$ GeV, i.e., 
the mass-shell conditions 
\begin{eqnarray}
m_t(m_t)=\bar g_t^2(m_t)v/\sqrt{2}\approx 173 {\rm GeV},
\quad m_H(m_H)=[2\tilde \lambda(m_H)]^{1/2} v\approx 126 {\rm GeV}
\label{thshell}
\end{eqnarray}
to determine the solutions for $\tilde Z_H(\mu)$ and $\tilde\lambda(\mu)$
in the IR-domain of the energy scale $v=239.5$ GeV. As a result, we 
obtained the unique solution (see Fig.~\ref{figyt}) for the composite Higgs-boson model (\ref{bhl}) or (\ref{eff}) as well as at the energy scale $\E$  
\begin{eqnarray}
\E \approx 5.1\,\, {\rm TeV},\quad \tilde Z_H \approx 1.26,\quad \tilde\lambda(\E)=0.
\label{thvari}
\end{eqnarray}
%and effective quartic coupling vanishes $\tilde\lambda(\E)=0$. 
More detailed discussions can be found in Ref.~\cite{xue2016}.
The interested readers are referred to Ref.~\cite{xue2013} for 
the resolution to drastically fine-tuning problem.

\subsubsection
{Experimental indications of composite Higgs boson ?}\label{expH}
\hskip0.1cm
To end this section, we discuss the experimental indications 
of composite Higgs boson. 
In the IR-domain, 
the dynamical symmetry breaking of four-fermion 
operator $G(\bar\psi^{ia}_Lt_{Ra})(\bar t^b_{R}\psi_{Lib})$ of 
the top-quark channel (\ref{bhlx}) accounts for the masses of 
top quark, $W$ and $Z$ bosons as well as a Higgs boson composed 
by a top-quark pair ($\bar t t$) \cite{bhl1990}. 
It is shown \cite{xue2013,xue2014} that   
this mechanism consistently gives rise to the top-quark 
and Higgs masses, 
provided the appropriate value of non-vanishing form-factor 
of composite Higgs boson at the high-energy scale 
$\E\gtrsim 5\,$ TeV.

Due to its finite form factor (\ref{thvari}), 
the composite Higgs boson behaves as if an elementary Higgs particle, 
the deviation from the SM is too small to be identified by 
the low-energy collider signatures at the present level \cite{xue2014}. More detailed analysis of the composite Higgs boson phenomenology 
is indeed needed. It deserves another lengthy article for this issue, 
nevertheless we present a brief discussion on this aspect.
The non-vanishing form-factor
$\tilde Z_H(\mu)$ means that after conventional wave-function and vertex 
renormalizations $Z^{1/2}_H H\rightarrow H$, 
$Z_{HY}g_{t0}\rightarrow g_{t0}$ and 
$Z_{4H}\lambda_{0}\rightarrow \lambda_{0}$ 
[see Eqs.~(\ref{eff}) and (\ref{boun0})], 
the composite Higgs boson behaves as an elementary particle. 
The non-vanishing form-factor of composite Higgs boson 
is in fact related to the effective 
Yukawa-coupling of Higgs boson and top quark, i.e., 
$\tilde Z^{-1/2}_H(\mu)=\bar g_t(\mu)$ of Eq.~(\ref{eff}). 
The effective Yukawa coupling $\bar g_t(\mu)$
%\in [0.96, 0.89]$ for $\mu\in [m_t, \E]$
and quartic coupling $\tilde\lambda(\mu)$ 
%\in [0.15, 0.0]$ for $\mu\in [m_{_H}, \E]$
monotonically decrease with the energy scale $\mu$ increasing 
in the range $m_{_H}< \mu <\E\approx 5$ TeV (see Fig.~\ref{figyl}). 
This means that the composite Higgs boson becomes more tightly bound as 
the  the energy scale $\mu$ increases.

On the other hand, that the 
effective Yukawa coupling $\bar g_t(\mu)$
%\in [0.96, 0.89]$ for $\mu\in [m_t, \E]$
and quartic coupling $\tilde\lambda(\mu)$ 
%\in [0.15, 0.0]$ for $\mu\in [m_{_H}, \E]$
decrease as the energy scale $\mu$ increases 
in the range $m_{_H}< \mu <\E$ implies 
some effects on the rates or cross-sections 
of the following three dominate processes of Higgs-boson production 
and decay \cite{ATLAS,CMS} or 
other relevant processes. Two-gluon fusion produces a Higgs boson via a top-quark loop, 
which is proportional to the effective Yukawa coupling $\bar g_t(\mu)$. Then,  
the produced Higgs boson decays into the two-photon state by coupling to a top-quark loop,
and into the four-lepton state by coupling to two massive $W$-bosons or two massive $Z$-bosons.
Due to the $\bar t\,t$-composite nature of Higgs boson, the one-particle-irreducible (1PI) 
vertexes of Higgs-boson coupling to a top-quark loop, two massive $W$-bosons 
or two massive $Z$-bosons are proportional to the effective 
Yukawa coupling $\bar g_t(\mu)$. 
%The effective Yukawa coupling $\bar g^2_t(\mu)$
%\in [0.96, 0.89]$ for $\mu\in [m_t, \E]$ and quartic coupling $\bar\lambda(\mu)$ 
%\in [0.15, 0.0]$ for $\mu\in [m_{_H}, \E]$.  
As a result, both the Higgs-boson decaying rate to each of these three channels and 
total decay rate are proportional to $\bar g^2_t(\mu)$, 
which does not affect on the branching ratio of each Higgs-decay channel. 
The energy scale $\mu$ is actually the 
Higgs-boson energy, representing the total energy of final states, 
e.g., two-photon state and four-lepton states, into which the produced Higgs boson decays. 

These discussions imply that the resonant 
amplitude (number of events) of two-photon invariant 
mass $m_{\gamma\gamma}\approx 126$ GeV and/or 
four-lepton invariant mass $m_{4l}\approx 126$ GeV is expected 
to become smaller as the produced Higgs-boson energy $\mu$ increases, i.e., the
energy of final two-photon and/or four-lepton states increases,   
when the CM energy $\sqrt{s}$ of LHC $p\,p$ collisions increases 
with a given luminosity. 
Suppose that the total decay rate or each channel 
decay rate of the SM Higgs boson is measured at the Higgs-boson energy $\mu=m_t$ 
and the SM value of Yukawa coupling 
$\bar g^2_t(m_t)=2m_t^2/v\approx 1.04$ (see Fig.~\ref{figyl}). 
In this scenario of composite Higgs boson, as the Higgs-boson energy $\mu$
increases to $\mu=2m_t$, the Yukawa coupling 
$\bar g^2_t(2m_t)\approx 0.98$ (see Fig.~\ref{figyl}), 
the variation of total decay rate or each channel 
decay rate is expected to be 
$6\%$ for $\Delta \bar g^2_t\approx 0.06$. Analogously, the
variation is expected to be $9\%$ at 
$\mu=3m_t$, $\bar g^2_t(3m_t)\approx 0.95$ or $11\%$ at $\mu=4m_t$, 
$\bar g^2_t(4m_t)\approx 0.93$ (see Fig.~\ref{figyl}).
These variations are still too small to be clearly 
distinguished by the present LHC experiments.  
Nevertheless, these effects are the nonresonant new signatures 
of low-energy collider that show the deviations of this scenario from the SM. We see that the induced (1PI) Yukawa couplings 
$\bar g_b(\mu)$ and $\bar g_\tau(\mu)$ \cite{xue2016}, as well as 
$\bar g_f(\mu)$ (the present article) of composite Higgs boson to the bottom-quark, tau-lepton and other fermions 
also weakly decrease with increasing Higgs-boson energy, 
this implies a slight decrease of
number of dilepton events in the Drell-Yan process. 

\section
{\bf Origins of explicit symmetry breaking}\label{esbS}
\hskip0.1cm
%\vskip0.1cm
We study in this section, once the top quark mass  
is generated by the SSB at the scale $\E$, other quarks and leptons acquire their masses 
by the explicit symmetry breaking (ESB), via both  
quark-lepton interactions (\ref{bhlqlm}) and fermion-family mixing. 
We henceforth indicate the SSB-generated top-quark  
mass $m^{\rm sb}_t$ and ESB-generated masses $m^{\rm eb}_f$ of other fermions, they
represent bare masses at the cutoff energy scale $\E$ of the symmetry breaking phase.

%Through mass-gap equations, the four-fermion operators (\ref{bhlqlm}) or (\ref{bhlqlm2}) 
%relate the lepton mass matrix (\ref{mlnu}) [(\ref{ml})] to the the quark mass matrix 
%(\ref{mqu})  [(\ref{mqd})], via explicit symmetry breaking discussed in the next section, 

\begin{figure}%[!h]
\begin{center}
\includegraphics[height=1.25in]{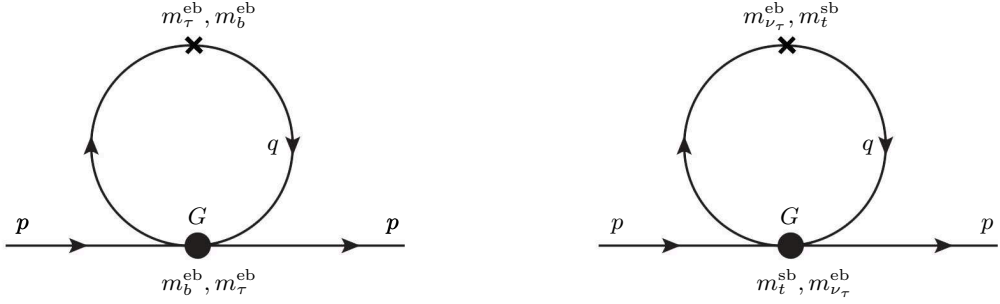}
\includegraphics[height=1.25in]{xue_fermi_5.eps}
\put(-85,45){\footnotesize $q$}
\put(-180,15){\footnotesize $p$}
\put(-40,15){\footnotesize $p$}
\put(-116,17){\footnotesize $G$}
\put(-130,93){\footnotesize $m^{\rm eb}_{\nu_\tau},m^{\rm sb}_t$}
\put(-125,-8){\footnotesize $m^{\rm sb}_t,m^{\rm eb}_{\nu_\tau}$}
\put(-310,45){\footnotesize $q$}
\put(-405,15){\footnotesize $p$}
\put(-265,15){\footnotesize $p$}
\put(-340,17){\footnotesize $G$}
\put(-350,93){\footnotesize $m^{\rm eb}_\tau,m^{\rm eb}_b$}
\put(-350,-8){\footnotesize $m^{\rm eb}_b,m^{\rm eb}_\tau$}
\put(-405,15){\footnotesize $p$}
\put(-265,15){\footnotesize $p$}
\caption{Using the third fermion family as an example, 
we show the tadpole diagrams of quark-lepton 
interactions (\ref{bhlql}) that contribute quark and lepton ESB masses $m^{\rm eb}$ 
to mass-gap equations or SD-equations (\ref{deq}-\ref{lb}). The 
mixing matrix element $U_L^{\tau b}U_R^{b\tau}$ or $U_L^{\tau b\dagger}U_R^{b\tau\dagger}$ associates to the interacting vertex $G$ in the left diagram. The mixing matrix element $U_L^{\nu_\tau t}U_R^{t \nu_\tau}$ or $U_L^{\nu_\tau t\dagger}U_R^{t \nu_\tau\dagger}$ associates to the interacting vertex $G$ in the right diagram. The mixing matrix elements 
with the first and second fermion family are neglected.
%$p$ is the external momentum and $q$ internal momentum integrated 
%up to $\E$ and the cross represents the fermion mass.
} \label{figt}
\end{center}
\end{figure}

\subsection{Quark-lepton interactions}\label{bare}

Once quarks acquire their bare masses $m^{\rm eb}_u$ and $m^{\rm eb}_d$, 
due to the ESB or the SSB for 
top quark only at the scale $\E$,
four-fermion operators (\ref{bhlqlm2}) 
contribute,   
via the tadpole diagram in Fig.~\ref{figt}, the bare mass terms 
$m^{\rm eb}_{\ell_\nu}$ and $m^{\rm eb}_\ell$ in mass-gap equations in
the lepton sector. {\it Vice versa} once leptons acquire their bare masses, 
via the same tadpole diagram in Fig.~\ref{figt},
four-fermion operators (\ref{bhlqlm2}) 
contribute the bare mass terms 
$m^{\rm eb}_u$ and $m^{\rm eb}_d$ in mass-gap equations 
in the quark sector. The superscript ``sb'' indicates the mass generated by the SSB.
The superscript ``eb'' indicates the mass generated by the ESB. These are bare fermion masses at the energy scale $\E$.  
As a result, from Eq.~(\ref{bhlqlm2}) 
we obtain the relationships between quark and lepton diagonal mass matrices, 
%up to the leading order in the large $N_f$-expansion 
%with $GN_f\E^2$ fixed, where $N_f$ is the number of fermions like color for example.
\begin{eqnarray}
[m^{\rm eb}_{e,\mu,\tau}]&=&(1/N_c)U_L^{ed}[m^{\rm eb}_{d,s,b}]U_R^{de},%\label{massql1}\\
%&=&(1/N_c){\mathcal U}_L^{e\dagger}M^{\rm eb}_{d,s,b}{\mathcal U}_R^{e}, \nonumber\\
%M^{\rm eb}_{d,s,b} &=& {\mathcal U}_L^{d} [m^{\rm eb}_{d,s,b}]{\mathcal U}_R^{d\dagger},\nonumber
%\end{eqnarray}
%and
%\begin{eqnarray}
\qquad
\big[m^{\rm eb}_{\nu_e,\nu_\mu,\nu_\tau}\big]=(1/N_c)U_L^{\nu u} [m^{\rm eb}_{u,c,t}] U_R^{u\nu}\label{massql2}
%&=&(1/N_c){\mathcal U}_L^{\nu_e\dagger}M^{\rm eb}_{u,c,t}{\mathcal U}_R^{\nu_e}, \nonumber\\
%M^{\rm eb}_{u,c,t} &=& {\mathcal U}_L^{u} [m^{\rm eb}_{u,c,t}]{\mathcal U}_R^{u\dagger},\nonumber
\end{eqnarray}
where the four diagonal matrices are
\begin{eqnarray}
[m^{\rm eb}_{e,\mu,\tau}]&\equiv &{\rm diag}(m^{\rm eb}_e,m^{\rm eb}_\mu,m^{\rm eb}_\tau),
\quad  
[m^{\rm eb}_{\nu_e,\nu_\mu,\nu_\tau}]\equiv {\rm diag}(m^{\rm eb}_{\nu_e},
m^{\rm eb}_{\nu_\mu},m^{\rm eb}_{\nu_\tau}),
\label{massld}\\
\big[m^{\rm eb}_{d,s,b}\big] &\equiv &{\rm diag}(m^{\rm eb}_d,m^{\rm eb}_s,m^{\rm eb}_b),\quad
[m^{\rm eb}_{u,c,t}]\equiv {\rm diag}(m^{\rm eb}_u,m^{\rm eb}_c,m^{\rm sb}_t),
\label{massqd}
\end{eqnarray} 
and their corresponding non-diagonal mass matrices are Eqs.~(\ref{mqu}), (\ref{mqd}), 
(\ref{mlnu}) and (\ref{ml}). The unitary quark-lepton mixing 
matrices (\ref{mql1}) make the transformations from 
lepton diagonal mass-matrices to quark diagonal mass-matrices, {\it vice versa}.  
%This is reminiscent of neutrino Dirac masses coinciding with the up-quark masses 
%in $SO(10)$ theories of unification of gauge interactions.
%$M^{\rm eb}_{e,\mu,\tau}$, 
%$M^{\rm eb}_{\nu_e,\nu_\mu,\tau}$, $M^{\rm eb}_{d,s,b}$ and $M^{\rm eb}_{u,c,t}$.

Apart from the SSB-generated top-quark mass $m^{\rm sb}_t$, all other fermion masses 
$m^{\rm eb}_f$ are ESB-generated and related to the top-quark mass $m^{\rm sb}_t$ by the mixing matrices (\ref{bhlqlm2}) or (\ref{mql1}). 
Analogously to Eq.~(\ref{tqmassd}) for the $\langle \bar t t\rangle$, in terms of two-fermion operators in mass eigenstates, we define
Dirac quark, lepton and neutrino bare masses at the energy scale $\E$, 
as well as Majorana mass $M$
\begin{eqnarray}
m^{\rm eb}_{\rm q}&=&-(1/2N_c)G\sum_a\langle \bar\psi^a \psi_a\rangle
=-G/N_c\sum_a\langle \bar\psi^a_L \psi_{aR}\rangle,\label{qmassd}\\
m^{\rm eb}_{\ell}&=&-(1/2)G\langle \bar\ell \ell\rangle
=-G\langle \bar\ell_L \ell_R\rangle,
\label{massd}\\
m^{\rm eb}_{\rm \nu^\ell} &=&-(1/2)G\langle \bar\nu^{\ell} \nu^{\ell}\rangle
=-G\langle \bar\nu^\ell_L \nu^\ell_R\rangle,
\label{massdnu}\\
m^{\rm M}_{\rm \nu} &=&
-G\sum_\ell\langle \bar\nu^{c\ell}_ R \nu^\ell_ R\rangle,
\label{mmassd}
\end{eqnarray}
where the color index $a$ is summed over in Eq.~(\ref{qmassd}) and the lepton-family index $\ell$ is summed over in Eq.~(\ref{mmassd}), whereas
in Eqs.~(\ref{massd}) and (\ref{massdnu}) $\ell=e,\mu,\tau$ respectively indicates each 
of three fermion families (mass eigenstates). In Eqs.~(\ref{qmassd}-\ref{mmassd}), 
the notation $\langle \cdot\cdot\cdot\rangle$ does 
not represent new SSB-condensates, but the 1PI functions of 
fermion mass operator $\bar\psi_L \psi_{R}$, i.e., 
the self-energy functions $\Sigma_f$ that satisfy the self-consistent SD 
equations or mass-gap equations.

We use the quark-lepton interaction of the third family as an example to show 
the quark-lepton interactions contribute to the SD-equations of fermion self-energy functions \cite{xue2016}. 
The quark-lepton interaction (\ref{bhlql}) of the third family reads
\begin{eqnarray}
G\left[(\bar\ell^{i}_L\tau_{R})(\bar b^a_{R}\psi_{Lia})
+(\bar\ell^{i}_L\nu^\tau_{R})(\bar t^a_{R}\psi_{Lia})\right],
\label{bhlqlt}
\end{eqnarray}
where $\ell^i_L=(\nu^\tau_L,\tau_L)$ and $\psi_{Lia}=(t_{La},b_{La})$. 
Once the top quark mass $m^{\rm sb}_t$ is generated by the SSB, 
the quark-lepton interactions (\ref{bhlqlt}) introduce the ESB terms
to the SD equations (mass-gap equations) for other fermions.

In order to show these ESB terms, we first approximate the SD equations to be self-consistent
mass gap-equations by neglecting perturbative gauge interactions and using the large $N_c$-expansion to the leading order, as indicated by  Fig.~\ref{figt}.
The quark-lepton interactions (\ref{bhlqlt}), 
via the tadpole diagrams in Fig.~\ref{figt}, 
contribute to the tau lepton mass 
$m^{\rm eb}_{\tau}$ and tau neutrino mass $m^{\rm eb}_{\tau_\nu}$, provided the 
bottom quark mass $m^{\rm eb}_{b}$ and top quark mass $m^{\rm sb}_{t}$ are not zero. 
The latter $m^{\rm sb}_{t}$ is generated by the SSB, see Sec.~\ref{SSBS}. 
The former $m^{\rm eb}_b$ is generated by 
the ESB due to the $W^\pm$-boson vector-like coupling and top-quark mass $m^{\rm sb}_{t}$, 
see next Sec.~\ref{wESBS}. 

Corresponding to 
the tadpole diagrams in Fig.~\ref{figt}, 
the mass-gap equations of tau lepton and tau neutrino 
are given by 
\begin{eqnarray}
m^{\rm eb}_{\nu_\tau}&=&(U_L^{\nu_\tau t}U_R^{t \nu_\tau})2Gm^{\rm sb}_{t} \frac{i}{(2\pi)^4}\int d^4l[l^2-(m^{\rm sb}_{t})^2]^{-1}=(U_L^{\nu_\tau t}U_R^{t \nu_\tau})(1/N_c)m^{\rm sb}_t,
\label{massql}\\
m^{\rm eb}_{\tau}&=&(U_L^{\tau b}U_R^{b\tau})2Gm^{\rm eb}_{b} \frac{i}{(2\pi)^4}\int d^4l[l^2-(m^{\rm eb}_{b})^2]^{-1}= (U_L^{\tau b}U_R^{b\tau})(1/N_c)m^{\rm eb}_{b}.\label{massql'}
\end{eqnarray}
%{\it Vice versa} once leptons are massive, 
%via the same tadpole diagram in Fig.~\ref{figt},
%quark-lepton interactions (\ref{bhlql}) 
%contribute to the quark bare masses 
%$m^{\rm eb}_t$ and $m^{\rm eb}_b$.  As a result, we have
%up to the leading order in the large $N_f$-expansion 
%with $GN_f\E^2$ fixed, where $N_f$ is the number of fermions like color for example. $U_L^{\nu_\tau t}U_R^{t \nu_\tau}$ or $U_L^{\nu_\tau t\dagger}U_R^{t \nu_\tau\dagger}$
Here we use the self-consistent 
mass-gap equations of the bottom and top quarks (see Eq.~2.1 and 2.2 in Ref.~\cite{bhl1990})
\begin{eqnarray}
m^{\rm eb}_{b}&=&2GN_cm^{\rm eb}_{b} \frac{i}{(2\pi)^4}\int d^4l[l^2-(m^{\rm eb}_{b})^2]^{-1},\label{massbt'}\\ 
m^{\rm sb}_{t}&=&2GN_cm^{\rm sb}_{t} \frac{i}{(2\pi)^4}\int d^4l[l^2-(m^{\rm sb}_{t})^2]^{-1},
\label{massbt}
\end{eqnarray}
and the definitions of Dirac quark, lepton and neutrino bare masses 
in Eqs.~(\ref{qmassd}-\ref{mmassd}). 
It is important to note the difference 
that Eq.~(\ref{massbt}) is the 
mass-gap equation for the top-quark mass $m^{\rm sb}_{t}$ generated by the SSB, 
while Eq.~(\ref{massbt'}) is just a self-consistent mass-gap equation  
for the bottom-quark mass $m^{\rm eb}_{b}\not=0$, as given by the 
tadpole diagram.
The tau-neutrino mass $m^{\rm eb}_{\nu_\tau}$ and 
tau-lepton mass $m^{\rm eb}_\tau$ are not zero, 
if the top-quark mass $m^{\rm sb}_t$ and bottom-quark mass $m^{\rm eb}_b$ are not zero. 
This is meant to the mass generation of tau neutrino and tau lepton due to
the ESB terms introduced by the quark-lepton interactions (\ref{bhlql}), 
quark masses $m^{\rm sb}_t$ and $m^{\rm eb}_b$. On the other hand, if  
the tau-neutrino mass $m^{\rm eb}_{\nu_\tau}$ and 
tau-lepton mass $m^{\rm eb}_\tau$ are not zero, they also contribute to the self-consistent mass-gap equations for $m^{\rm sb}_t$ and $m^{\rm eb}_b$. 

These discussions can be generalized to the three-family case by replacing $t\rightarrow t,c,u$ and 
$\nu_\tau\rightarrow \nu_\tau,\nu_\mu,\nu_e$ in Eqs.~(\ref{massql}) 
and (\ref{massbt}); $b\rightarrow b,s,d$ and 
$\tau\rightarrow \tau,\mu,e$ in Eqs.~(\ref{massql'}) 
and (\ref{massbt'}), and summing all contributions.  
All these self-consistent mass-gap equations are coupled together.

\subsection{$W^\pm$-boson coupling to right-handed fermions}\label{wESBS}
\hskip0.1cm
In addition to the ESB terms due to quark-lepton interactions, 
the effective vertex 
of $W^\pm$-boson coupling to right-handed fermions \cite{xue2016},
\begin{eqnarray}
\Gamma_\mu^W(p,p')&=& i\frac{g_2}{\sqrt{2}}\gamma_\mu P_R\,\Gamma^W(p,p')
\label{vr}
\end{eqnarray}
at the energy scale $\E$, also introduces the ESB terms to the 
Schwinger-Dyson equations. This is the main reason 
for the nontrivial bottom-quark mass $m_b$, once 
the top-quark mass $m_t$ is generated by the SSB \cite{xue2016}. 
This will be generalized 
to other fermions in Sec.~\ref{sdS}.

Before leaving this Section, 
we would like to mention that 
the vector-like feature of $W^\pm$-boson coupling at high energy $\E$
is expected to have some collider signatures (asymmetry)
on the decay channels of $W^\pm$-boson into both left- and right-handed 
helicity states of two high-energy leptons or quarks \cite{xueparity,xue2013}. 
\comment{NPB The branching ratios of
different helicity states are expected to be almost the same, given the qualitative 
estimation of Eqs.~(\ref{alphaw}) and (\ref{alphav}) below. 
This contrasts to the helicity suppression  
in the low-energy SM due to its $W^\pm$-boson coupling being purely left-handed, recalling 
the helicity suppression of pion decay into an electron and the corresponding electron antineutrino. In order to see this, it was suggested \cite{xueparity,xue2013} to measure the asymmetry 
\begin{eqnarray}
A_{L,R}&=&\frac{\sigma_L-\sigma_R}{\sigma_L+\sigma_R},
\label{as0}
\end{eqnarray} 
of cross-sections of left- and right-handed polarized particles in collisions, and compare the result with the theoretical results of the SM and the vector-like
feature (\ref{vr}).
} 
The collider signatures %of the asymmetry (\ref{as0}) 
should be more evident in high energies, where 
heavier fermions are produced.
In fact, at the Fermilab Tevatron $p\bar p$ collisions 
the CDF \cite{asCDF} and D0 \cite{asD0} experiments measured
the forward-backward asymmetry in top-quark pair production
\begin{eqnarray}
A_{FB}&=&\frac{N_t(\cos\theta >0)-N_t(\cos\theta <0)}
{N_t(\cos\theta >0)+N_t(\cos\theta <0)}
=0.19 \pm 0.065({\rm stat}) \pm 0.024({\rm syst}),
\label{as1}
\end{eqnarray} 
where the number $N_t(\cos\theta)$ of outgoing top quarks in the direction $\theta$ w.r.t.~the incoming proton beam. 
This is larger than the asymmetry within the SM.
%which is only up to $5\%$. 
In addition to the
$s$-channel of one gauge boson $(\gamma, g,Z^0)$ exchange, 
the process $d(p_1)d(p_2)\rightarrow t(k_1)t(k_2)$, i.e., down-quark
pair to top-quark pair, has the $t$-channel of one SM W-boson exchange. 
Its contributions to the asymmetry (\ref{as1}) and total $t\bar t$-production rate were studied \cite{cky2009} by assuming a new massive
boson $W^\prime$ with left- and right-handed couplings $(g_L,g_R)$ to the top and down quarks.
Performing the same analysis as that in Ref.~\cite{cky2009}, 
we can explain the asymmetry (\ref{as1}) by using the SM boson masses 
$(\approx M_z)$ and renormalized $SU_L(2)$-coupling $\bar g^2_2(M_z) \approx 0.45$ with 
$(g_L = 1, g_R = \Gamma^W\approx \gamma_w\approx 0.57)$. %their g'=g_2/\sqrt{2}=0.474 in their fig 1%
The detailed analysis will %is too long to
be presented somewhere else. However, we want to point out that the analogous asymmetry should be also
present in the $b\bar b$ channel, 
since the vector-like coupling (\ref{vr}) is approximately universal for all fermions \cite{xue2016}.

\section
{\bf  Schwinger-Dyson equations for fermion self-energy functions}\label{sdS}
\hskip0.1cm
In order to understand how fermion masses are generated 
by the ESB and obey their RG equations,  
we are bound to study the Schwinger-Dyson (SD) equations 
for fermion self-energy functions $\Sigma_f$. The SD equations are
generalized from the third family \cite{xue2016} to the three families.  

%\vskip0.1cm
%\noindent
\subsection{Chiral symmetry-breaking terms in SD equations}
\hskip0.1cm
In a vector-like gauge theory, SD equations for fermion
self-energy functions were intensively studied in Ref.~\cite{kogut}.  
In the Landau gauge, SD equations for quarks are given by
\begin{eqnarray}
\Sigma_f(p)&=& m^{\rm eb}_f+m^{\rm sb}_f \delta_{ft}+
3\int_{p'}\frac{V_{2/3}(p,p')}{(p-p')^2}\frac{\Sigma_f(p')}{p'^2
+\Sigma_f(p')}\quad\quad f=u,c,t\label{23}\\
%&+&3(g_2/\sqrt{2})^2\sum_{f'}|U_{ff'}|^2\int_{p'} \frac{\Gamma^W(p,p')}{(p-p')^2}
%\frac{\Sigma_{f'}(p')}{p'^2+\Sigma^2_{f'}(p')},\label{23+}\\
\Sigma_{f'}(p)&=& m^{\rm eb}_{f'}+m^{\rm sb}_{f'} + 3
\int_{p'}\frac{V_{-1/3}(p,p')}{(p-p')^2}\frac{\Sigma_{f'}(p')}{ p'^2
+\Sigma_{f'}(p')}\quad\quad f'=d,s,b,\label{13}
%\\
%&+&3(g_2/\sqrt{2})^2\sum_{f}|U_{f'f}|^2\int_{p'} \frac{\Gamma^W(p,p')}{(p-p')^2}
%\frac{\Sigma_{f}(p')}{ p'^2+\Sigma^2_{f}(p')},
%\label{13+}
\end{eqnarray}
where the integration $\int_{p'}\equiv \int d^4p'/(2\pi)^4$ is 
up to the cutoff $\E$.
$V_{2/3}(p,p')$ and $V_{-1/3}(p,p')$ are the vertex-functions of
vector-like gauge theories. We neglect corrections to vertex-functions of vector-like gauge 
interactions, for example, $V_{{2/3}}=(2e/3)^2$ and $V_{-1/3}=(e/3)^2$ 
in the QED case.

In Eqs.~(\ref{23}) and (\ref{13}), only for the top quark the SSB-generated 
mass term $m^{\rm sb}_t\not=0$, see the simplest mass-gap equation (\ref{massbt}) and discussions in Sec.~\ref{SSBS}, while for all other quarks the SSB-generated 
mass term $m^{\rm sb}_f=m^{\rm sb}_{f'}=0$, see the discussions in Sec.~\ref{onlyt}.

Instead, the bare mass terms
$m^{\rm eb}_f$ and $m^{\rm eb}_{f'}$ in Eqs.~(\ref{23}) and (\ref{13}) 
come from the ESB terms due to the quark-lepton interactions 
(\ref{bhlqlm}),  see %for example Fig.~\ref {figt} and 
the self-consistent mass-gap equations (\ref{massql}-\ref{massbt'}), 
and the effective $W^\pm$-boson coupling vertex (\ref{vr}).  
%at the scale $\E$, as (\ref{massql},\ref{massbt}) and Fig.~\ref{figt}
Since the vertex function $\Gamma^W(p,p')$ 
in Eq.~(\ref{vr}) does not vanish only for high energies, 
we approximately treat it
as a boundary value at the scale $\E$
\begin{eqnarray}
\alpha_w=\alpha_2(\E)(\gamma_w/\alpha_c\sqrt{2}),\quad \alpha_2(\E)=\bar g_2^2(\E)/4\pi,\quad
\gamma_w=\Gamma^W(p,p')|_{p, p'\rightarrow \E},
\label{alphaw}
\end{eqnarray}
where $\alpha_c=\pi/3$, the $W$-contributions %of Fig.~\ref{figs} %vertex function $\Gamma^W(p,p')$ 
are approximately boundary terms in the integral SD equations 
(\ref{23}) and (\ref{13}), see Fig.~4 in Ref.~\cite{xue2016} for the third family.  %see the following Equations (\ref{deq}-\ref{boundaryb}). 
\comment{NPB
Thus we neglect possible right-handed couplings of the
would-be Nambu-Goldstone bosons in the Landau gauge, which could have effects 
on the explicit gauge symmetry breaking. %In future we will study these effects in some more details.
}

We recall that in the SM the $W^\pm$ boson does not contribute 
to the SD equations for fermion self-energy functions $\Sigma_f$. 
However, due to the nontrivial vertex 
function (\ref{vr}), the $W^\pm$ gauge boson has the vector-like 
contributions to SD equations \cite{xue1997mx,xue1999nu}. 
%, as shown in Fig.~\ref{figs}.
These contributions not only introduce additional ESB terms,
but also mix up SD equations 
for self-energy functions of different fermion fields 
via the CKM mixing matrix $U_{ff'}=({\mathcal U}^{u\dagger}_L{\mathcal U}^d_L)_{ff'}$
and the PMNS mixing matrix $U^\ell_{ff'}=({\mathcal U}^{\nu\dagger}_L{\mathcal U}^\ell_L)_{ff'}$.

%\vskip0.1cm
%\noindent
\subsection{Twelve coupled SD equations for SM quark and lepton masses}
\hskip0.1cm
Following the approach of Ref.~\cite{kogut}, we convert integral 
equations (\ref{23}) and (\ref{13}) to the following boundary value 
problems ($x=p^2$, $\alpha=e^2/4\pi$):
\begin{eqnarray}
&&\frac{d}{dx}\left(x^2\Sigma_f'(x)\right)+\frac{\alpha_f}{4\alpha_c}\frac{x\Sigma_f(x)}{ x+
\Sigma_f^2(x)}=0,\quad \quad f=u,c,t\label{deq}\\
&& \E^2\Sigma_f'(\E^2)+\Sigma_f(\E^2)-\delta_{ft} m^{\rm sb}_t=\alpha_w\sum_{f'=d,s,b}|U_{ff'}|^2\Sigma_{f'}(\E^2)+m^{\rm eb}_{f},
\label{boundary}
\end{eqnarray}
and
\begin{eqnarray}
&&\frac{d}{dx}\left(x^2\Sigma_{f'}'(x)\right)+\frac{\alpha_{f'}}{4\alpha_c}\frac{x\Sigma_{f'}(x)}{ x+
\Sigma^2_{f'}(x)}=0,\quad \quad f'=d,s,b\label{deqb}\\
&&\E^2\Sigma_{f'}'(\E^2)+\Sigma_{f'}(\E^2)=\alpha_w\sum_{f=u,c,t}|U_{f'f}|^2\Sigma_f(\E^2)
+m^{\rm eb}_{f'},
\label{boundaryb}
\end{eqnarray}
where the fine structure constant $\alpha_{f} (\alpha_{f'})$ corresponds 
to the quark sector $2/3 (-1/3)$ and the QCD contributions are not explicitly shown. 

Analogously, we obtain the following boundary value problems in the lepton sector:
\begin{eqnarray}
\frac{d}{dx}\left(x^2\Sigma_{\ell_{\nu}}'(x)\right)&=&0,\quad\quad\quad 
\ell_{\nu}=\nu_e,\nu_\mu,\nu_\tau\label{ldeq}\\
\E^2\Sigma_{\ell_{\nu}}'(\E^2)+\Sigma_{\ell_{\nu}}(\E^2)
&=&\alpha_w\sum_{\ell'=e,\mu,\tau}|U^\ell_{\ell_{\nu}\ell'}|^2\Sigma_{\ell'}(\E^2)+m^{\rm eb}_{\ell_{\nu}},
\label{nub}
\end{eqnarray}
and
\begin{eqnarray}
\frac{d}{dx}\left(x^2\Sigma_\ell'(x)\right)
+\frac{\alpha}{ 4\alpha_c}\frac{x\Sigma_\ell(x)}{ x+
\Sigma^2_\ell(x)}&=&0,\quad\quad \ell=e,\mu,\tau \label{ldeqb}\\
\E^2\Sigma_\ell'(\E^2)+\Sigma_\ell(\E^2)
&=&\alpha_w\sum_{\ell'_{\nu}}|U^\ell_{\ell\ell'_{\nu}}|^2\Sigma_{\ell'_{\nu}}(x)+m^{\rm eb}_\ell,
\label{lb}
\end{eqnarray}
where $U^\ell_{\ell\ell'_{\nu}}$ is the PMNS mixing matrix 
$U^\ell={\mathcal U}^{\nu_e\dagger}_L{\mathcal U}^e_L$ of CKM-type 
in the lepton sector. The boundary conditions 
(\ref{boundary}), (\ref{boundaryb}), (\ref{nub}) and (\ref{lb}) are actually 
the mass-gap equations of quarks and leptons 
at the scale $\E$.  Note that the quark-lepton interactions (\ref{bhlqlm}) have 
the contributions to the 
$m^{\rm eb}$-term in mass-gap equations (\ref{boundary}), (\ref{boundaryb}), 
(\ref{nub}) and (\ref{lb}), see also Fig.~\ref{figt}, 
therefore the quark and lepton mass-gap equations are coupled together.
In total, these are twelve coupled and mixed SD equations of three quark and lepton families. 

These twelve inhomogeneous SD equations admit massive 
solutions \cite{kogut,xue2000fi}
\begin{equation}
\Sigma_f(p)\propto m_f\big(\frac{p^2}{ m^2_f}\big)^{\gamma},\quad m_f\leq p\leq \E,
\label{exp}
\end{equation}
where $\gamma\ll 1$ is the anomalous dimension of fermion mass 
operators, and running 
fermion masses at an infrared scale $\mu$ and mass-shell conditions read
\begin{eqnarray}
m_{_f}(\mu)=\Sigma_f(\mu)=\bar g_{_f}(\mu)v/\sqrt{2},
\quad m_{_f}=\Sigma_f(m_{_f})(\mu)=\bar g_{_f}(m_{_f})v/\sqrt{2},
\label{yukawa0} 
\end{eqnarray}
where $\bar  g_{_f}(\mu)$ is the corresponding Yukawa coupling,  
see more discussions in Ref.~\cite{xue2016}.

\subsection{Realistic massive solutions}
\hskip0.1cm
Once the top-quark mass $m^{\rm sb}_t$ in Eq.~(\ref{boundary}) 
is generated by the SSB, 
the SD equations (\ref{boundaryb}) for $d,s,b$ quarks
acquire inhomogeneous $\alpha_w$-terms via flavor mixing.  
{\it Vice versa}, once $d,s,b$ quarks are massive, the SD equations (\ref{boundary}) 
for $u,c,t$ quarks acquire inhomogeneous $\alpha_w$-terms via flavor mixing as well.
In the same way for the lepton sector, $\alpha_w$-terms 
due to the lepton-flavor mixing in Eqs.~(\ref{nub}) and (\ref{lb}) relate the massive
solutions of charged leptons $e,\mu,\tau$ and neutrinos $\nu_e,\nu_\mu,\nu_\tau$.
These inhomogeneous $\alpha_w$-terms are the ESB terms,
in additional to the ESB mass terms 
$m_f^{\rm eb},m_{f'}^{\rm eb}, m_{\ell_{\nu}}^{\rm eb}$ and $m_\ell^{\rm eb}$ 
generated by quark-lepton interaction discussed in Sec.~\ref{bare}. 

\comment{
\begin{figure}%[!h]
\begin{center}
\includegraphics[height=1.50in]{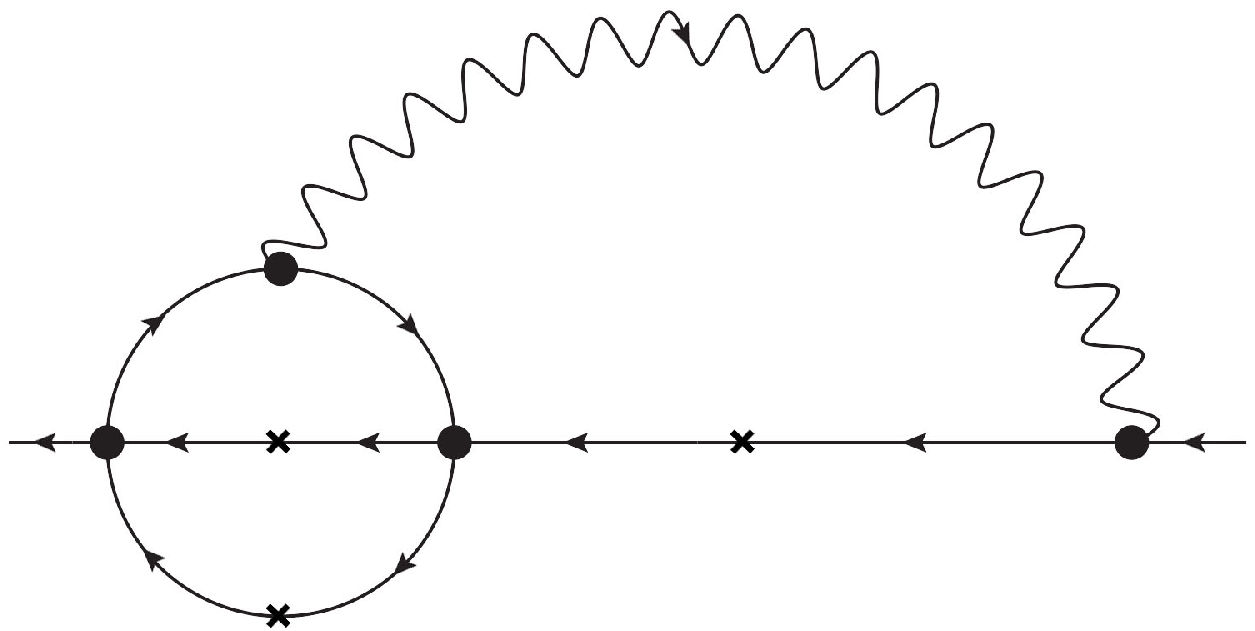}
%\put(-105,70){\footnotesize $q=p+p'$}
\put(-130,110){\footnotesize $W^+(p'-p)$}
\put(-177,52){\footnotesize $g_{_2}$}
\put(-40,13){\footnotesize $g_{_2}|U_{tb}|^2$}
%\put(-157,40){\footnotesize $p'$}
%\put(-144,40){\footnotesize $p$}
\put(-130,42){\footnotesize $\bar t_{_R}(p')$}
\put(-70,42){\footnotesize $t_{_L}(p')$}
\put(-250,29){\footnotesize $b_{_R}(p)$}
\put(-3,29){\footnotesize $\bar b_{_L}(p)$}
\put(-140,22){\footnotesize $G$}
\put(-213,22){\footnotesize $G$}
\caption{We adopt the third quark family $(t,b)$
as an example to illustrate the general $W^\pm$-boson vector-like 
contribution to the fermion self-energy function $\Sigma(p)$. The bottom quark 
self-energy function $\Sigma_b(p)$ is related to the top-quark one 
$\Sigma_t(p')$.} \label{figs}
\end{center}
\end{figure}
}

These discussions show that the SSB generated top-quark mass $m^{\rm sb}_t$ 
introduces the ESB terms into the SD equations of both quark and lepton sectors, which 
become inhomogeneous, completely
coupled together and have nontrivial massive solutions. For instance, the top-quark mass
$m_t$ introduces neutrino bare masses
$m_{\ell_{\nu}}^{\rm eb}$ of Eq.~(\ref{massql2}).   

%The ESB is not only associated to fermion chiral symmetry
%but also fermion flavor symmetry. 

At the end of this section, it is worth noting that the 
top-quark mass $m_t$ is generated by the SSB with three Goldstone bosons, which become
the longitudinal modes of massive $W^\pm$ and $Z^0$ gauge bosons, and 
the ESB for generating masses does not associate with Goldstone bosons. 
The top-quark mass $m_t$ is the unique origin of the ESB for generating 
all other fermion masses, therefore there is no any extra Goldstone boson 
in addition to those in the SSB channel of the top quark.

\section
{\bf The hierarchy spectrum of SM fermion masses}\label{hiS}
\hskip0.1cm
In this section, we focus on 
approximately finding the qualitative fermion masses first for the third family 
$(\nu_\tau, \tau,t, b)$, then for the second family 
$(\nu_\mu,\mu,c,s)$ and the first family $(\nu_e,e,u, d)$, in order to understand what 
is the dominate contribution to each fermion mass and how the hierarchy of fermion masses 
is built in by the fermion-family mixing. 

%\vskip0.1cm
\subsection{The third fermion family}\label{3family}
\hskip0.1cm
This family is much more massive than the first and second fermion families
in coupled SD equations. Therefore, we treat the massive 
solution ($m_{\nu_\tau},m_\tau, m_b,m_t$) 
of the third fermion family as a leading term and 
those for the first and second fermion families
as perturbations in SD equations. 

\subsubsection{Approximate fermion mass-gap equations for the third family}

In Eqs.~(\ref{boundary}), (\ref{boundaryb}), (\ref{nub}) and (\ref{lb}), 
we use Eq.~(\ref{exp}) to calculate the term
$\E^2\Sigma_i'(\E^2)=\gamma\Sigma_i(\E^2)\ll \Sigma_i(\E^2)$, thus 
we neglect the term $\E^2\Sigma_i'(\E^2)$ in these equations. 
At the scale $\E$, the top-quark bare mass 
$\Sigma_t(\E^2)\equiv m^0_t
%=\bar g_t(\E) v/\sqrt{2}$ due to the SSB, and $m^0_t 
\approx m^{\rm sb}_t$ introduces an explicit symmetry breaking term into 
SD equations for other fermions, and we define bare fermion masses  
$\Sigma_f(\E^2)\equiv m^0_f\approx m^{\rm eb}_f, ~(f=\nu_\tau,\tau,b)$ 
%$\Sigma_b(\E^2)\equiv m^0_b\approx m^{\rm eb}_b$, 
%$\Sigma_\tau(\E^2)\equiv m^0_\tau\approx m^{\rm eb}_\tau$, 
%$\Sigma_{\nu_\tau}(\E^2)\equiv m^0_{\nu_\tau}\approx m^{\rm eb}_{\nu_\tau}$ 
due to the ESB. Neglecting the contributions from the first and second fermion families,
we approximately obtain the gap-equations
(\ref{boundary}), (\ref{boundaryb}), (\ref{nub}) 
and (\ref{lb}) as follow,
\begin{eqnarray}
m^0_{\nu_\tau}&\approx&\alpha_w|U^\ell_{\tau\nu_\tau}|^2m_\tau^0
+U^{\nu_\tau t}_L U^{t\nu_\tau}_R m^0_t/N_c\approx {\mathcal M_1}
 m^0_t/N_c%\lesssim m^0_t/N_c
\label{gapnu}\\
%\nonumber\\
m^0_\tau&\approx&\alpha_w|U^\ell_{\tau\nu_\tau}|^2m^0_{\nu_\tau}
+ U^{\tau b}_L U^{b\tau}_Rm^0_b/N_c
\approx {\mathcal M}_2(1/2)\alpha_w m_t^0
%\lesssim 2 m^0_b/N_c
\label{gapta}\\
%\nonumber\\
m^0_t&\approx&\alpha_w|U_{tb}|^2m_b^0
+ %N_c
U^{\nu_\tau t\dagger}_L U^{t\nu_\tau \dagger}_R m^0_{\nu_\tau} + m^{\rm sb}_t\approx m^{\rm sb}_t,
\label{gapt}\\
%\nonumber\\
m^0_b&\approx&\alpha_w|U_{bt}|^2m_t^0
+ %N_c
U^{\tau b\dagger}_L U^{b\tau \dagger}_R m^0_\tau 
\approx  {\mathcal M_0} (N_c/2)\alpha_w  m_t^0 
%\approx (N_c/2)\alpha_w  m_t^0
%\approx \alpha_w m_t^0
\label{gapb}
\end{eqnarray}
where $|U_{tb}|\approx 1.03$ \cite{pdg2012}, 
$|U^\ell_{\tau\nu_\tau}|\approx (0.590\rightarrow 0.776)$ \cite{PMNS}. The dominate contributions in the RHS of these equations can be figured out. 
We obtain the approximate solution to Eqs.~(\ref{gapnu}) and (\ref{gapt}), % neglecting small $\alpha_w$-terms, the 
as well as the approximate solution to Eqs.~(\ref{gapta}) and (\ref{gapb}),
which are given in the last step with
\begin{eqnarray}
{\mathcal M_0}&\equiv& |U_{bt}|^2+ N_c^{-1}
|U^\ell_{\tau\nu_\tau}|^2(U^{\nu_\tau t}_L U^{t\nu_\tau}_R)(U^{\tau b}_L U^{b\tau}_R)\approx |U_{bt}|^2\approx 1\nonumber\\
{\mathcal M_1}&\equiv& U^{\nu_\tau t}_L U^{t\nu_\tau}_R\gg \alpha_w|U^\ell_{\tau\nu_\tau}|^2
(m^0_\tau N_c/m^0_t)\sim {\mathcal O}(10^{-5})\nonumber\\
{\mathcal M_2}&\equiv& |U_{bt}|^2U^{\tau b}_L U^{b\tau}_R+|U^\ell_{\tau\nu_\tau}|^2U^{\nu_\tau t}_L U^{t\nu_\tau}_R<2.
\label{pm1}
\end{eqnarray}
\comment{In the following, the diagonal elements of mixing matrix CMK/PMNS for the same quark/lepton family are approximately treated as an order of unit. 
As indicated by the definitions (\ref{mql1}) of mixing matrices, 
we assume that the diagonal elements of 
these mixing matrices for the same fermion family are of order of unit, e.g. 
\begin{eqnarray}
|U^{\nu_\tau t}_L U^{t\nu_\tau}_R|\lesssim 1, \qquad |U^{\tau b}_L U^{t\nu_\tau}_R|\lesssim 1,
\label{lqm1}
\end{eqnarray} 
in Eqs.~(\ref{gapnu})-(\ref{gapb}).}  
Equations (\ref{gapnu}-\ref{gapb}) show that at the energy scale $\E$, 
the bare masses $m^0_{\nu_\tau}$, 
$m^0_\tau$ and $m^0_b$ are related to the bare mass $m^0_t$ from the SSB. The dominate contributions
follow the following way: (i) the $\tau$-neutrino 
acquires its mass $m^0_{\nu_\tau}$ from the top-quark mass $m^0_t$ via the quark-lepton mixing 
(\ref{bhlql}) and Fig.~\ref{figt} (right), 
(ii) the bottom-quark 
acquires its mass $m^0_b$ from the top-quark mass $m^0_t$ via the CKM mixing, (iii) 
the $\tau$-lepton 
acquires its mass $m^0_\tau$ from the bottom-quark mass $m^0_b$ via the quark-lepton mixing
and $\tau$-neutrino mass $m^0_{\nu_\tau}$ via the PMNS mixing.
These fermion bare masses $m^0_f$ due to the ESB at the scale $\E$ are in terms of the top-quark mass $m^0_t$ due to the SSB.

\subsubsection{Fermion masses and running Yukawa couplings}

These fermion bare masses $m^0_f$ evolve to their infrared masses
$m_f(\mu)$, mainly follow the top-quark one $m_t(\mu)$, apart from the energy-scale 
evolutions of the SM gauge interactions.
In order to qualitatively calculate the infrared scale $m_f=m_f(\mu)$ as 
functions of the running scale $\mu$ in Eq.~(\ref{yukawa0})
%Eqs.~(\ref{msol}) or (\ref{msoln}) 
for each fermion ``$f$'',  we neglect the corrections from perturbative gauge interactions %$(\mu^2/m_f^2)^{(\alpha_f/4\alpha_c)}$, 
and define the effective Yukawa couplings   
\begin{eqnarray}
m_b(\mu)=\bar g_b(\mu)v/\sqrt{2},\quad m_\tau(\mu)=\bar g_\tau(\mu)v/\sqrt{2},\quad
m_{\nu_\tau}(\mu)=\bar g_{\nu_\tau}(\mu)v/\sqrt{2},
\label{yukawa3}
\end{eqnarray} 
analogously to the top-quark mass $m_t(\mu)=\bar g_t(\mu) v/\sqrt{2}$. This means that
effective fermion Yukawa couplings $\bar g_f(\mu)$ are functions of 
the top-quark one $\bar g_t(\mu)$.   
\comment{The physical solution to Eq.~(\ref{ldeq}) and boundary condition (\ref{nub}) or 
(\ref{gapnu}) is $\Sigma_{\ell_\nu}(\mu)\sim m_{\ell_\nu}(\mu)$ 
(see Ref.~\cite{xue2000fi}), where the infrared scale for neutrino masses 
$m_{\ell_\nu}(\mu)$ are related to the ESB at the scale $\E$ and 
their RG flows. Namely, the neutrino mass $m_{\nu_\tau}(\mu)$ and Yukawa coupling
$\bar g_{\nu_\tau}(\mu)$ are mainly related to the top-quark mass $m_t(\mu)$ and Yukawa coupling
$\bar g_t(\mu)$ via the quark-lepton interactions (\ref{bhlql}) and Fig.~\ref{figt} (right). 
%The SM neutrinos have not 
%vector-like gauge interactions, see SD equations (\ref{ldeq}), 
%the neutrino masses $m_{\ell_\nu}(\mu)$ are given by 
%other fermion masses $m_f(\mu)$, as shown below.
}
%Using Eqs.~(\ref{gapnu}-\ref{gapb}) and Eqs.~(\ref{trg}-\ref{lrg}), 
Equations 
(\ref{gapnu}-\ref{gapb}) become
\begin{eqnarray}
m_{_{\nu_\tau}}(\mu)&\approx& {\mathcal M}_1m_t(\mu)/N_c
%(U^{\nu_\tau t}_L U^{t\nu_\tau}_R) [\bar g_1(\mu)]^{1/10}[\bar g_3(\mu)]^{-8/7},
\label{rtaunu}\\
m_\tau(\mu)&\approx& {\mathcal M}_2 (1/2) \alpha_w m_t(\mu)% [\bar g_1(\mu)]^{-9/20}
%[\bar g_1(\mu)]^{-1/2}[\bar g_3(\mu)]^{-8/7},
\label{rtaub}\\
%m_t(\mu)&\approx& m_t(\mu),\label{rtm}\\
m_b(\mu)&\approx& (N_c/2) \alpha_w  m_t(\mu).%[\bar g_3(\mu)]^{8/7} [\bar g_1(\mu)]^{1/20}
%[\bar g_1(\mu)]^{3/20},
\label{rbm}
\end{eqnarray}
\comment{
and Yukawa couplings 
\begin{eqnarray}
\bar g_{_{\nu_\tau}}(\mu)&\approx&N_c^{-1} \bar g_t(\mu)
%(U^{\nu_\tau t}_L U^{t\nu_\tau}_R)[\bar g_1(\mu)]^{1/10}\bar g_t(\mu)[\bar g_3(\mu)]^{-8/7} ,
\label{ntaum}\\
\bar g_\tau(\mu)&\approx& 2 N_c^{-1} \bar g_b(\mu)%  [\bar g_1(\mu)]^{-9/20}
%[\bar g_1(\mu)]^{-1/2}[\bar g_3(\mu)]^{-8/7}, 
\label{taum}\\
\bar g_t(\mu)&\approx& \bar g_t(\mu) ,
\label{tm}\\
\bar g_b(\mu)&\approx&  \alpha_w\bar g_t(\mu). %[\bar g_3(\mu)]^{8/7} [\bar g_1(\mu)]^{1/20}
%[\bar g_1(\mu)]^{3/20},
\label{bm}
\end{eqnarray}
}
Then, based on the top-quark mass-shell condition $m_t=\bar g_t(m_t)v/\sqrt{2}$ and 
experimental values of top and bottom quark masses: 
$m_t=m_t(m_t)\approx 173$ GeV and  $m_b=m_b(m_b)\approx 4.2$ GeV,  
as well as the Yukawa-coupling values 
$\bar g_t(m_t)=1.02$ and $\bar g_t(m_b)=1.29$ (Fig.~\ref{figyt}), 
Eq.~(\ref{rbm}) determines the approximately ``universal'' value in 
Eq.~(\ref{alphaw}) \cite{xue2016} 
\begin{eqnarray}
\alpha_w \approx  (2/N_c)
\Big(\frac{m_b}{m_t}\Big) \Big[\frac{\bar g_t(m_t)}{\bar g_t(m_b)}\Big]
\approx 1.9\times 10^{-2}~(2/N_c).
%\alpha_w = \Big(\frac{m_b}{m_t}\Big) \Big[\frac{g_1(m_b)}{g_1(m_t)}\Big]^{1/10}
%\Big[\frac{g_1(\E)}{g_1(m_b)}\Big]^{3/20}\Big[\frac{[g_3(m_t)}{g_3(m_b)}\Big]^{8/7}
\label{alphav}
\end{eqnarray}
Equation (\ref{alphaw}) gives $\gamma_w\approx 0.85~(2/N_c)\sim {\mathcal O}(1)$, where 
the value $\bar g^2_2(\E)\approx 0.42$. In this way, we approximately 
determine the finite part of vertex function 
$\Gamma^W(p,p')$ (\ref{vr}) or (\ref{alphaw}).%, which has not yet been calculated. 
%In an opposite way, we can treat  as a parameter $\alpha_w$ is  fix its value (\ref{alphav}) 
%such that we obtain the experimental value of the $b$-quark mass $m_b$. 

\begin{figure}%[!h]
\begin{center}
\includegraphics[height=1.40in]{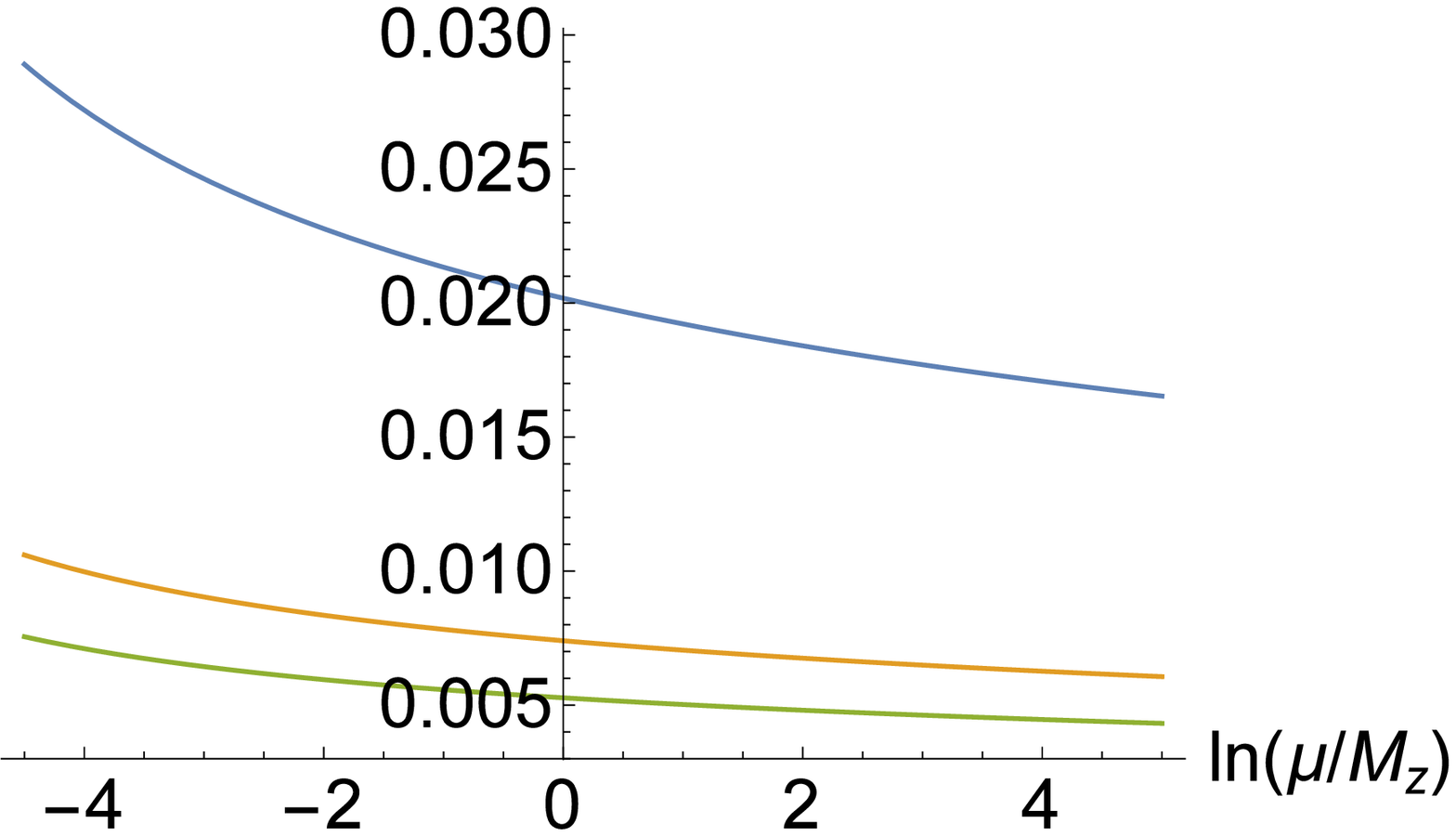}
\put(-115,105){\footnotesize $\bar g_{b,\tau}(\mu)$}
%\put(-300,105){\footnotesize $\bar g_b(\mu)$}
\put(-160,35){\footnotesize $\bar g_\tau(\mu)$}
\put(-160,85){\footnotesize $\bar g_b(\mu)$}
\put(-195,15){\footnotesize $5\times \bar g_{\nu_\tau}(\mu)$}
\caption{The Yukawa couplings $\bar g_b(\mu)$, $\bar g_\tau(\mu)$ and 
$\bar g_{\nu_\tau}(\mu)$ in the range 
$1.0 ~{\rm GeV}\lesssim \mu \lesssim 13.5~ {\rm TeV}$ for ${\mathcal M}_2\approx 1.1$ 
(\ref{masstau}) and ${\mathcal M}_1\approx 3\times 10^{-3}$ (\ref{masstaun}).
Note that $m_{b,\tau,\nu_\tau}(\mu)=\bar g_{b,\tau,\nu_\tau}(\mu)v/\sqrt{2}$.} \label{figyl}
\end{center}
\end{figure}

Using the Yukawa coupling $\bar g_\tau(\mu)$ (\ref{yukawa3}) 
and $\bar g_t(\mu)$ (Fig.~\ref{figyt}),
we numerically solve Eq.~(\ref{rtaub}) and mass-shell condition $m_\tau=\bar g_\tau(m_\tau)v/\sqrt{2}$, 
\begin{eqnarray}
m_\tau\approx 1.69~ {\rm GeV},\quad {\rm for} \quad {\mathcal M}_2\approx 1.1
\label{masstau}
\end{eqnarray}
which is qualitatively consistent with the experimental value. 
Some contributions from the first and second fermion families should be expected.
Analogously, using the Yukawa coupling $\bar g_{\nu_\tau}(\mu)$ (\ref{yukawa3}) 
and $\bar g_t(\mu)$ (Fig.~\ref{figyt}), we numerically calculate Eq.~(\ref{rtaunu}) at $\mu=2~$GeV
and obtain the neutrino Dirac mass
\begin{eqnarray}
m_{\nu_\tau}\approx 235.8~ {\rm MeV}, \quad {\rm for}\quad {\mathcal M}_1=U^{\nu_\tau t}_L U^{t\nu_\tau}_R \approx 3\times 10^{-3} .
\label{masstaun}
\end{eqnarray}
Figure \ref{figyl} shows the Yukawa couplings 
$\bar g_{\nu_\tau}(\mu)$, $\bar g_\tau(\mu)$ and $\bar g_b(\mu)$, which are 
functions of $\bar g_t(\mu)$, see Fig.~\ref{figyt}.  
The variations of Yukawa couplings $\bar g_{b,\tau,\nu_\tau}(\mu)$ are very small over 
the energy scale $\mu$. 
%The variations of Yukawa couplings $\bar g_b(\mu)$ 
%and $\bar g_\tau(\mu)$ 
%in the range $\nu\in (1, 5)$ GeV are about $10^{-3}$, indicating 
%that in this energy range the variations of masses $m_\tau(\mu)$ and $m_b(\mu)$ 
%are about $0.17$ GeV.    
%Fig.~\ref{figyl'} shows that the $\bar g_{\nu_\tau}(\mu)$ varies slowly as the scale
%$\mu$ so that the $\tau$-neutrino mass $m_{\nu_\tau}$ should not 
%change very much with the scale $\mu$.   
%In Eqs.~(\ref{alphav}), (\ref{masstau}) and (\ref{massntau}), we include the factor $[\bar g_3(\mu)]^{-8/7}$ from perturbative QCD corrections.

Equations (\ref{reg1}) and (\ref{reg2}) show that 
$\bar g_t(\mu)$ has received the contributions 
from gauge interactions $g_{1,2,3}(\mu)$ of the SM. This means that the RG-equations of 
these Yukawa couplings calculated are only valid in the high-energy region where the $\bar g_3(\mu)$- and $\bar g_2(\mu)$-perturbative contributions to $\bar g_t(\mu)$ are taken into account.
This is the reason that we adopt the point $\mu=2$~GeV to calculate $m_{\nu_\tau}$
(\ref{masstaun}), instead of using the mass-shell condition. The same reason 
will be for calculating at $\mu=2$~GeV the light fermion masses 
of the second and first families.

\subsection
{The second fermion family}\label{2family}
\hskip0.1cm  
In this section, we examine how the masses 
$m_{\nu_\tau,\tau, t,b}$ of the third fermion family introduce ESB terms 
into the SD equations of the second fermion family via
SM gauge interactions and four-fermion interactions, leading to the mass generation 
of the second fermion family. 

\begin{figure}%[!h]
\begin{center}
\includegraphics[height=1.50in]{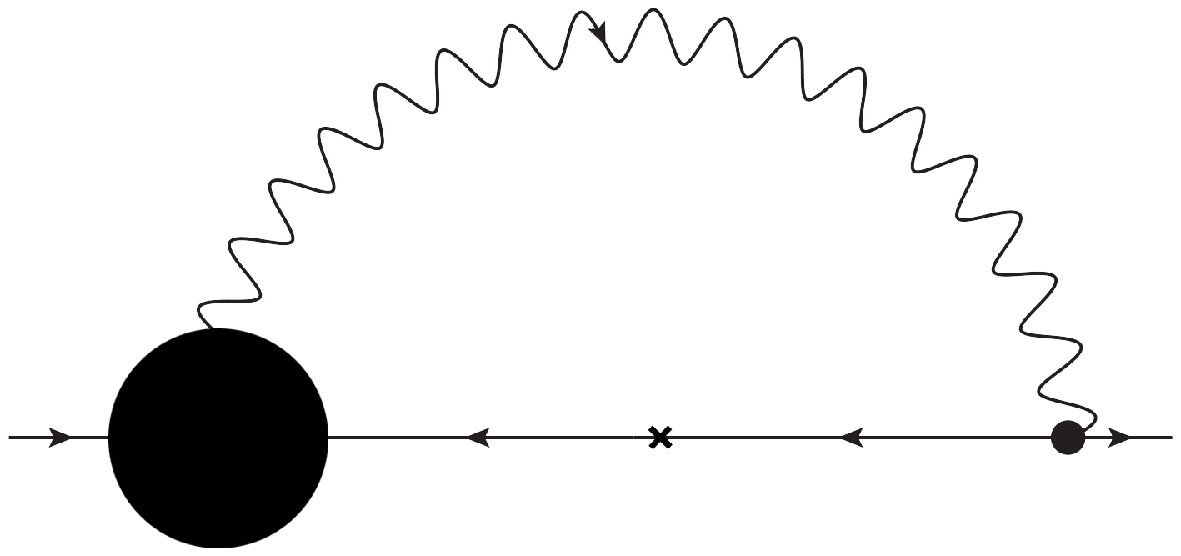}
%\put(-105,70){\footnotesize $q=p+p'$}
\put(-130,110){\footnotesize $\gamma(p'-p)$}
%\put(-177,52){\footnotesize $g_{_2}$}
%\put(-40,13){\footnotesize $g_{_2}|U_{tb}|^2$}
%\put(-157,40){\footnotesize $p'$}
%\put(-144,40){\footnotesize $p$}
\put(-130,42){\footnotesize $\bar t_{_R}(p')$}
\put(-70,42){\footnotesize $t_{_L}(p')$}
\put(-250,29){\footnotesize $c_{_R}(p)$}
\put(-3,29){\footnotesize $\bar c_{_L}(p)$}
%\put(-140,22){\footnotesize $G$}
%\put(-213,22){\footnotesize $G$}
\caption{We adopt quarks $(t,c)$
as an example to illustrate a neutral gauge-boson $\gamma$ contribution 
to the fermion self-energy function $\Sigma_c(p)$ in terms of $\Sigma_t(p)$, 
the same diagrams for other quarks $(u,c,t)$ of  $q=2/3$ charged sector, 
$(d,s,b)$ of $q=-1/3$ charged sector, as well as for other leptons 
$(e,\mu,\tau)$ of $q=-1$ charged sector,  $(\nu_e,\nu_\mu,\nu_\tau)$ of $q=0$ 
neutral sector.
} \label{fexchange}
\end{center}
\end{figure}

It is worthwhile to mention that at the lowest order (tree-level),
SM neutral gauge-bosons ($\gamma$ and $Z^0$) 
interactions and four-fermion interactions (\ref{q}) and (\ref{l})  
do not give rise to a 1PI vertex function of 
the interactions among three fermion families with the same electric charge 
$q=0,-1,2/3,-1/3$, as an example, the black blob in Fig.~\ref{fexchange}. 
This indicates the separate conservations of $u$-quark,  $c$-quark 
and $t$-quark numbers for the $q=2/3$ sector, and the same for other charged 
sectors $q=0,-1,-1/3$.
As a result, the contributions of the 1PI self-energy functions, 
as shown in Fig.~\ref{fexchange}, to SD equations for fermion self-energy functions 
are negligible.  

\subsubsection{Approximate fermion mass-gap equations of the second family}

Neglecting the contributions from the first 
fermion family, we assume that fermions in the second family mainly acquire 
their masses by ESB terms relating to fermion masses of the third family by 
the following ways: (i) family-mixing diagram, Fig.~4 in Ref.~\cite{xue2016}, 
%Fig.~\ref{figs} 
via $W^\pm$-boson exchange at high-energy scale $\E$; 
(ii) Eq.~(\ref{massql2}) via tadpole diagrams Fig.~\ref{figt} 
of quark-lepton interactions (\ref{bhlql}) or (\ref{bhlqlm}). 
%In the view of this consideration,
%we approximately solve the SD-equations 
%and their boundary conditions (\ref{deq}-\ref{lb}) at the scale $\E$ with 
Defining bare fermion masses $\Sigma_f(\E^2)\equiv m^0_f\approx m^{\rm eb}_f, ~(f
=\nu_\mu,\mu,s,c)$, 
%$\Sigma_s(\E^2)\equiv m^0_s\approx m^{\rm eb}_s$, 
%$\Sigma_c(\E^2)\equiv m^0_c\approx m^{\rm eb}_c$, 
%$\Sigma_\mu(\E^2)\equiv m^0_\mu\approx m^{\rm eb}_\mu$, and 
%$\Sigma_{\nu_\mu}(\E^2)\equiv m^0_{\nu_\mu}\approx m^{\rm eb}_{\nu_\mu}$.
%Neglecting the contributions from the first fermion family, 
mass-gap equations (\ref{boundary}), (\ref{boundaryb}), (\ref{nub}) 
and (\ref{lb}) for the second fermion family can be approximately 
written as follow, 
\begin{eqnarray}
m^0_{\nu_\mu}&\approx&\alpha_w|U^\ell_{\mu\nu_\mu}|^2m_\mu^0
+\alpha_w|U^\ell_{\tau\nu_\mu}|^2m_\tau^0\nonumber\\
&+&U^{\nu_\mu c}_L U^{c\nu_\mu}_R m^0_c/N_c+
U^{\nu_\mu t}_L U^{t\nu_\mu}_R m^0_t/N_c 
%\nonumber\\
\approx 
%\alpha_w|U^\ell_{\tau\nu_\mu}|^2m_\tau^0+
%\alpha_w|U^\ell_{\mu\nu_\mu}|^2m_\mu^0+
%U^{\nu_\mu c}_L U^{c\nu_\mu}_R m^0_c/N_c+
{\mathcal M}_3 m^0_t, % \ll m^0_t/N_c
\label{2gapnu}\\
%\nonumber\\
m^0_\mu&\approx&\alpha_w\left(|U^\ell_{\mu\nu_\mu}|^2m^0_{\nu_\mu}
+|U^\ell_{\mu\nu_\tau}|^2m^0_{\nu_\tau}\right)+ U^{\mu s}_L U^{s\mu}_Rm^0_s/N_c + U^{\mu b}_L U^{b\mu}_Rm^0_b/N_c\nonumber\\ 
&\approx&\alpha_w\left(|U^\ell_{\mu\nu_\mu}|^2m^0_{\nu_\mu}
+|U^\ell_{\mu\nu_\tau}|^2m^0_{\nu_\tau}\right)+(4/N_c){\mathcal M}_6 m_b^0
%\approx \alpha_w|U^\ell_{\mu\nu_\tau}|^2m^0_{\nu_\tau}
\label{2gapta}\\
%\nonumber\\
m^0_c&\approx&\alpha_w|U_{cs}|^2m_s^0 + \alpha_w|U_{cb}|^2m_b^0\nonumber\\
&+& U^{\nu_\mu c\dagger}_L U^{c\nu_\mu \dagger}_R m^0_{\nu_\mu}
+ U^{\nu_\tau c\dagger}_L U^{c\nu_\tau \dagger}_R m^0_{\nu_\tau}
\approx
%\alpha_w|U_{cb}|^2m_b^0 + 
%U^{\nu_\tau c\dagger}_L U^{c\nu_\tau \dagger}_R m^0_{\nu_\tau}
%&\approx&  N_cU^{\nu_\mu c\dagger}_L U^{c\nu_\mu \dagger}_R m^0_{\nu_\mu}+ N_cU^{\nu_\tau c\dagger}_L U^{c\nu_\tau \dagger}_R m^0_{\nu_\tau}\nonumber\\
%\approx %(U^{\nu_\mu t}_L U^{t\nu_\mu}_R +U^{\nu_\tau c\dagger}_L U^{c\nu_\tau \dagger}_R)
(4/N_c){\mathcal M_4}  m^0_t 
%\approx U^{\nu_\mu t}_L U^{t\nu_\mu}_R U^{\nu_\mu c\dagger}_L U^{c\nu_\mu \dagger}_R m^0_t,
\label{2gapt}
\\
%\nonumber\\
m^0_s&\approx&\alpha_w\left(|U_{sc}|^2m_c^0+|U_{st}|^2m_t^0\right)+ U^{\mu s\dagger}_L U^{s\mu \dagger}_R m^0_\mu +U^{\tau s\dagger}_L U^{s\tau \dagger}_R m^0_\tau\nonumber\\
 &\approx& \alpha_w\left(|U_{sc}|^2m_c^0+ |U_{st}|^2m_t^0\right)+{\mathcal M}_5 m_b^0%\approx {\mathcal M}_5 (N_c/2)\alpha_w m_t^0,
\label{2gapb}
\end{eqnarray}
where $|U_{cs}|\approx 0.986$, $|U_{cb}|\approx 4.1\times 10^{-2}$  and 
$|U_{ts}|\approx 4.0\times 10^{-2}$ \cite{pdg2012}, as well as 
$|U^\ell_{\tau\nu_\mu}|\approx (0.614 \rightarrow 0.699)$,
$|U^\ell_{\mu\nu_\tau}|\approx (0.464 \rightarrow 0.713)$ 
and $|U^\ell_{\mu\nu_\mu}|\approx (0.441 \rightarrow 0.699)$ \cite{PMNS}, and we use their central value for approximate calculations.  
The dominate contributions in the RHS of these equations can be figured out. 
We obtain the approximate solution to Eqs.~(\ref{2gapnu}) and (\ref{2gapt}), % neglecting small $\alpha_w$-terms, the 
as well as the approximate solution to Eqs.~(\ref{2gapta}) and (\ref{2gapb}),
which are given in the last step with
\begin{eqnarray}
{\mathcal M_3}&\equiv& \frac{1}{2}\left[U^{\nu_\mu t}_L U^{t\nu_\mu}_R + (U^{\nu_\tau t}_L U^{t\nu_\tau}_R)(U^{\nu_\tau c\dagger}_L U^{c\nu_\tau\dagger}_R)(U^{\nu_\mu c}_L U^{c\nu_\mu}_R) \right]=U^{\nu_\mu t}_L U^{t\nu_\mu}_R \nonumber\\
{\mathcal M_4} &\equiv& 
\frac{N_c}{8}\left[\frac{5}{N_c}(U^{\nu_\tau c\dagger}_L U^{c\nu_\tau\dagger}_R)(U^{\nu_\tau t}_L U^{t\nu_\tau}_R) +(U^{\nu_\mu c\dagger}_L U^{c\nu_\mu\dagger}_R)(U^{\nu_\mu t}_L U^{t\nu_\mu}_R) \right]
=(U^{ct}_L U^{tc}_R)\nonumber\\
{\mathcal M_5} &\equiv& 
(U^{s \tau\dagger}_L U^{\tau s\dagger}_R)(U^{\tau b}_L U^{b\tau}_R)
=(U^{sb}_L U^{bs}_R),\nonumber\\
{\mathcal M_6} &\equiv& 
(U^{\mu b}_L U^{b\mu}_R),
\label{pm2}
\end{eqnarray}
where Eq.~(\ref{mql1}) is used.

The dominate contributions in mass-gap equations (\ref{2gapnu})-(\ref{2gapb}) to the 
fermion masses are: 
(i) the $\nu_\mu$-neutrino and $c$-quark acquire their ESB masses $m^0_{\nu_\mu}$ and 
$m^0_c$ from the top-quark mass $m^0_t$ via the quark-lepton interactions (\ref{bhlqlm2}) 
between the third and second families, i.e., ${\mathcal M_3}$ and ${\mathcal M_4}$; 
(ii) the $s$-quark
acquires it ESB mass $m^0_s$ via the CKM mixing and the quark-lepton interactions 
${\mathcal M_5}$;
(iii) the $\mu$-lepton  acquires its ESB mass $m^0_\mu$ via the PMNS mixing and the quark-lepton interactions 
${\mathcal M_6}$.

\comment{Equations 
(\ref{2gapt}-\ref{2gapta}) show that at the energy scale $\E$, 
how the bare masses $m^0_{\nu_\mu}$, 
$m^0_\mu$, $m^0_s$ and $m^0_c$ of the second fermion family are 
related to the bare mass $m^0_{\tau_\mu}$, 
$m^0_\tau$, $m^0_t$ and $m^0_b$ of the second fermion family, where $m^0_t$ comes 
from the SSB. The dominate contributions
follow the following way: (i) the $\nu_\mu$-neutrino 
acquires its mass $m^0_{\nu_\mu}$ from the top-quark mass $m^0_t$ via the quark-lepton interaction, 
(ii) the $c$-quark acquires its mass $m^0_c$ from $\nu_\mu$-neutrino mass $m^0_{\nu_\mu}$ 
via the quark-lepton interaction, (iii) the $\mu$-lepton 
acquires its mass $m^0_\mu$ from the $\nu_\tau$-neutrino mass $m^0_{\nu_\tau}$
and $\nu_\mu$-neutrino mass $m^0_{\nu_\mu}$ via the PMNS mixing, 
(iv) the $s$-quark acquires its mass $m^0_s$ from the $\mu$-lepton mass $m^0_\mu$ 
via the quark-lepton interaction and $c$-quark mass $m_c^0$ via the CKM mixing.}

\subsubsection{Running fermion masses and Yukawa couplings} 

Analogously to the discussion for the third  fermion family from 
Eqs.~(\ref{gapnu}-\ref{gapb}) to Eqs.~(\ref{rtaunu}-\ref{rbm}), 
neglecting the perturbative corrections from the SM gauge interactions, and  
defining running fermion masses and Yukawa couplings 
\begin{eqnarray}
m_{\nu_\mu}(\mu)&=&\bar g_{\nu_\mu}(\mu)v/\sqrt{2},\quad 
m_\mu(\mu)=\bar g_\mu(\mu)v/\sqrt{2},\nonumber\\
m_c(\mu)&=&\bar g_c(\mu)v/\sqrt{2},\quad
m_s(\mu)=\bar g_s(\mu)v/\sqrt{2},
\label{yukawa2}
\end{eqnarray}
and the mass-gap equations at the scale $\mu$ are obtained by replacing 
$m_f^0\rightarrow m_f(\mu)$ in Eqs.~(\ref{2gapnu}-\ref{2gapb}).

\begin{figure}%[!h]
\begin{center}
\includegraphics[height=1.40in]{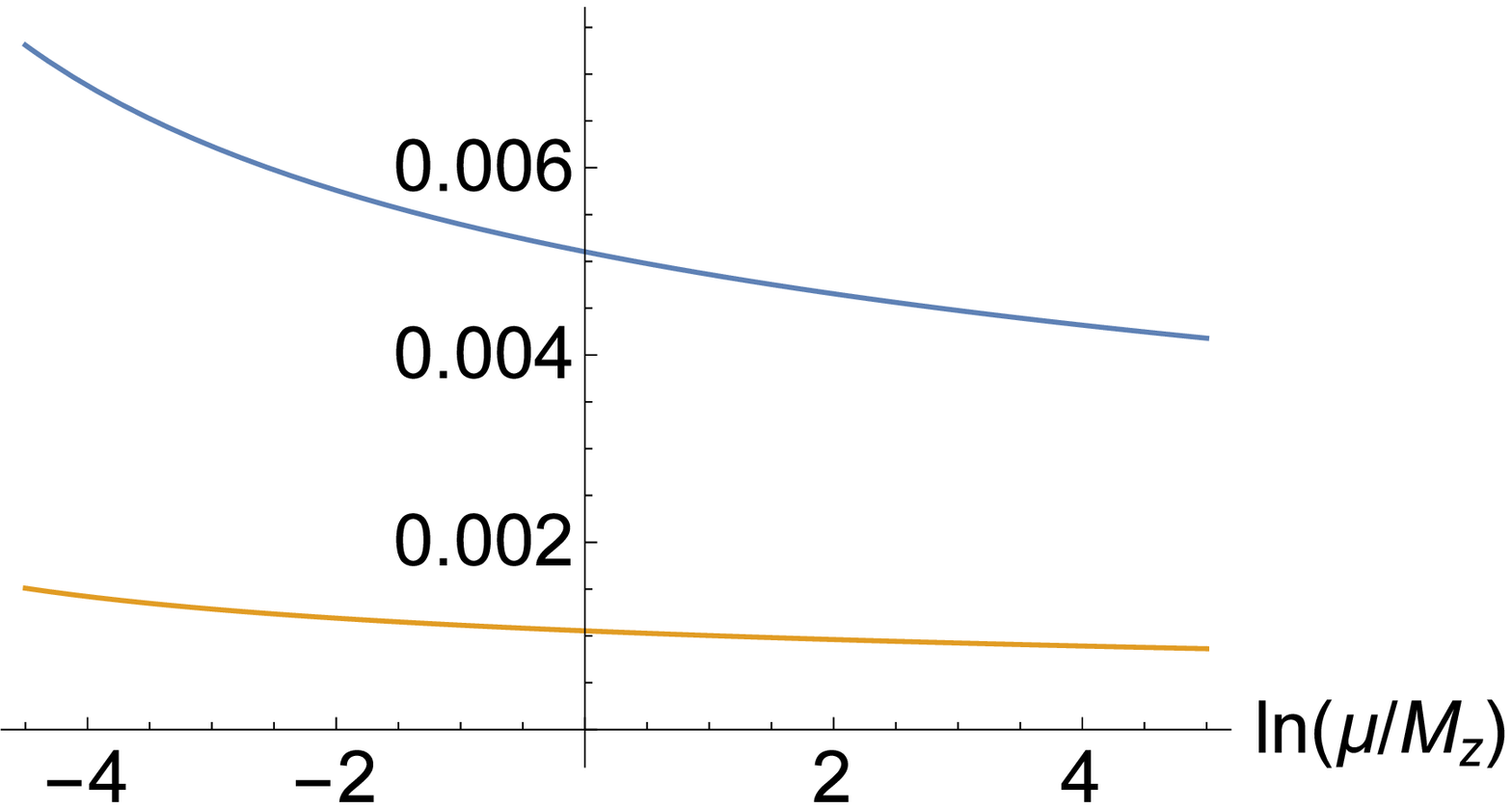}
\put(-185,100){\footnotesize $\bar g_c(\mu)$}
\put(-195,35){\footnotesize $100\times \bar g_{\nu_\mu}(\mu)$}
\caption{The Yukawa couplings $\bar g_c(\mu)$ and 
$\bar g_{\nu_\mu}(\mu)$ are plotted in the range 
$1.0 ~{\rm GeV}\lesssim \mu \lesssim 13.5~ {\rm TeV}$ 
for ${\mathcal M}_4\approx 3.6\times 10^{-3}$ (\ref{2mixtmu}) and 
${\mathcal M}_3\approx 1.0\times 10^{-5}$ (\ref{sm3}). 
Note that $m_{c,\nu_\mu}(\mu)=\bar g_{c,\nu_\mu}(\mu)v/\sqrt{2}$.} \label{figyl2}
\end{center}
\end{figure}

\begin{figure}%[!h]
\begin{center}
\includegraphics[height=1.40in]{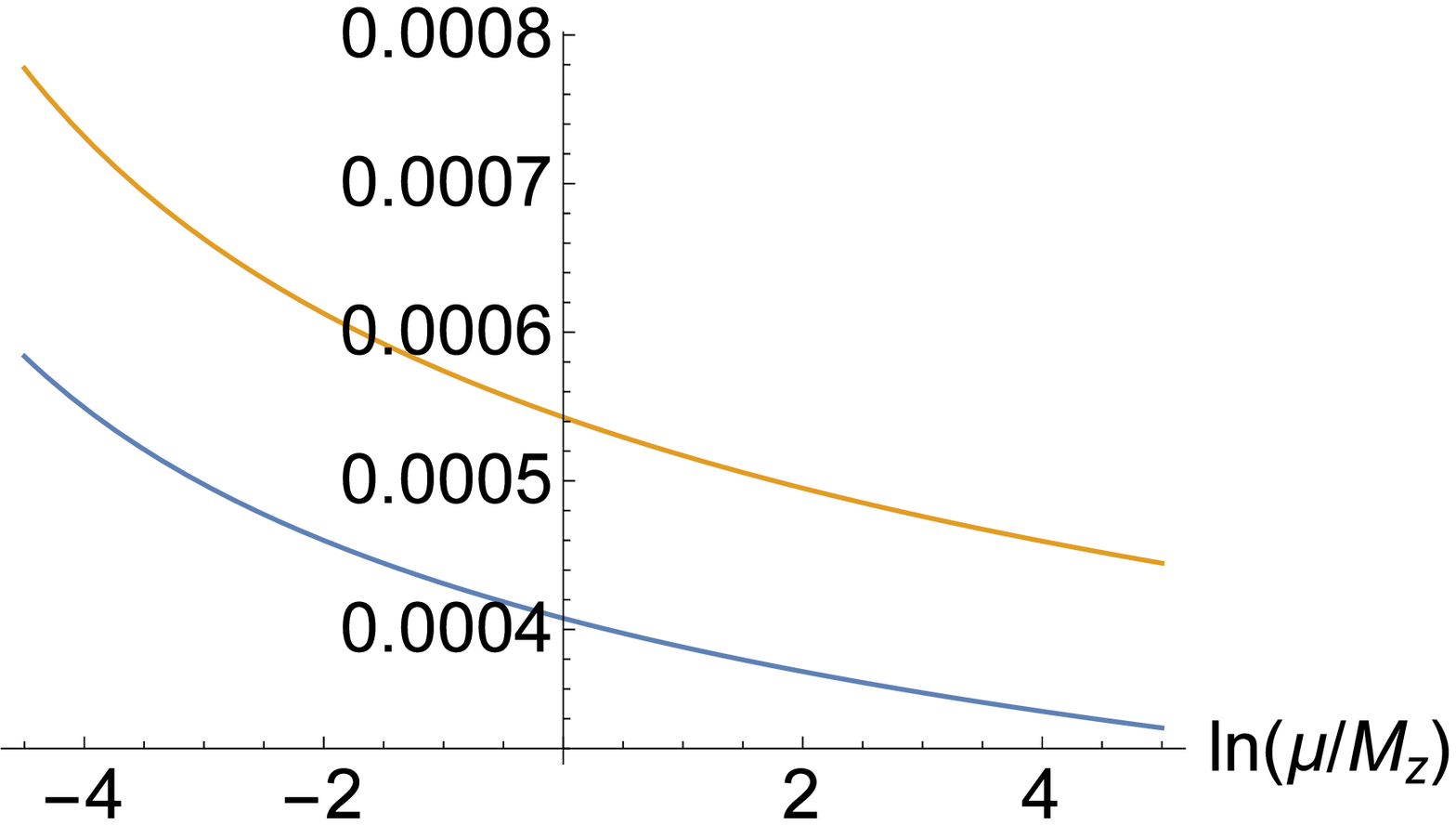}
\put(-165,85){\footnotesize $\bar g_s(\mu)$}
\put(-170,60){\footnotesize $\bar g_\mu(\mu)$}
\caption{The Yukawa couplings $\bar g_s(\mu)$ and 
$\bar g_\mu(\mu)$ are plotted in the range 
$1.0 ~{\rm GeV}\lesssim \mu \lesssim 13.5~ {\rm TeV}$ 
for ${\mathcal M}_5$ (\ref{sm5}) and 
${\mathcal M}_6$ (\ref{sm6}). 
Note that $m_{s,\mu}(\mu)=\bar g_{s,\mu}(\mu)v/\sqrt{2}$.} \label{figyl2'}
\end{center}
\end{figure}

On the basis of Eqs.~(\ref{2gapt}) and (\ref{yukawa2}) at the scale $\mu$ and 
the $c$-quark mass-shell condition $m_c=\bar g_c(m_c)v/\sqrt{2}$, as well as the results
of the third family in Sec.~\ref{3family}, we numerically obtain  
\begin{eqnarray}
m_c\approx 1.2~{\rm GeV},\quad {\rm for}\quad {\mathcal M_4} =(U^{ct}_L U^{tc}_R) \approx 3.6 \times 10^{-3}.
\label{2mixtmu}
\end{eqnarray}
%$\bar g_t(m_c) \approx 1.48$, the values $g^2_1(m_c)=0.124$ and $g^2_3(m_c)=3.53$,
Using Eqs.~(\ref{2gapnu},\ref{2gapta},\ref{2gapb}) and (\ref{yukawa2}) at the scale $\mu$, 
we calculate the $\nu_\mu$-neutrino, light $s$-quark mass and $\mu$-muon mass at the scale $\mu=2$ GeV,
\begin{eqnarray}
m_{\nu_\mu}&\approx& 2.4 ~{\rm MeV}, \quad {\rm for}\quad {\mathcal M}_3=U^{\nu_\mu t}_L U^{t\nu_\mu}_R \approx 1.0\times 10^{-5},\label{sm3}\\
m_\mu&\approx& 121.5 ~{\rm MeV}, \quad {\rm for}\quad {\mathcal M_6} 
=U^{\mu b}_L U^{b\mu}_R \approx 2.0\times 10^{-2},\label{sm6}\\
m_s&\approx& 91.2 ~{\rm MeV}, \quad {\rm for}\quad {\mathcal M}_5=U^{s b}_L U^{bs}_R \approx 1.6\times 10^{-2}.
\label{sm5}
\end{eqnarray}
As a result, the Yukawa couplings $\bar g_c(\mu)$  
and $\bar g_{\nu_\mu}(\mu)$ are shown in Fig.~\ref{figyl2},
the Yukawa coupling $\bar g_s(\mu)$ and $\bar g_\mu(\mu)$ are shown in Fig.~\ref{figyl2'}.
The variations of Yukawa couplings $\bar g_{c,s,\mu,\nu_\mu}(\mu)$ are very small over 
the energy scale $\mu$.

In summary, the preliminary study (\ref{2mixtmu}-\ref{sm5}) 
shows that the pattern of fermion masses in the second family can be consistently 
obtained by the pattern (${\mathcal M}_{3,4,5,6}$) of quark-lepton interactions and mixing between the third and second families. 
The scale $\mu$-evolution of masses and Yukawa couplings are functions of the top-quark 
one $\bar g_t(\mu)$, see Fig.~\ref{figyt}. 

%\vskip0.1cm
\subsection
{The first fermion family}
\hskip0.1cm
We turn to the masses and Yukawa couplings of the first fermion family. 
The coupled SD gap-equations receive ESB contributions 
from the third and second families, through 
the CKM and PMNS mixing as well as quark-lepton interactions between 
fermion families. As a result, the fermion masses of the first family
are generated.  Analogously to the calculations of the second family case, 
we neglect the perturbative contributions from gauge 
interactions and calculate the fermion masses at the scale $\mu=2$~GeV.
 
\subsubsection{Approximate mass-gap equations of the first fermion family}

Analogously to Eqs.~(\ref{gapnu}-\ref{gapb}) and 
Eqs.~(\ref{2gapnu}-\ref{2gapb}) respectively 
for the third and second fermion family,   
Equations (\ref{boundary}), (\ref{boundaryb}), (\ref{nub}) 
and (\ref{lb}) for the first fermion family read, 
\begin{eqnarray}
m^0_{\nu_e}&\approx&\alpha_w|U^\ell_{e\nu_e}|^2m_e^0+
\alpha_w|U^\ell_{\mu\nu_e}|^2m_\mu^0+\alpha_w|U^\ell_{\tau\nu_e}|^2m_\tau^0\nonumber\\
&+&U^{\nu_e u}_L U^{u\nu_e}_R m^0_u/N_c+U^{\nu_e c}_L U^{c\nu_e}_R m^0_c/N_c +U^{\nu_e t}_L U^{t\nu_e}_R m^0_t/N_c\nonumber\\
&\approx& 
(11/2N_c) U^{\nu_e t}_L U^{t\nu_e}_R m^0_t=(11/2N_c) {\mathcal M}_7 m^0_t
\label{1gapnu0}
\\
%\nonumber\\
m^0_e&\approx&\alpha_w|U^\ell_{e\nu_e}|^2m_{\nu_e}^0+
\alpha_w|U^\ell_{e\nu_\mu}|^2m_{\nu_\mu}^0+\alpha_w|U^\ell_{e\nu_\tau}|^2m_{\nu_\tau}^0\nonumber\\
&+&U^{e d}_L U^{de}_R m^0_d/N_c
+U^{e s}_L U^{se}_R m^0_s/N_c +U^{e b}_L U^{be}_R m^0_b/N_c\nonumber\\
&\approx& (N_c/2)\alpha_w\Big[|U^\ell_{e\nu_e}|^2m_{\nu_e}^0+
|U^\ell_{e\nu_\mu}|^2m_{\nu_\mu}^0+|U^\ell_{e\nu_\tau}|^2m_{\nu_\tau}^0\Big]
%+(15/4N_c)U^{e b}_L U^{be}_R m^0_b
\label{1gapta0}
\\
%\nonumber\\
m^0_u&\approx&\alpha_w|U_{ud}|^2m_d^0+\alpha_w|U_{us}|^2m_s^0 + \alpha_w|U_{ub}|^2m_b^0 \nonumber\\
&+& U^{\nu_e u\dagger}_L U^{u\nu_e \dagger}_R m^0_{\nu_e}+U^{\nu_\mu u\dagger}_L U^{u\nu_\mu \dagger}_R m^0_{\nu_\mu}
+U^{\nu_\tau u\dagger}_L U^{u\nu_\tau \dagger}_R m^0_{\nu_\tau}\nonumber\\
%&\approx& U^{\nu_e u\dagger}_L U^{u\nu_e \dagger}_R m^0_{\nu_e}+U^{\nu_\mu u\dagger}_L U^{u\nu_\mu \dagger}_R m^0_{\nu_\mu}+U^{\nu_\tau u\dagger}_L U^{u\nu_\tau \dagger}_R m^0_{\nu_\tau}\nonumber\\
&\approx& (19/2N_c) U^{u t}_L U^{tu}_R m^0_t=(19/2N_c) {\mathcal M}_8 m^0_t,
\label{1gapt0}
\\
%\nonumber\\
m^0_d&\approx&\alpha_w|U_{du}|^2m_u^0+\alpha_w|U_{dc}|^2m_c^0+ \alpha_w|U_{dt}|^2m_t^0\nonumber\\
&+& U^{e d\dagger}_L U^{de \dagger}_R m^0_e
+ U^{\mu d\dagger}_L U^{d\mu \dagger}_R m^0_\mu+ U^{\tau d\dagger}_L U^{d\tau \dagger}_R m^0_\tau
\nonumber\\
&\approx&(N_c/2)\alpha_w\Big[|U_{du}|^2m_u^0+|U_{dc}|^2m_c^0+ |U_{dt}|^2m_t^0\Big]
%+(N_c/2)U^{\tau d\dagger}_L U^{d\tau \dagger}_R m^0_\tau
+(3/2){\mathcal M}_9 m^0_b,
\label{1gapb0}
\end{eqnarray}
where the CKM matrix elements $|U_{ud}|\approx 0.974$, $|U_{us}|\approx 0.225$, $|U_{cd}|\approx 0.225$, 
$|U_{ub}|\approx 4.1\times 10^{-3}$, $|U_{td}|\approx 8.4\times 10^{-3}$ \cite{pdg2012},
as well as the PMNS matrix elements  
$|U^\ell_{e\nu_e}|\approx (0.801 \rightarrow 0.845)$, $|U^\ell_{\mu\nu_e}|\approx (0.514 \rightarrow 0.580)$, $|U^\ell_{\tau\nu_e}|\approx (0.137 \rightarrow 0.158)$,
$|U^\ell_{e\nu_\mu}|\approx (0.225 \rightarrow 0.517)$,
$|U^\ell_{e\nu_\tau}|\approx (0.246 \rightarrow 0.529)$,
$|U^\ell_{\mu\nu_\tau}|\approx (0.464 \rightarrow 0.713)$ 
and $|U^\ell_{\mu\nu_\mu}|\approx (0.441 \rightarrow 0.699)$ \cite{PMNS}.
The dominate contributions in the RHS of these equations can be figured out. 
We obtain the approximate solution to Eqs.~(\ref{1gapnu0}) and (\ref{1gapt0}),
as well as the approximate solution to Eqs.~(\ref{1gapta0}) and (\ref{1gapb0}),
which are given in the last step with
\begin{eqnarray}
{\mathcal M}_7\equiv U^{\nu_e t}_L U^{t\nu_e}_R, \quad {\mathcal M}_8\equiv U^{u t}_L U^{tu}_R
, \quad {\mathcal M}_9\equiv U^{bd}_L U^{db}_R. 
\label{1gapb0m}
\end{eqnarray}
The dominate contributions are: (i) the $\nu_e$-neutrino 
acquires its mass $m^0_{\nu_\mu}$ from the $t$-quark mass $m^0_t$ via the quark-lepton
interaction ${\mathcal M}_7$; 
(ii) the $u$-quark acquires its mass $m^0_u$ from the $t$-quark mass $m^0_t$ 
via the quark-lepton interaction ${\mathcal M}_8$; (iii) the $e$-lepton 
acquires its mass $m^0_e$ from the neutrino masses $m^0_{\nu_e}$, $m^0_{\nu_\mu}$
and $m^0_{\nu_\tau}$ via the PMNS mixing, %as well as from the quark-lepton interaction ${\mathcal M}_9$
which implies the approximate relation of light lepton masses and 
PMNS mixing angles; 
(iv) the $d$-quark dominantly acquires its mass $m^0_d$ from quark masses $m_u,m_c$ and $m_t$
via the CKM mixing, as well as a small contribution from 
the quark-lepton interaction ${\mathcal M}_9$,
which implies the approximate relations of light quark masses and CKM mixing angles.

\subsubsection{Running fermion masses and Yukawa couplings}

\begin{figure}%[!h]
\begin{center}
\includegraphics[height=1.40in]{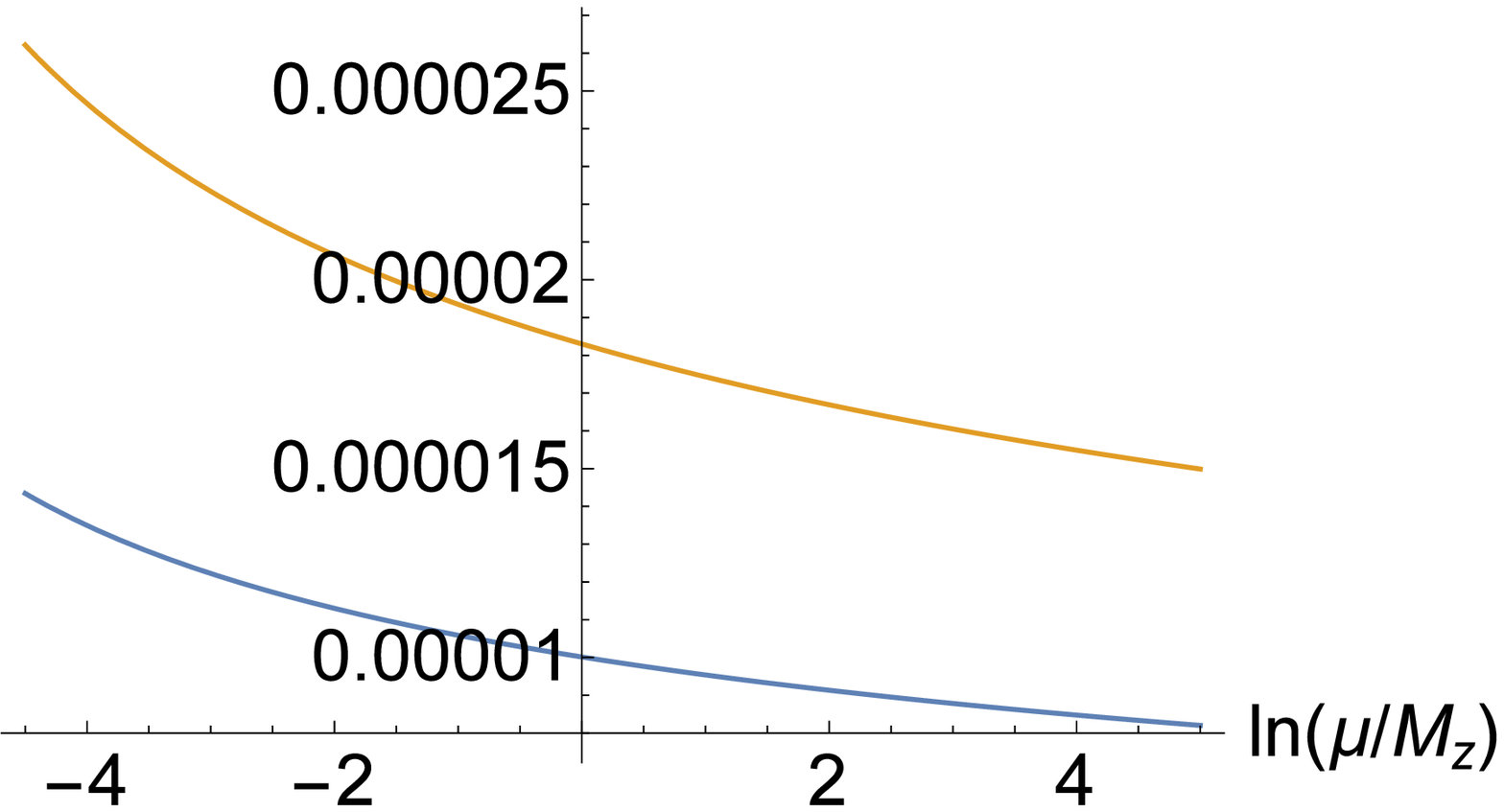}
\put(-120,105){\footnotesize $\bar g_{u,d}(\mu)$}
\put(-110,65){\footnotesize $\bar g_d(\mu)$}
\put(-110,25){\footnotesize $\bar g_u(\mu)$}
\caption{The Yukawa couplings $\bar g_u(\mu)$ and 
$\bar g_d(\mu)$ are plotted in the range 
$1.0 ~{\rm GeV}\lesssim \mu \lesssim 13.5~ {\rm TeV}$ 
for ${\mathcal M}_8$ (\ref{1mud8}) and ${\mathcal M}_9$ (\ref{1mud9}). Note that 
$m_{u,d}(\mu)=\bar g_{u,d}(\mu)v/\sqrt{2}$.} \label{figyl1}
\end{center}
\end{figure}

\begin{figure}%[!h]
\begin{center}
\includegraphics[height=1.40in]{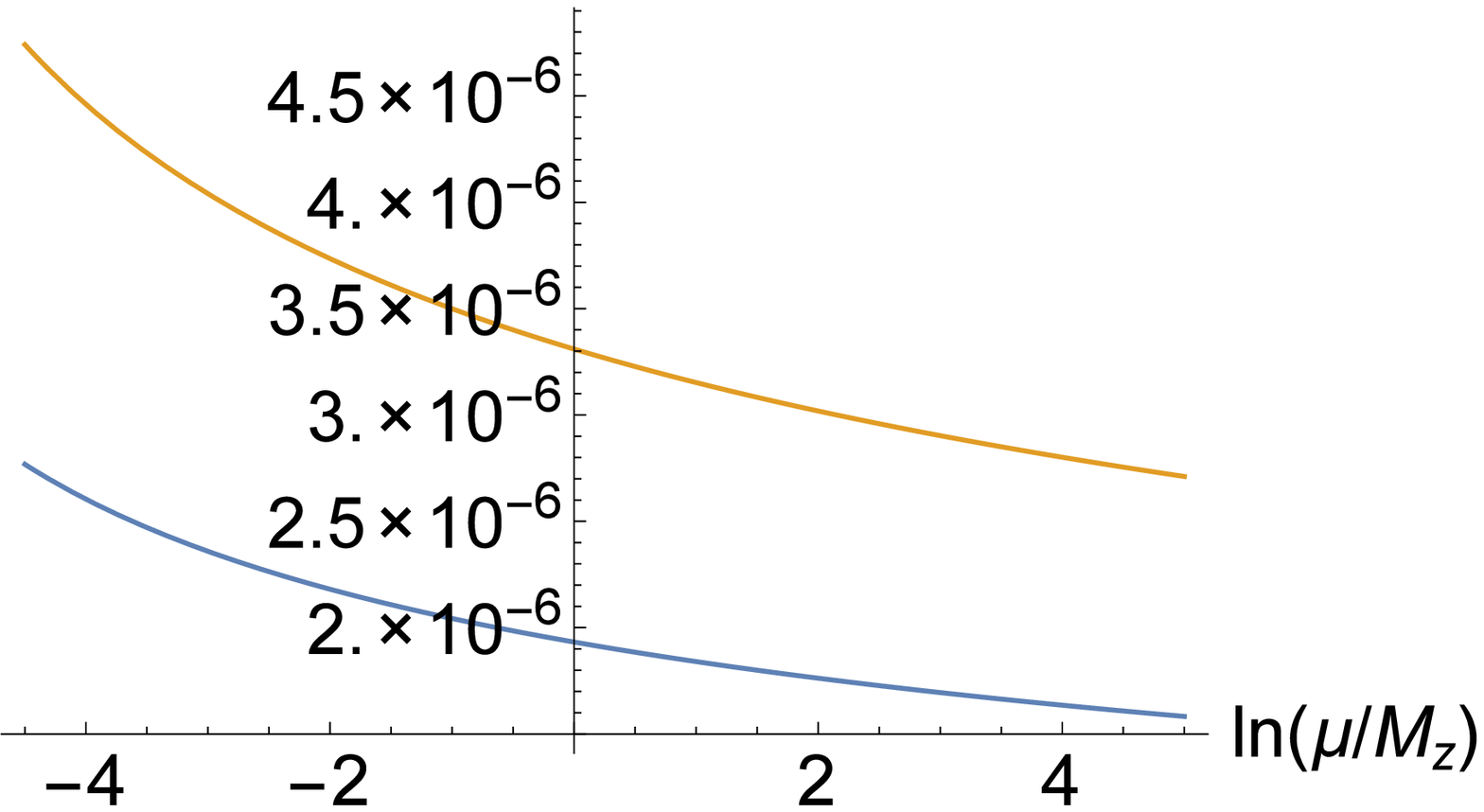}
\put(-150,105){\footnotesize $\bar g_{e,\nu_e}(\mu)$}
\put(-85,55){\footnotesize $\bar g_e(\mu)$}
\put(-90,22){\footnotesize $10\times \bar g_{\nu_e}(\mu)$}
\caption{The Yukawa couplings $\bar g_e(\mu)$ and 
$\bar g_{\nu_e}(\mu)$ are plotted in the range 
$1.0 ~{\rm GeV}\lesssim \mu \lesssim 13.5~ {\rm TeV}$ 
for ${\mathcal M}_7$ (\ref{1mud7}). Note that 
$m_{e,\nu_e}(\mu)=\bar g_{e,\nu_e}(\mu)v/\sqrt{2}$.} \label{figyl1'}
\end{center}
\end{figure}

Analogously to the discussion for the third  fermion family from 
Eqs.~(\ref{gapnu}-\ref{gapb}) to Eqs.~(\ref{rtaunu}-\ref{rbm}), 
neglecting the perturbative corrections from the SM gauge interactions, and  
defining running fermion masses and Yukawa couplings 
\begin{eqnarray}
m_{\nu_e}(\mu)&=&\bar g_{\nu_e}(\mu)v/\sqrt{2},\quad 
m_e(\mu)=\bar g_e(\mu)v/\sqrt{2},\nonumber\\
m_u(\mu)&=&\bar g_u(\mu)v/\sqrt{2},\quad
m_d(\mu)=\bar g_d(\mu)v/\sqrt{2},
\label{yukawa1}
\end{eqnarray}
and the gap-equations at the scale $\mu$ are obtained by replacing 
$m_f^0\rightarrow m_f(\mu)$ in Eqs.~(\ref{1gapnu0}-\ref{1gapb0}).
On the basis of Eqs.~(\ref{1gapnu0},\ref{1gapt0}) and (\ref{yukawa1}) at the scale $\mu$, 
we numerically calculate the $\nu_e$, $e$, 
$u$- and $d$-quark masses at $\mu=2$~GeV  
\begin{eqnarray}
m_{\nu_e}&\approx& 4.3 ~{\rm KeV}, \quad {\rm for}\quad {\mathcal M}_7=U^{\nu_e t}_L U^{t\nu_e}_R \approx 1.0 \times10^{-7}\label{1mud7}\\ 
m_u&\approx& 2.2~{\rm MeV},\quad {\rm for}\quad {\mathcal M_8} =(U^{ut}_L U^{tu}_R) \approx 3.0 \times 10^{-6}\label{1mud8}\\
m_d&\approx& 4.1~{\rm MeV},\quad {\rm for}\quad {\mathcal M_9} =(U^{bd}_L U^{db}_R) \approx 4.0 
\times 10^{-4}\label{1mud9}\\
m_e&\approx& 0.7 ~{\rm MeV}, %\quad {\mathcal M}_{10}=U^{e b}_L U^{be}_R \approx 1.0 \times10^{-5}
\label{1mud10}
\end{eqnarray}
and the Yukawa couplings $\bar g_u(\mu)$ 
and $\bar g_d(\mu)$, see Fig.~\ref{figyl1}, and $\bar g_e(\mu)$ 
and $\bar g_{\nu_e}(\mu)$, see Fig.~\ref{figyl1'}. 
The variations of Yukawa couplings $\bar g_{u,d,e,\nu_e}(\mu)$ are very small over 
the energy scale $\mu$.

%\vskip0.9cm
\subsection
{Summary and discussion}
\hskip0.1cm 
We show that the top-quark mass and Yukawa coupling $m_t(\mu)=\bar g_t(\mu)v/\sqrt{2}$, which 
is originated from the SSB, inevitably introduce the inhomogeneous (ESB) terms 
into the SD equations for other fermion masses via the fermion-family mixing 
due to the quark-lepton interactions and the $W^\pm$-boson vector-like vertex  
(CKM and PMNS mixing) at high energies.
As a consequence, this leads to the generations of 
other fermion masses by the ESB mechanism, and 
their Yukawa couplings ($m_f(\mu)=\bar g_f(\mu)v/\sqrt{2}$) are functions of 
the top-quark Yukawa coupling $\bar g_t(\mu)$, Fig.~\ref{figyt}. 
We approximately analyze the coupled SD gap-equations for 
the fermion masses and Yukawa couplings of the third, 
second and the first family of the SM. With the knowledge of the CKM and PMNS matrices, 
as well as the fermion mass spectra, we try to identify
the dominate ESB contributions to the SD gap-equations, and approximately find 
their masses, consistently with the fermion-family mixing parameters ${\mathcal M}_i$. 
We have checked that the contributions from perturbative gauge interactions are negligible, 
compared with the essential contributions due to the fermion-family mixing.
As qualitative and preliminary results, without any drastic fine-tuning
we approximately obtain the hierarchy pattern of 12 SM-fermion masses, 
see Table \ref{spectrum}, and their Yukawa couplings,
consistently with the parameter $\alpha_w$ (\ref{alphaw}) 
and the hierarchy pattern of 9 family-mixing parameters ${\mathcal M}_i$.  

\begin{table}[t]
\begin{center}
\begin{tabular}{|c|c|c|c|}
\hline %\backslashbox[15mm]
$m_{\nu_e}\approx 4.3\times 10^{-6}$ GeV & $m_{\nu_\mu}\approx 2.4\times 10^{-3}$ GeV & 
$m_{\nu_\tau}\approx 2.4\times 10^{-1}$ GeV  \\
\hline  
$m_e\approx 7.0\times 10^{-4}$ GeV &$m_\mu\approx 1.2\times 10^{-1}$ GeV    
& $m_\tau\approx 1.7\times 10^{0}$ GeV \\
\hline
$m_u\approx 2.2\times 10^{-3}$ GeV
& $m_c\approx 1.3\times 10^{0}$ GeV &  $m_t\approx 1.7\times 10^{2}$ GeV \\
\hline
$m_d\approx 4.1\times 10^{-3}$ GeV
&$m_s\approx 9.1\times 10^{-2}$ GeV  & $m_b\approx 4.2\times 10^{0}$ GeV \\
\hline
\end{tabular}
\end{center}
\caption{We present our qualitative result of the hierarchy spectrum of 
12 SM-fermion masses, which seems to be consistent with the SM. The top-quark mass
is generated by the SSB, and others by the ESB attributed to the top-quark mass 
and family-mixing. All masses are calculated at $\mu=2$~GeV, except $m_t$, $m_b$ and 
$m_\tau$ calculated by their mass-shell conditions.} 
\label{spectrum}
\end{table}

It is energetically favorable that the SSB solely occurs 
for the $\bar t t$-channel (\ref{bhl}) generating the top-quark mass and three Goldstone modes only. The SSB realizes the approximate ground state (vacuum), 
in which the pattern of fermion masses is $m_t\not=0$ and $m_{f\not=t}=0$. However, 
this SSB generated vacuum alignment is re-arranged to
the real ground states, where the real hierarchy pattern (table \ref{spectrum}) is realized. 
Such rearrangement is due to the nontrivial ESB terms in the SD gap-equations for
fermion masses, so that fermions become massive $m_t\gg m_{f\not=t}\not=0$. 
These ESB terms are introduced by the top-quark mass and 
fermion-family mixing matrices in the two ways: (i) 
the fermion-family-mixing matrices (\ref{mmud}) and (\ref{mmnul}) including the CKM and PMNS matrices introduce the ESB terms, due to the vector-like coupling 
$\alpha_w$ (\ref{vr}) and (\ref{alphaw}) of the $W^\pm$-boson at high energies ($\E$)
(see preliminary study \cite{xue1997mx,xue1999nu}); 
(ii) the quark-lepton-family mixing matrices
(\ref{mql1}) introduce the ESB terms, due to the quark-lepton interactions (\ref{bhlqlm2})
at high energies ($\E$). 
It is expected that the ESB terms perturbatively re-arrange 
the SSB generated vacuum alignment,
because of the small coupling $\alpha_w$ and fermion-family-mixing matrix elements. 
The Table (\ref{spectrum}) shows that the following 
relations between (i) neutrino Dirac masses and charged (2/3) quark masses; (ii)   
charged lepton and charged (-1/3) quark masses;
\begin{eqnarray}
m_{\nu_\tau}:m_{\nu_\mu}:m_{\nu_e}&\approx &m_t:m_c:m_u\approx 10^{-5}:10^{-2}:1,\nonumber\\
m_{\tau}:m_{\mu}:m_{e}&\approx & m_b:m_s:m_d\approx 10^{-4}:10^{-2}:1.
\label{rel1}
\end{eqnarray}

In conclusion, the spectrum of fermion masses, i.e., the structure of eigenvalues 
of fermion mass matrices mainly depends on the ESB terms that relats to 
the unitary matrices or mixing matrices between three fermion-flavor families 
and four families of fermions with different charges. We cannot theoretically determine 
these matrices, except for adopting those CKM- and PMNS-matrix elements 
already experimentally measured. If these fermion-family mixing-matrix elements
are small deviations from triviality, namely the hierarchy pattern likes the 
observed CKM matrix, the pattern of fermion masses is hierarchy, and {\it vice versa}.
In this article, the hierarchy pattern of fermion masses (Yukawa couplings) 
is obtained consistently with the hierarchy pattern of fermion-family 
mixing-matrix elements. 
It should be mentioned that both of them are equally the basic parameters of the Nature, 
and they are closely related each other by the symmetries and/or dynamics 
of the fundamental theory. For the light quarks and leptons, they acquire their masses
dominantly from the ESB terms of the $W^\pm$-boson coupling $\alpha_w$-terms associating with 
either CKM or PMNS matrix. This implies that there are the approximate relations 
of light quark/lepton masses and CKM/PMNS mixing angles.
Some other relations between them are given in this article, 
however more fundamental relations 
are expected in the framework of unification theories, e.g., $SO(10)$-theory.

It should be emphasized that we have at the infrared scale %``$\mu$'' 
12 SD equations for 12 SM fermion masses coupled together via the 
fermion-family-mixing matrices (\ref{mmud}), (\ref{mmnul}) and (\ref{mql1}), which are unknown except the well (poor) known CKM (PMNS) matrix. These mixing matrices have to be understood in a UV-fundamental theory symmetrically unifying not only gauge interactions but also three fermion families. In this sense, fermion mixing matrices are even more fundamental than fermion masses. Their values, mixing matrix elements 
and fermion masses in unit of the top-quark mass, are related and determined upon the chiral-symmetry-breaking ground state of 
the UV-fundamental theory. The presented results only 
show that the known hierarchical masses ($10^{0}-10^{-8}$) 
of 12 SM Dirac fermions are related to 
the hierarchical pattern of 9 fermion-family mixing parameters ${\mathcal M}_i$ ($10^{0}-10^{-7}$) of Eqs.~(\ref{pm1}), (\ref{pm2}) and (\ref{1gapb0m}). Since we 
have not understood the hierarchical mixing-matrix pattern of the UV-fundamental theory, the hierarchical fermion masses are not ultimately explained.
It should be also emphasized that the presented results are preliminarily qualitative, and far from 
being quantitatively compared with the SM fermion masses and precision tests of e.g., Yukawa couplings. Due to the fact that 12 coupled SD 
mass-gap equations depend on not only poorly known and totally unknown family-mixing parameters, but also running gauge couplings,     
the quantitative study of solving these SD equations is a difficult and challenging task. These results could be quantitatively improved, if
one would be able to solve coupled SD equations 
by using a numerical approach in future. 
Our goal in this article is to present an insight into a possible scenario and understanding of 
the origins and hierarchy spectrum of fermion masses in the SM without drastic fine-tuning.

In the next section, we will relabel neutrino Dirac mass $m_\nu$ by $m^D_\nu$,
discuss three heavy sterile Majorana 
neutrinos ($\nu^f_R+\nu_R^{fc}$) and three light gauged Majorana neutrinos 
($\nu^f_L+\nu_L^{fc}$) in terms of their Dirac masses $m^{\rm D}_{\nu}$ 
and Majorana masses $m^{\rm M}_{\nu}$. 

\section
{\bf Neutrino sector}\label{neutrinoS}
\hskip0.1cm 
%The some preliminary material of this section was presented in Ref.~\cite{xue2015}.
On the basis of Dirac neutrino mass eigenstates and masses calculated (see table \ref{spectrum}) 
in previous sections, as well as some experimental results of neutrino oscillations, 
we calculate the mass-spectra of gauged and sterile neutrinos  
by taking into account the Majorana masses generated by
the spontaneous symmetry breaking of the global $U_{\rm lepton}(1)$ symmetry 
for the lepton-number conservation. 

\subsection{Spontaneous symmetry breaking of $U_{\rm lepton}(1)$ symmetry}

In the four-fermion operators (\ref{l}) of the lepton sector, the last term reads
\begin{eqnarray}
-G\, \sum _{ff'}\left[ 
(\bar\nu^{f\, c}_R\nu^{f'\,}_{R})(\bar \nu^{f'}_{R}\nu^{f c}_{R})\right],
\label{ln}
\end{eqnarray}
where the conjugate fields of sterile Wely neutrinos $\nu^f_R$ are given by
$\nu_R^{f c}=i\gamma_2(\nu_R^{f})^*$. This four-fermion operator 
preserves the global $U_{\rm lepton}(1)$-symmetry for the lepton-number conservation.
Similarly to the discussions of the SSB mechanism for the generation of top-quark mass 
in Sec.~\ref{SSBS}, the four-fermion operator (\ref{ln}) can 
generate a mass term of Majorana type, since the family index ``$f$'' 
is summed over as the color index ``$a$'' and the family number 
$N_f=3$ plays the similar role as the color number $N_c$ in the 
$\langle\bar tt\rangle$-condensate (\ref{tqmassd}). We notice that the lepton-number 
is conserved in the ground state (vacuum state) realized by the SSB of the SM 
chiral gauge symmetries, whereas the lepton-number is not conserved in 
the ground state realized by the spontaneous symmetry breaking of the 
global $U_{\rm lepton}(1)$-symmetry of the Lagrangian (\ref{ln}).

On the basis of the mass eigenstates, the spontaneous symmetry breaking of the 
$U_{\rm lepton}(1)$-symmetry generates the masses of Majorana type 
\begin{eqnarray}
m^M=\sum_{f=1,2,3} m^M_f,\quad  m^M_f=-G\langle\bar\nu^{f\, c}_R\nu^{f\,}_{R}\rangle, 
\label{mam}
\end{eqnarray}
together with a sterile massless Goldstone boson, i.e.~the pseudoscalar bound 
state 
\begin{eqnarray}
\phi^M=\sum_{f=1,2,3}\langle\bar\nu^{f\,c}_R\gamma_5\nu^f_R\rangle, 
\label{mps}
\end{eqnarray}
and a sterile massive scalar particle, i.e.~the scalar bound state 
\begin{eqnarray}
\phi^M_{_H}=\sum_{f=1,2,3}\langle\sum_f\bar\nu^{f\,c}_R\nu^f_R\rangle, 
\label{mhs}
\end{eqnarray}
both of them carry two units of the lepton number.
The sterile neutrino mass $m^M$ and sterile scalar 
particle mass $m^M_{_H}$ satisfy the mass-shell conditions, 
\begin{eqnarray}
m^M=\bar g_{\rm sterile}(m^M)v_{\rm sterile}/\sqrt{2},\quad (m^M_{_H})^2/2
=\tilde \lambda_{\rm sterile} (m^M_{_H}) v_{\rm sterile}^2,
\label{smass}
\end{eqnarray}
where $\bar g_{\rm sterile}(\mu^2)$ and $\tilde \lambda_{\rm sterile} (\mu^2)$ obey the same 
RG equations (absence of gauge interactions) of Eqs.~(\ref{boun0}), (\ref{reg1}) and (\ref{reg2}), 
as well as the boundary conditions (\ref{smass}).   
However, we cannot determine the solutions $\bar g_{\rm sterile}(\mu^2)$ and $\tilde \lambda_{\rm sterile} (\mu^2)$, since the energy scale $v_{\rm sterile}$ 
of boundary conditions (\ref{smass}) are unknown.  
The electroweak scale $v$ is determined by the 
gauge-boson masses $M_{_W}$ and $M_{_Z}$ experimentally 
measured, the scale $v_{\rm sterile}$ needs to be determined by the sterile neutrino 
masses $m^M_i$ and the sterile scalar particle mass $m^M_{_H}$. In fact,
the scale $v_{\rm sterile}$ represents the energy scale of the lepton-number violation.

\subsection{Gauged and sterile Majorana neutrino masses}

The SSB and ESB of the SM chiral gauge symmetries, as well as
the spontaneous symmetry breaking of the 
$U_{\rm lepton}(1)$-symmetry result in
the following bilinear Dirac and Majorana mass terms 
\begin{eqnarray}
m^{D}_f\bar\nu^{f}_L\nu^f_{R}+ m^{M}_f\bar \nu^{f c}_{R}\nu^{f}_{R}+ {\rm h.c.},
\label{lmass}
\end{eqnarray}
in terms of neutrino mass eigenstates $\nu^f_L$ and $\nu^f_R$ in the $f$-th fermion family, see 
Eqs.~(\ref{mlnu}) and (\ref{mm}).
Following the usual approach \cite{zli,mohapatra},  diagonalizing 
the $2\times 2$ mixing matrix (\ref{lmass}) in terms
of the neutrino and sterile neutrino mass eigenstates of the family ``$f=1,2,3$'', 
we obtains two mass eigenvalues
\begin{eqnarray}
M_f^{g} &=& \frac{1}{2} \Big\{m^M_f- \left[(m^M_f)^2 
+ (m^D_f)^2\right]^{1/2}\Big\},
\label{g2lmass}\\
M_f^{s} &=& \frac{1}{2} \Big\{m^M_f + \left[(m^M_f)^2 
+ (m^D_f)^2\right]^{1/2}\Big\}.
\label{s2lmass}
\end{eqnarray}
This corresponds to two mass eigenstates: three light gauged Majorana neutrinos 
(four components)
\begin{eqnarray}
\nu^f_g=\nu^f_L+\nu_L^{fc},\quad E_g^f=[p^2+(M_f^g)^2]^{1/2},\quad M_f^{g}\approx (m^D_f)^2/4m^M_f,
\label{gmm}
\end{eqnarray}
and three heavy sterile Majorana 
neutrinos (four components)
\begin{eqnarray}
\nu_s^f=\nu^f_R+\nu_R^{fc},\quad E_s^f=[p^2+(M_f^s)^2]^{1/2},\quad M_f^{s}\approx m^M_f.
\label{smm}
\end{eqnarray}
where $p$ stands for neutrino momentum, corresponding velocity $v_p$.
The mixing angles between gauged and sterile Majorana neutrinos are 
\begin{eqnarray}
2\theta_f =\tan ^{-1} (m^D_f/m^M_f)\approx (m^D_f/m^M_f)
\ll 1,
\label{mmd}
\end{eqnarray}
The previously obtained Dirac masses $m_f\equiv m^{D}_f$ have the structure of 
hierarchy (see Table \ref{spectrum}).   
The discussions after Eq.~(\ref{mm1}) show that the Majorana masses $m^M_f$ 
are expected to have a hierarchy structure relating 
to the one of Dirac masses $m^{D}_f$ \footnote{In Ref.~\cite{xue2015}, we assume that the Majorana masses $m^M_f$ are
approximately equal (degenerate) $m^{M}_i\approx m^{M}/3$ , 
since there is not any preferential 
$i$-th component of the condensate 
$\langle\bar\nu^{i\, c}_R \nu^{i}_{R}\rangle$. This is not correct because of the
non-trivial chiral transformation ${\mathcal U}^\nu_R$ in Eqs.~(\ref{mlnu}) and (\ref{mm}).}.
This indicates the normal hierarchy structure of neutrino mass spectrum: Dirac neutrino masses
$m^D_1 < m^D_2 < m^D_3$, sterile Majorana neutrino masses  
$m^M_1 < m^M_2 < m^M_3$ (\ref{smm}) and gauged Majorana neutrino masses
$M^g_1 < M^g_2 < M^g_3$ (\ref{gmm}), i.e.,
\begin{eqnarray}
(m^D_1)^2/4m^M_1< (m^D_2)^2/4m^M_2< (m^D_3)^2/4m^M_3.
\label{mmh}
\end{eqnarray}
Moreover, due to the absence of observed lepton-violating processes 
up to the electroweak scale and the smallness of gauged neutrino masses, 
it is nature to assume that the neutrino Majorana masses are much larger than their 
Dirac masses $m^M_f\gg m_f^D$, i.e., 
the energy scale $v_{\rm sterile}$ of the lepton-number violation
is much larger than the electroweak scale $v$. 

%Due to the replicative symmetry of fermion-flavor families, 
%it is nature to expect that such oscillations between the sterile and gauged should be almost 
%independent of the fermion-flavor family. This is another point of view for Eq.~(\ref{mmd})

\subsection{Flavor oscillations of gauged Majorana neutrinos} 

We first discuss the family-flavor oscillations of three light gauged Majorana neutrinos  
(\ref{gmm}) in the usual framework. They are described by the PMNS mixing matrix 
$U^\ell_L={\mathcal U}^{\nu\,\dagger}_L{\mathcal U}^\ell_L$, the mass and
mass-squared differences of gauged Majorana neutrino mass-eigenstates ($f, f'=1,2,3$),
%. The former is not discussed in this article. The latter
which are calculated by using Eq.~(\ref{g2lmass}) 
\begin{eqnarray}
\Delta M^{g}_{ff'}&\equiv& M_f^{g}-M_{f'}^{g}\approx  \Big[\frac{(m^{D}_{f'})^2}{4m^M_{f'}}\Big]-
\Big[\frac{(m^{D}_{f})^2}{4m^M_{f}}\Big],\label{gmdmass}\\
\Delta M^{g2}_{ff'}&\equiv& (M_f^{g})^2-(M_{f'}^{g})^2 \approx 
\Big[\frac{(m^{D}_{f})^2}{4m^M_f}\Big]^2-
\Big[\frac{(m^{D}_{f'})^2}{4m^M_{f'}}\Big]^2,
\label{gmdmass'}
\end{eqnarray}
Equation (\ref{gmdmass}) 
is up to the order ${\mathcal O}\{(m^D_f)^2/4m^M_f\}$, and 
Eq.~(\ref{gmdmass'}) 
is up to the order ${\mathcal O}\{[ (m^D_f)^2/4m^M_f]^2\}$. 
The oscillating probability from the flavor $\nu^g_\alpha$ to the flavor $\nu^g_\beta$ reads
\begin{eqnarray}
P_{\nu^g_\alpha\longleftrightarrow\nu^g_\beta}(t)
=\sum_{ff'}(U^\ell_L)^*_{\alpha f}(U^\ell_L)_{\beta f}(U^\ell_L)_{\alpha f'}(U^\ell_L)^*_{\beta f'}\exp[-i(E_g^f-E_g^{f'})t].
\label{oggp}
\end{eqnarray}
The large oscillating lengths of relativistic and non-relativistic gauged neutrinos are given by 
\begin{eqnarray}
L^{ff'}_{g}&=&\frac{2\pi}{(E^f_g-E^{f'}_g)}\approx \frac{2\pi(2p)}{\Delta M^{g2}_{ff'}} \quad {\rm for}\quad
p\gg \frac{(m^D_f)^2}{4m^M_f},~~(f\rightarrow f'),
\label{glmm1}\\
L^{ff'}_{g}&=&\frac{2\pi}{(E^f_g-E^{f'}_g)}\approx \frac{2\pi}{\Delta M^{g}_{ff'}}\left(\frac{v_p}{c}\right), \quad {\rm for}\quad p \ll \frac{(m^D_f)^2}{4m^M_f},~~(f\rightarrow f'),
\label{glmm1'}
\end{eqnarray}
where the second line (\ref{glmm1'}) 
may be used for the case of cosmic neutrino background of temperature 
${\mathcal O}(10^{-4})$ eV.

\begin{table}[t]
\begin{center}
\begin{tabular}{|c|c|c|c|}
\hline %\backslashbox[15mm]
Dirac mass &
$m^D_1\approx 4.3\times 10^{-6}$ GeV & $m^D_2\approx 2.4\times 10^{-3}$ GeV & 
$m^D_3\approx 2.4\times 10^{-1}$ GeV  \\
\hline 
Majorana mass & 
$m_1^M \approx 1.7\times 10^{2}$ GeV   & 
$m^M_2 \approx 1.7\times 10^{5}$ GeV    & 
$m^M_3 \approx 2.9\times 10^{8}$ GeV \\
\hline
sterile neutrino mass & 
$M_1^{s} \approx 1.7\times 10^{2}$ GeV   & 
$M_2^{s} \approx 1.7\times 10^{5}$ GeV    & 
$M_3^{s} \approx 2.9\times 10^{8}$ GeV \\
\hline
gauged neutrino mass & 
$M_1^{g} \approx 2.8\times 10^{-5}$ eV    & 
$M_2^{g} \approx 8.4\times 10^{-3}$ eV    & 
$M_3^{g} \approx 5.0\times 10^{-2}$ eV \\
\hline
\end{tabular}
\end{center}
\caption{Spectra of neutrino Dirac masses $m_f^D$, neutrino Majorana masses $m_f^M$, sterile Majorana neutrino masses $M^s_f\approx m_f^M$ and gauged Majorana neutrino masses 
$M_f^{g}\approx (m^D_f)^2/4m^M_f$. Here we label three flavor-families by 
using notation $f=1,2,3$ instead of 
$f=\nu_e,\nu_\mu,\nu_\tau$.} \label{nspectrum}
\end{table}
 
These oscillations between the family flavors of gauged Majorana neutrinos
have been important for experiments performed in ground and underground laboratories. 
Using Eq.~(\ref{gmdmass'}), neutrino Dirac masses $m_f^D$
(Table \ref{spectrum}) and experimental values \cite{pdg2012}:
\begin{eqnarray}
%\Delta M^{-}_{21}\approx 9.3\times 10^{-4} {\rm eV},\quad
|\Delta M^{g2}_{21}|\approx 7.5 \times 10^{-5}({\rm eV})^2,
\quad
%\Delta M^{-}_{31}\approx  2.3\times 10^{-1}{\rm eV},
|\Delta M^{g2}_{31}|\approx 2.5\times 10^{-3}({\rm eV})^2,
\label{s2mass'}
\end{eqnarray}
neglecting the term $[(m^D_1)^2/4m_1^M]^2$ we obtain ratios $m^D_3/m_3^M=8.3\times 10^{-10}$ 
and $m^D_2/m_2^M=1.4\times 10^{-8}$, and the Majorana masses 
$m_3^M\approx 2.9\times 10^{8}$ GeV and $m_2^M\approx 1.7\times 10^{5}$ GeV. As a result, 
two sterile Majorana neutrino masses (\ref{smm}) $M_3^s\approx 2.9\times 10^{8}$ GeV 
and $M_2^s\approx 1.7\times 10^{5}$ GeV, 
two gauged Majorana neutrino masses (\ref{gmm}) $M_3^g\approx 5.0\times 10^{-2}$ eV
and $M_2^g\approx 8.4\times 10^{-3}$ eV, see Table \ref{nspectrum}. 

Among the three neutrino mass-squared differences (\ref{gmdmass'}), only two of them
are independent 
for $\Delta M^{g2}_{32}= \Delta M^{g2}_{31}-\Delta M^{g2}_{21}\approx \Delta M^{g2}_{31}$.  
In principle we cannot determine the values $m^M_1$, $m^D_1/m^M_1$,
and $[(m^D_1)^2/4m_1^M]^2$. 
However, we infer the hierarchy structure of Majorana masses $m^M_{2}$ and $m^M_{1}$
\begin{eqnarray}
(m^M_2/m^M_1)\approx (m^M_3/m^M_2)\approx 1.0\times 10^3, \quad {\rm then} \quad m^M_1\approx 1.7\times 10^{2} {\rm GeV}, 
\label{hmm}
\end{eqnarray}
on the basis of the reasons we discussed in the paragraph of Eq.~(\ref{mm1}). 
This inference (\ref{hmm}) leads to the ratio $m_1^D/m_1^M\approx 2.5\times 10^{-8}$ 
and the lowest lying Majorana neutrino mass
\begin{eqnarray}
M^g_1\approx (m_1^D)^2/4m^M_1\approx 2.8\times 10^{-5} {\rm eV}. 
\label{lmm}
\end{eqnarray}
Thus we tabulate the values (\ref{hmm}) and (\ref{lmm}) in the first column of Table \ref{nspectrum}.
These results satisfy the recent cosmological constrain \cite{ccon} on the total mass of three
light gauged neutrinos (\ref{gmm}),
\begin{eqnarray}
\sum_{f=1,2,3} M^g_f \approx \sum_{f=1,2,3} (m^D_f)^2/4m^M_f\approx 5.8\times 10^{-2} {\rm eV}
%\approx (A/4)\sum_f m_f^D \approx (A/4) m_3^D 
< 2.3\times 10^{-1} {\rm eV}.
\label{tmc}
\end{eqnarray}
Needless to say, it is important that the sensitivity of experiments and observations 
on neutrino masses can be reached at least to the level ${\mathcal O}(10^{-2})$ eV.  

\subsection{Flavor oscillations of sterile Majorana neutrinos}

We turn to discuss the family-flavor oscillations of three heavy sterile 
Majorana neutrinos (\ref{smm}). They are described by the mixing matrix 
$U_R^\ell={\mathcal U}^{\nu\,\dagger}_R{\mathcal U}^\ell_R$ (\ref{rtan}), the mass and
mass-squared differences of sterile Majorana neutrino mass-eigenstates ($i, j=1,2,3$),
which are calculated by using Eq.~(\ref{s2lmass}) 
\begin{eqnarray}
\Delta M^{s}_{ff'}&\equiv& M_f^{s}-M_{f'}^{s}\approx 
m^{M}_{f}-m^{M}_{f'}
%\approx A^{-1}\Delta m^{D}_{ff'} 
\label{sdmass}\\
\Delta M^{s2}_{ff'}&\equiv& (M_f^{s})^2-(M_{f'}^{s})^2 \nonumber\\
&\approx& \frac{1}{2} 
(2\Delta m^{2M}_{ff'} +  \Delta m^{2D}_{ff'})
\approx \Delta m^{2M}_{ff'}.
\label{s2mass}
\end{eqnarray}
Equation (\ref{sdmass}) 
is up to the order ${\mathcal O}\{(m^D_i)^2/4m^M_i\}$,
Eq.~(\ref{s2mass}) 
is up to the order ${\mathcal O}\{[ (m^D_i)^2/4m^M_i]^2\}$ and the 
definitions are
\begin{eqnarray}
%\Delta m^{M}_{ff'}&\equiv& (m^{M}_{f})-(m^{M}_{f'}),\quad 
\Delta m^{2M}_{ff'}\equiv (m^{M}_{f})^2-(m^{M}_{f'})^2,
%\label{mdmass}\\
%\Delta m^{D}_{ff'}&\equiv& (m^{D}_{f})-(m^{D}_{f'}),
\quad 
\Delta m^{2D}_{ff'}\equiv (m^{D}_{f})^2-(m^{D}_{f'})^2.
\label{ddmass}
\end{eqnarray}
We $\Delta m^{2M}_{ff'}\gg\Delta m^{2D}_{ff'}$ and $\Delta m^{M}_{ff'}\gg\Delta m^{D}_{ff'}$, 
see Table \ref{nspectrum}. The oscillating probability from the sterile 
flavor $\nu^s_\alpha$ to the sterile flavor $\nu^s_\beta$ reads
\begin{eqnarray}
P_{\nu^s_\alpha\longleftrightarrow\nu^s_\beta}(t)
=\sum_{ff'}(U^\ell_R)^*_{\alpha f}(U^\ell_R)_{\beta f}(U^\ell_R)_{\alpha f'}(U^\ell_R)^*_{\beta f'}\exp[-i(E_s^f-E_s^{f'})t].
\label{ossp}
\end{eqnarray}
The oscillating lengths of non-relativistic and relativistic sterile neutrinos are given by
\begin{eqnarray}
L^{ff'}_{s}&=&\frac{2\pi}{(E^f_s-E^{f'}_s)}\approx \frac{2\pi}{\Delta M^{s}_{ff'}}\left(\frac{v_p}{c}\right), \quad {\rm for}\quad
m^D_f\ll p \ll m_f^M,~~(f\rightarrow f')
\label{slmm1}\\
L^{ff'}_{s}&=&\frac{2\pi}{(E^f_s-E^{f'}_s)}\approx \frac{2\pi(2p)}{\Delta M^{s2}_{ff'}}, \quad {\rm for}\quad
m^D_f\ll m_f^M < p,~~(f\rightarrow f').
\label{slmm1'}
\end{eqnarray}
Table \ref{nspectrum} shows the large mass and mass-squared 
differences (\ref{sdmass}) and (\ref{s2mass}), therefore in addition to their sterility  the 
oscillating lengths between the flavors of sterile Majorana neutrinos are too small to 
be relevant for experiments in ground and underground laboratories. However 
these oscillations could be relevant in early universe evolution, depending on the Majorana masses
$m^M$ or the energy scale $v_{\rm sterile}$ of the lepton-number violation. 

\subsection{Oscillations between gauged and sterile Majorana neutrinos}

Following Eqs.~(\ref{g2lmass}-\ref{mmd}), the oscillating probability between two mass eigenstates 
of gauged Majorana neutrino $\nu^f_g$ and sterile Majorana neutrino $\nu^f_s$ reads
\begin{eqnarray}
P_{\nu^f_g\longleftrightarrow\nu^f_s}(t)
=1-2^{-1}\sin^22\theta_f [1-\cos(E_g^f-E_s^f)t],
\label{omm}
\end{eqnarray}
where  $f=1,2,3$ and 
\begin{eqnarray}
(E^f_s-E^f_g)\approx m^M_f, \quad 
(m^D_f)^2/4m^M_f\ll p \ll m_f^M,
\label{dlmm1}\\
(E^f_s-E^f_g)\approx (m^M_f)^2/(2p), \quad
(m^D_f)^2/4m^M_f\ll m_f^M < p .
\label{dlmm1'}
\end{eqnarray}
for non-relativstic and relativistic cases. The oscillating lengths read 
\begin{eqnarray}
L^f_{sg}&=&\frac{2\pi}{(E^f_s-E^f_g)}\approx \frac{2\pi}{m^M_f}\left(\frac{v_p}{c}\right), \quad {\rm for}\quad
\frac{(m^D_f)^2}{4m^M_f}\ll p \ll m_f^M,
\label{lmm1}\\
L^f_{sg}&=&\frac{2\pi}{(E^f_s-E^f_g)}\approx \frac{2\pi(2p)}{(m^M_f)^2}, \quad {\rm for}\quad
\frac{(m^D_f)^2}{4m^M_f}\ll m_f^M < p~ .
\label{lmm1'}
\end{eqnarray}
The large  values of Majorana mass $m^M_f$ 
and mass-squared $(m^M_f)^2$ [see Table \ref{nspectrum}] 
show the small oscillating lengths. 
The small mixing angle (\ref{mmd}) indicates the small oscillating probabilities (\ref{omm}) between gauged and sterile Majorana neutrinos.

The oscillating probability between the sterile flavor $\nu^s_\alpha$ and the gauged 
flavor $\nu^g_\beta$ reads
\begin{eqnarray}
P_{\nu^s_\alpha\longleftrightarrow\nu^g_\beta}(t)
=\sum_{ff'}(U^\ell_R)^*_{\alpha f}(U^\ell_L)_{\beta f}(U^\ell_R)_{\alpha f'}(U^\ell_L)^*_{\beta f'}\exp[-i(E_s^f-E_g^{f'})t].
\label{osgp}
\end{eqnarray}
Apart from mixing matrices, via the oscillatory factor $\exp[-i(E_s^f-E_g^{f'})t]$ 
the oscillation probability  
depends on the sum over the mass differences $\Delta m_{ff'}$ or 
mass-squared differences $\Delta m^2_{ff'}$ of mass-eigenstates ($f\not= f'$) 
of two flavor neutrinos $\nu^s_\alpha$ and $\nu^g_\beta$. Given neutrino energies, 
and their masses or mass-squared differences, one can select an oscillating length $L_{ff'}$ 
that is relevant for a possible observation or effect.  
The mass spectra (Table \ref{nspectrum}) 
of gauged and sterile neutrinos show a large difference of their mass scales, indicating
oscillations between them at short distances. For the 
example $M_1^s$ and $M_1^g$ cases, the oscillating length is at least $10^{-1}~{\rm GeV}^{-1}(v_p/c)$ for $p\ll 10^{2}~{\rm GeV}$ 
or $10^{-1}~{\rm GeV}^{-1}(p/10^{2}~{\rm GeV})$ for 
$p\gg 10^{2}~{\rm GeV}$,  as shown in Eqs.~(\ref{lmm1}) and (\ref{lmm1'}). The latter implies 
the possibility (\ref{osgp}) for very 
high-energy electron neutrinos converting themselves into sterile neutrinos.   
It seems to be hard to detect the oscillations between gauged and sterile Majorana neutrinos
in experiments performed in space, ground and underground laboratories. However 
these oscillations could be important in early universe evolution, depending on the energy scale $v_{\rm sterile}$ or $m^M$ of the lepton-number violation. 
  
\comment{
one keeps the leading term by neglecting the
small mass-squared differences for short baseline experiments.
Since $\Delta m^{2D}_{ff'}\gg \Delta m^{2M}_{ff'}$, the neutrino mass-squared difference
$\Delta m^{2M}_{ff'}$ accounts for  
neutrino flavor oscillations with $E_\nu/L\sim\Delta m^{2M}_{12}\simeq 5
\times 10^{-3}{\rm eV^2}$ in the long-baseline experiments, where $E_\nu$ and $L$ 
respectively are neutrino energy and travel distance from a 
source to a detector.
Whereas the neutrino mass-squared difference 
$\Delta m^{2D}_{ff'}$ may accounts for neutrino 
oscillations $E_\nu/L\sim\Delta m^{2D}_{ff'}\gg 10^{-3}{\rm eV^2}$ in short 
baseline experiments \cite{Aguilar:2001ty}. 
}
 
Actually, the probabilities of three flavor oscillations  
(\ref{oggp}), (\ref{ossp}) and (\ref{osgp}) are described by the following $6\times 6$ mixing matrix, see Eq.~(\ref{mmnul}))
\begin{equation}
\frac{1}{\sqrt{2}}
\begin{pmatrix}
 ~~~~~~   U^\ell_L ~~~~~~ e^{i\varphi_1}{\mathcal U}^{\nu\dagger}_L{\mathcal U}^{\ell}_R\\
    e^{i\varphi_2}{\mathcal U}^{\nu\dagger}_R{\mathcal U}^{\ell}_L ~~~~~~U^\ell_R~~~~~~ 
\end{pmatrix},
\label{66}
\end{equation}
where $e^{i\varphi_1}$ is a relative phase between ${\mathcal U}^{\nu\dagger}_L$ and 
${\mathcal U}^{\ell}_R$, and 
$e^{i\varphi_2}$ is another relative phase between ${\mathcal U}^{\nu\dagger}_R$ and 
${\mathcal U}^{\ell}_L$. The maxing matrix (\ref{66}) is unitary, if $\varphi_2-\varphi_1=n\pi, ~~n=1,2,3,\cdot\cdot\cdot$. The diagonal parts $U^\ell_L$ (PMNS) and $U^\ell_R$ 
respectively represent the mixing matrices for the gauged flavor oscillations (\ref{oggp}) 
and sterile flavor oscillations (\ref{ossp}), and the off-diagonal parts represent the mixing matrices for the gauged-sterile flavor oscillations (\ref{osgp}). 
   
%\vskip0.1cm
\section
{\bf A summary and some remarks}\label{darkmatter}
\hskip0.1cm  
We end this lengthy article by making some relevant remarks and preliminary discussions on possible 
consequences of SM gauged particle, Majorana sterile and gauged neutrino spectra, 
Tables \ref{spectrum} and \ref{nspectrum} qualitatively obtained in this article.  

\subsection{SM fermion Dirac masses and Yukawa couplings}

\comment{NPB Our goal in this article is to present a possible framework and understanding of 
the hierarchy spectrum of fermion masses and Yukawa couplings in the SM.
Since decades, there have been many studies made 
for this long standing issue,
as very briefly described in the introductory and other sections 
with a long list of references, which is certainly not complete. 
Indeed, there are different theoretical approaches for instance 
the string-theory or the phenomenological approach to the issue. 
Thought the issue has not been satisfactorily solved, 
in particularly for the neutrino sector, 
some relations between light-fermion masses and mixing angles are found. In the theoretical scenario of the EC Lagrangian and SM content, 
we point out the SSB-generated top-quark mass from 
which all other fermion masses are ESB-generated via 
both the quark-lepton and fermion-family mixing, especially the
top-quark-neutrino interaction. 
To show this including Dirac neutrino masses,
the twelve  SD equations are approximately solved. 
This gives rise to the light-neutrino masses, provided 
the neutrino Majorana mass is SSB-generated by the interaction 
of four sterile neutrinos at the scale of lepton-number violation. 
As a result, though preliminarily 
we achieve the hierarchy patterns of the SM three-family 
fermion masses and mixing,
as well as the energy-scale dependent Yukawa couplings, 
the latter is related to experimental measurements.
These definitely add valuable additions to the old studies of this 
issue by different approaches adopted.}

Due to the ground-state (vacuum) alignment of the effective theory of relevant 
four-fermion operators, the top-quark mass is generated by the SSB, 
and other fermion masses are originated from the ESB terms, which are induced by 
the top-quark mass via the fermion-family mixing, quark-lepton interactions and 
vector-like $W^\pm$-boson coupling at high energies. As a consequence, the fermion
masses are functions of the top-quark mass and 
the fermion Yukawa couplings are functions of the top-quark Yukawa coupling. 
Based on the approach adopted and the results obtained in Ref.~\cite{xue2016}, 
we study the inhomogeneous SD-equations for all SM fermion masses with the ESB terms
and obtain the hierarchy patter of fermion masses and Yukawa couplings, consistently 
with the hierarchy patter of the fermion-family mixing matrix elements. 
However, we do not discuss the detailed properties of the 
quark-flavor mixing matrices (\ref{mmud}), 
the lepton-flavor mixing matrices (\ref{mmnul}) [or (\ref{66})], where the CKM matrix 
$U^q={\mathcal U}^{u\dagger}_L{\mathcal U}^{d}_L$ and PMNS matrix 
$U^\ell_L={\mathcal U}^{\nu\dagger}_L{\mathcal U}^{\ell}_L$ are particular examples relating to
the coupling vertex of $W^\pm$-bosons. Also we do not discuss the
quark-lepton flavor mixing matrices (\ref{mql1}) relates 
to the quark-lepton interactions. These unitary matrices are composed by the 
eigenvectors corresponding to eigenvalues (fermion-mass spectra) of fermion-mass matrices. 
They code all information about mixing angles and CP-violations. 
%lepton CP-angle comparing with CP-angle in quark sector in the work 1997.

\comment{NPB The top-quark Yukawa coupling $\bar g_{t}(\mu)$ in fact relates to 
the non-vanishing form-factor $\tilde Z_H(\mu)$ of composite 
Higgs boson, see Eq.~(\ref{boun0}). Both of them, as shown in Fig.~\ref{figyt},
are of the order of unity and slowly vary from $1$ GeV to $5$ TeV. This means that 
the composite Higgs boson is a tightly bound state, as if an elementary Higgs boson.}
Relating to the slowly varying Yukawa coupling $\bar g_{t}(\mu)$ of the top quark, 
see Sec.~\ref{expH}, 
all fermion Yukawa couplings %$\bar g_{b}(\mu)$ and $\bar g_{\tau}(\mu)$ also 
obtained slowly vary from $1$ GeV to $13.5$ TeV. These features 
imply that it should be hard to have any
detectable nonresonant signatures in the LHC $pp$-collisions, showing the deviations 
from the SM with the elementary Higgs boson. 
All these results are preliminarily qualitative, and they should receive the high-order corrections and some non-perturbative contributions. It should be emphasized that these qualitative results cannot to be quantitatively compared with the SM precision tests. 
The quantitative study is a difficult and challenging task and 
one will probably be able to carry on it by using a numerical approach in future.
Nevertheless, these qualitative results may give us some insight into the 
long-standing problem of fermion-mass origin and hierarchy.

\subsection{Neutrinos and dark-matter particles}

The values of three light gauged Majorana neutrino masses $M^g_f$ give some insight into the neutrino problems that directly relate to the absolute values of neutrino masses. The $\beta$-decay rate depends on $m^2_\beta=\sum_f|U^\ell_{ef}|^2(M^{g}_f)^2$. 
The double $\beta$-decay rate depends on
$m_{2\beta} = \sum_f |U^\ell_{ef}|^2M^g_f\tilde\eta_f$, where $\tilde \eta_f=\pm 1$ 
is the CP eigenvalue of the mass eigenstate $\nu^g_f$ (\ref{gmm}). The $M^g_3$ and $M^g_2$ 
values, as well as $M^g_1$ range in Table \ref{nspectrum} seem to be in agreement with the analysis of using experimental data of mass-squared differences (\ref{s2mass'}) and the PMNS
mixing matrix $U^\ell_{L}$
in the normal hierarchy case (see for example Ref.~\cite{giuntibook}). In addition to the measurements of neutrino mixing angles, it is obviously important to 
experimentally measure neutrino masses with a sensitivity below $10^{-2}$ eV  so as to
determine the neutrino features. 

The very massive sterile neutrinos (\ref{smm}) of Majorana type, 
whose masses
$M^s_1\sim 10^{2}$GeV, $M^s_2 \sim 10^{5}$GeV and $M^s_3 \sim 10^8$GeV (see Table 
\ref{nspectrum}), could be candidates for very massive cold dark-matter (DM) particles. While, 
the right-handed sterile neutrinos $\nu_R^f$ of Dirac type,  whose Dirac masses
$M^D_1\sim 10^{-6}$GeV, $M^D_2 \sim 10^{-3}$GeV and $M^D_3 \sim 10^{-1}$GeV (see Table 
\ref{nspectrum}), could be considered as light, weak-interacting  ``warm'' DM particles, in particular the one $\nu_R^1$ with a few KeV mass. Moreover,
the sterile composite scalar particle (\ref{mhs}) could be 
probably a candidate for a massive cold DM particle, 
thought we do not know its mass $m^M_{_H}$ (\ref{smass}), 
i.e., the scale of lepton-number non-conservation.
What is then the candidate for light, non-interacting warm DM particle? 
We expect that it should be the 
pseudoscalar boson $\phi^M$ (\ref{mps}), which acquires a small mass $m_\phi$ 
by the analogy of the PCAC (partially conserved axial-vector current) 
and soft pion theorems,
\begin{eqnarray}
\partial^\mu A_\mu =f_\phi m^2_\phi \phi^M, \quad A_\mu=\sum_f\bar\nu^{fc}_R\gamma^\mu\gamma_5\nu^f_R.
\label{pcac}
\end{eqnarray} 
This is due to the presence of soft explicit $U_{\rm lepton}(1)$-symmetry breaking terms 
$\tilde m^s_f\bar\nu^{fc}_R\nu^f_R$ ($\tilde m^s_f\ll m^M_f$) in Eq.~(\ref{lmass}).
The $f_\phi$ is the pseudoscalar boson $\phi^M$ decay constant relating to the processes 
$\phi^M\rightarrow \nu^{fc}_R+\nu^f_R$.  Both mass $m_\phi$ and decay constant 
$f_\phi$ depend on the soft explicit breaking scale $\tilde m^s_f$ 
of $U_{\rm lepton}(1)$-symmetry.
It is worthwhile to notice that both sterile Majorana neutrinos (the candidates of 
cold DM particles) and the sterile pseudoscalar boson (the candidate of warm 
DM particle) carry two units of
lepton number. This implies that the relevant processes of these sterile particles interacting 
with the SM particles, thought very weak, should violate the lepton-number conservation and 
lead to the asymmetry of matter and anti-matter.  At the end we mention that for strong coupling $G$ the relevant four-fermion operators (\ref{bhlqlm}) and (\ref{l}) 
present the interactions of DM and SM particles, and   
form gauged and neutral composite particles as resonances of masses at TeV scale, 
%which could be the candidates of dark-matter particles, and 
then these composite particles (resonances) decay into their 
constitutes -- SM and/or DM particles \cite{xue2014,xue2015,xuejpg2003,xue2015_1}.      
     
\comment{
Apart from the Majorana neutrino electromagnetic form factors, 
the two rates of heavy neutrinos $\nu^g_3$
and $\nu^g_2$ decaying to light neutrino $\nu^g_1$ and emitting a photon are proportional to   
\begin{eqnarray}
\Gamma_3\propto\left(|\Delta M^{g2}_{31}|/M_3^g\right)^3,
\quad \Gamma_2\propto\left(|\Delta M^{g2}_{21}|/M_2^g\right)^3.
\label{ndecay}
\end{eqnarray}
The measurements of these rates and their ratio $\Gamma_3/\Gamma_2\sim 1.95\times 10^2$ could 
give constrains on Majorana neutrinos CP-properties.
}

%\vskip0.1cm
\section
{\bf Acknowledgment.}  
\hskip0.1cm The author thanks 
Prof.~Hagen Kleinert for discussions on 
the IR- and UV-stable fixed points of quantum field theories, and Prof.~Zhiqing Zhang for discussions on the LHC physics.
The author also thanks the anonymous referee for his/her effort of reviewing this lengthy article.

\comment{
For all numerical evaluations made in this work, we stick to the updated
values of the quark mixing matrix~, PDG Review 2014~\cite{Agashe:2014kda}
\begin{equation}\label{eq:CKMPDG}
  |V_{\text{CKM}}| =
  \begin{pmatrix}
    0.97427 \pm 0.00014 & 0.22536 \pm 0.00061 & 0.00355\pm 0.00015 \\
    0.22522 \pm 0.00061 & 0.97343 \pm 0.00015 & 0.0414 \pm 0.0012 \\
    0.00886^{+0.00033}_{-0.00032} & 0.0405^{+0.0011}_{-0.0012} &
    0.99914\pm 0.00005 \end{pmatrix},
\end{equation}
The most recent update on the $3\sigma$ allowed ranges of the elements
of the PMNS mixing matrix are given by~\cite{Gonzalez-Garcia:2014bfa},
\begin{equation}
  |U_{PMNS}| =
  \begin{pmatrix}
    0.801\rightarrow 0.845 & 0.514 \rightarrow 0.580 & 0.137 \rightarrow 0.158 \\
    0.225 \rightarrow 0.517 & 0.441 \rightarrow 0.699 & 0.614 \rightarrow 0.793 \\
    0.246 \rightarrow 0.529 & 0.464 \rightarrow 0.713 & 0.590
    \rightarrow 0.776
  \end{pmatrix}.
\end{equation}
}


\begin{thebibliography}{99}

\bibitem{njl} Y.~Nambu and G.~Jona-Lasinio, Phys.~Rev.~122 (1961) 345.
%\bibitem{wilson1}K.~G.~Wilson, Phys. Rev. D 10 2445 (1973).

\bibitem{higgs}
F.~Englert, R.~Brout, Phys.~Rev.~Lett. 13 (1964) 321;\\
P.~W.~Higgs, Phys. Lett. 12 (1964) 132;
Phys.~Rev.~Lett.~13 (1964) 508; Phys.~Rev.~145 (1966) 1156;\\
G.~S.~ Guralnik, C.~R.~Hagen, T.~W.~B.~Kibble, Phys.~Rev.~Lett. 13 (1964) 585;
and T.~W.~B.~ Kibble, Phys.~Rev.~155 (1967) 1554.

\bibitem{ATLASCMS} ATLAS Collaboration, Phys.~Lett.~B 716 (2012) 1;\\
%\bibitem{CMS} 
CMS Collaboration, Phys.~Lett.~B 716 (2012) 30-61.

\bibitem{CDFD0} CDF Collaboration, Phys.~Rev.~Lett.~ 74 (14): 2626–2631 
(1995);\\
%\bibitem{D0} 
D0 Collaboration, Phys.~Rev.~Lett.~ 74 (13): 2422–2426 (1995).

%------------------------------review----------------------------------------------%
\bibitem{1}
R. Gatto, G. Sartori, and M. Tonin, “Weak Selfmasses, Cabibbo Angle, and Broken SU(2) x
SU(2),” Phys.Lett. B28 (1968) 128–130.

\bibitem{2}
N. Cabibbo and L. Maiani, “Dynamical interrelation of weak, electromagnetic and strong
interactions and the value of theta,” Phys.Lett. B28 (1968) 131–135.

\bibitem{3}
K. Tanaka and P. Tarjanne, “Cabibbo angle and selfconsistency condition,”
Phys.Rev.Lett. 23 (1969) 1137–1139.

\bibitem{4}
R. Oakes, “SU(2) x SU(2) breaking and the Cabibbo angle,” Phys.Lett. B29 (1969) 683–685.

\bibitem{5}
H. Genz, J. Katz, L. Ram Mohan, and S. Tatur, “Chiral symmetry breaking and the cabibbo
angle,” Phys.Rev. D6 (1972) 3259–3265.

\bibitem{6}
H. Pagels, “Vacuum Stability and the Cabibbo Angle,” Phys.Rev. D11 (1975) 1213.

\bibitem{7}
A. Ebrahim, “Chiral SU(4) x SU(4) Symmetry Breaking and the Cabibbo Angle,”
Phys.Lett. B69 (1977) 229–230.

\bibitem{8}
H. Fritzsch, “Calculating the Cabibbo Angle,” Phys.Lett. B70 (1977) 436.

\bibitem{9}
S. Weinberg, “The Problem of Mass,” Trans.New York Acad.Sci. 38 (1977) 185–201.

\bibitem{10}
A. De Rujula, H. Georgi, and S. Glashow, “A Theory of Flavor Mixing,”
Annals Phys. 109 (1977) 258.

\bibitem{11}
F. Wilczek and A. Zee, “Discrete Flavor Symmetries and a Formula for the Cabibbo Angle,”
Phys.Lett. B70 (1977) 418.

\bibitem{12}
R. N. Mohapatra and G. Senjanovic, “Cabibbo Angle, CP Violation and Quark Masses,”
Phys.Lett. B73 (1978) 176.

\bibitem{13}
F. Wilczek and A. Zee, “Horizontal Interaction and Weak Mixing Angles,”
Phys.Rev.Lett. 42 (1979) 421.

\bibitem{14}
D. Wyler, “The Cabibbo Angle in the SU(2)L× U(1) Gauge Theories,”
Phys.Rev. D19 (1979) 330.

\bibitem{15}
H. Fritzsch, “Hierarchical Chiral Symmetries and the Quark Mass Matrix,”
Phys.Lett. B184 (1987) 391.

\bibitem{16}
P. H. Frampton and Y. Okada, “Simplified symmetric quark mass matrices and flavor mixing,”
Mod.Phys.Lett. A6 (1991) 2169–2172.

\bibitem{17}
J. L. Rosner and M. P. Worah, “Models of the quark mixing matrix,”
Phys.Rev. D46 (1992) 1131–1140.

\bibitem{18}
S. Raby, “Introduction to theories of fermion masses,” arXiv:hep-ph/9501349 [hep-ph].

\bibitem{19}
T. Ito and M. Tanimoto, “The Unitarity triangle and quark mass matrices on the NNI basis,”
Phys.Rev. D55 (1997) 1509–1514, arXiv:hep-ph/9603393 [hep-ph].

\bibitem{20}
Z.-Z. Xing, “Implications of the quark mass hierarchy on flavor mixings,”
J.Phys. G23 (1997) 1563–1578, arXiv:hep-ph/9609204 [hep-ph].

\bibitem{xue1997mx}
S.-S. Xue, %“Quark masses and mixing angles,” 
Phys.Lett. B398 (1997) 177–186,
arXiv:hep-ph/9610508 [hep-ph].

\bibitem{xue1999nu}
S.-S.~Xue, Modern Physics Letters A, Vol.~14 (1999) 2701, hep-ph/9706301.

\bibitem{22}
R. Barbieri, L. J. Hall, S. Raby, and A. Romanino, 
%“Unified theories with U(2) flavor symmetry,” 
Nucl.Phys. B493 (1997) 3–26. %, arXiv:hep-ph/9610449 [hep-ph].

\bibitem{23}
D. Falcone and F. Tramontano, % “Relation between quark masses and weak mixings,”
Phys.Rev. D59 (1999) 017302, arXiv:hep-ph/9806496 [hep-ph].

\bibitem{24}
A. Mondragon and E. Rodriguez-Jauregui, 
%“The Breaking of the flavor permutational symmetry: Mass textures and the CKM matrix,” 
Phys.Rev. D59 (1999) 093009,
arXiv:hep-ph/9807214 [hep-ph].

\bibitem{25}
A. Mondragon and E. Rodriguez-Jauregui, 
%“The CP violating phase delta(13) and the quark mixing angles theta(13), theta(23) and theta(12) from flavor permutational symmetry breaking,” 
Phys.Rev. D61 (2000) 113002, arXiv:hep-ph/9906429 [hep-ph].

\bibitem{26}
H. Fritzsch and Z.-Z. Xing, %“Mass and flavor mixing schemes of quarks and leptons,”
Prog.~Part.~Nucl.~Phys.` 45 (2000) 1–81, arXiv:hep-ph/9912358 [hep-ph].

\bibitem{27}
G. Branco, D. Emmanuel-Costa, and C. Simoes, 
%“Nearest-Neighbour Interaction from an Abelian Symmetry and Deviations from Hermiticity,” 
Phys.Lett. B690 (2010) 62–67, arXiv:1001.5065 [hep-ph].

\bibitem{28}
F. Gonz´alez Canales, A. Mondrag´on, M. Mondrag´on, U. J. Salda˜na Salazar, and
L. Velasco-Sevilla, 
%“Quark sector of S3 models: classification and comparison with experimental data,” 
Phys.Rev. D88 (2013) 096004, arXiv:1304.6644 [hep-ph].

\bibitem{29} L. J. Hall and A. Rasin, “On the generality of certain predictions for quark mixing,”
Phys.Lett. B315 (1993) 164–169, arXiv:hep-ph/9303303 [hep-ph].
\bibitem{30} A. Rasin, “Diagonalization of quark mass matrices and the Cabibbo-Kobayashi-Maskawa
matrix. ,” arXiv:hep-ph/9708216 [hep-ph].
\bibitem{31} A. Rasin, “Hierarchical quark mass matrices,” Phys.Rev. D58 (1998) 096012,
arXiv:hep-ph/9802356 [hep-ph].
\bibitem{32} H. Fritzsch and Z.-Z. Xing, “The Light quark sector, CP violation, and the unitarity triangle,”
Nucl.Phys. B556 (1999) 49–75, arXiv:hep-ph/9904286 [hep-ph].
\bibitem{33} Z.-z. Xing, “Model-independent access to the structure of quark flavor mixing,”
Phys.Rev. D86 (2012) 113006, arXiv:1211.3890 [hep-ph].
\bibitem{34} H. Ishimori, T. Kobayashi, H. Ohki, Y. Shimizu, H. Okada, et al., “Non-Abelian Discrete
Symmetries in Particle Physics,” Prog.Theor.Phys.Suppl. 183 (2010) 1–163,
arXiv:1003.3552 [hep-th].
\bibitem{35} S. F. King and C. Luhn, “Neutrino Mass and Mixing with Discrete Symmetry,”
Rept.Prog.Phys. 76 (2013) 056201, arXiv:1301.1340 [hep-ph].
\bibitem{36} J. Schechter and J. Valle, “Neutrino Masses in SU(2) x U(1) Theories,”
Phys.Rev. D22 (1980) 2227.

\bibitem{37} H. Fritzsch, %“Quark Masses and Flavor Mixing,” 
Nucl.Phys. B155 (1979) 189.

\bibitem{faraggi2014} 
A.~E.~Faraggi, Galaxies 2014, 2, 223-258, http://arxiv.org/abs/1404.5180,
and the references therein.

\bibitem{Dine2015} 
Michael Dine, ``Supersymmetry and String Theory: Beyond the Standard Model'' Cambridge University Press, ISBN 978-1-107-04838-6 (2015), the
refereces therein.


\bibitem{38} H. Fritzsch, %“Weak Interaction Mixing in the Six - Quark Theory,”
Phys.Lett. B73 (1978) 317–322.

\bibitem{39} P. Ramond, R. Roberts, and G. G. Ross, “Stitching the Yukawa quilt,”
Nucl.Phys. B406 (1993) 19–42, arXiv:hep-ph/9303320 [hep-ph].
\bibitem{40} G. Branco, D. Emmanuel-Costa, and R. Gonzalez Felipe, “Texture zeros and weak basis
transformations,” Phys.Lett. B477 (2000) 147–155, arXiv:hep-ph/9911418 [hep-ph].
\bibitem{41} R. Roberts, A. Romanino, G. G. Ross, and L. Velasco-Sevilla, “Precision test of a fermion mass
texture,” Nucl.Phys. B615 (2001) 358–384, arXiv:hep-ph/0104088 [hep-ph].
\bibitem{42} H. Fritzsch and Z.-z. Xing, “Four zero texture of Hermitian quark mass matrices and current
experimental tests,” Phys.Lett. B555 (2003) 63–70, arXiv:hep-ph/0212195 [hep-ph].
\bibitem{43} M. Gupta and G. Ahuja, “Flavor mixings and textures of the fermion mass matrices,”
Int. Jour. Mod. Phys. A, 27 (2012) 1230033, arXiv:1302.4823 [hep-ph].
\bibitem{44} S. Pakvasa and H. Sugawara, “CP Violation in Six Quark Model,” Phys.Rev. D14 (1976) 305.

\bibitem{45} S. Pakvasa and H. Sugawara, “Discrete Symmetry and Cabibbo Angle,”
Phys.Lett. B73 (1978) 61.
\bibitem{46} E. Derman, “Flavor Unification, τ Decay and b Decay Within the Six Quark Six Lepton
Weinberg-Salam Model,” Phys.Rev. D19 (1979) 317–329.
\bibitem{47} Y. Yamanaka, H. Sugawara, and S. Pakvasa, “Permutation Symmetries and the Fermion Mass
Matrix,” Phys.Rev. D25 (1982) 1895.
\bibitem{48} R. Yahalom, “Horizontal Permutation Symmetry, Fermion Masses and Pseudogoldstone Bosons
in SU(2)L× U(1),” Phys.Rev. D29 (1984) 536.
\bibitem{49} G. Altarelli, “Status of Neutrino Mass and Mixing,” Int.J.Mod.Phys. A29 (2014) 1444002,
arXiv:1404.3859 [hep-ph].
\bibitem{50} S. F. King, A. Merle, S. Morisi, Y. Shimizu, and M. Tanimoto, “Neutrino Mass and Mixing:
from Theory to Experiment,” New J.Phys. 16 (2014) 045018, arXiv:1402.4271 [hep-ph].
\bibitem{51} H. Fritzsch and Z.-z. Xing, “Lepton mass hierarchy and neutrino mixing,”
Phys.Lett. B634 (2006) 514–519, arXiv:hep-ph/0601104 [hep-ph].
\bibitem{52} H. Fritzsch and Z.-z. Xing, “Relating the neutrino mixing angles to a lepton mass hierarchy,”
Phys.Lett. B682 (2009) 220–224, arXiv:0911.1857 [hep-ph].

\bibitem{qlmixing2015}
Wolfgang Gregor Hollik, Ulises Jesus Saldana Salazar,
Nucl. Phys. B 892 (2015), 364-389
%DOI:	10.1016/j.nuclphysb.2015.01.019
arXiv:1411.3549 [hep-ph]

%------------------so(10)-----------------------------------%

\bibitem{so10_0} G.~Lazarides; Q.~Shafi; C.~Wetterich, 	
Nuclear Physics, Section B, Volume 181, Issue 2, p. 287-300.

\bibitem{so10_1} G.~Altarelli, G.~Blankenburg, JHEP 1103:133,2011, arXiv:1012.2697 [hep-ph].
\comment{Different SO(10) Paths to Fermion Masses and Mixings
Guido Altarelli, Gianluca Blankenburg
(Submitted on 13 Dec 2010 (v1), last revised 9 Mar 2011 (this version, v3))
Recently SO(10) models with type-II see-saw dominance have been proposed as a promising framework for obtaining Grand Unification theories with approximate Tri-bimaximal (TB) mixing in the neutrino sector. We make a general study of SO(10) models with type-II see-saw dominance and show that an excellent fit can be obtained for fermion masses and mixings, also including the neutrino sector. To make this statement more significant we compare the performance of type-II see-saw dominance models in fitting the fermion masses and mixings with more conventional models which have no built-in TB mixing in the neutrino sector. For a fair comparison the same input data and fitting procedure is adopted for all different theories. We find that the type-II dominance models lead to an excellent fit, comparable with the best among the available models, but the tight structure of this framework implies a significantly larger amount of fine tuning with respect to other approaches. 
Comments: 	24 pages, References and minor wording changes added
Subjects: 	High Energy Physics - Phenomenology (hep-ph)
Journal reference: 	JHEP 1103:133,2011
DOI: 	10.1007/JHEP03(2011)133
Report number: 	RM3-TH/10-13, CERN-PH-TH/2010-293
Cite as: 	arXiv:1012.2697 [hep-ph]
}
%------------------------
\bibitem{so10_2} A.~S.~ Joshipura, K.~M.~ Patel, 
Phys.Rev.D83:095002,2011, arXiv:1102.5148 [hep-ph]
\comment{
Fermion Masses in SO(10) Models
Anjan S. Joshipura, Ketan M. Patel, 
(Submitted on 25 Feb 2011)
    We examine many SO(10) models for their viability or otherwise in explaining all the fermion masses and mixing angles. This study is carried out for both supersymmetric and non-supersymmetric models and with minimal (10+126¯) and non-minimal (10+126¯+120) Higgs content. Extensive numerical fits to fermion masses and mixing are carried out in each case assuming dominance of type-II or type-I seesaw mechanism. Required scale of the B-L breaking is identified in each case. In supersymmetric case, several sets of data at the GUT scale with or without inclusion of finite supersymmetric corrections are used. All models studied provide quite good fits if the type-I seesaw mechanism dominates while many fail if the type-II seesaw dominates. This can be traced to the absence of the b-τ unification at the GUT scale in these models. The minimal non-supersymmetric model with type-I seesaw dominance gives excellent fits. In the presence of a 45H and an intermediate scale, the model can also account for the gauge coupling unification making it potentially interesting model for the complete unification. Structure of the Yukawa coupling matrices obtained numerically in this specific case is shown to follow from a very simple U(1) symmetry and a Froggatt-Nielsen singlet. 
Comments: 	31 pages, 9 Tables, 4 figures
Subjects: 	High Energy Physics - Phenomenology (hep-ph)
Journal reference: 	Phys.Rev.D83:095002,2011
DOI: 	10.1103/PhysRevD.83.095002
Cite as: 	arXiv:1102.5148 [hep-ph]
(or arXiv:1102.5148v1 [hep-ph] for this version)
}		
		
%--------------------------------------------------------------
\bibitem{so10_3} Walter Grimus, , Helmut Kühböck, Phys.~Let.~B10 (2006) 038.
\comment{Physics Letters B
Volume 643, Issues 3–4, 14 December 2006, Pages 182–189
Cover image
Fermion masses and mixings in a renormalizable SO(10)×Z2SO(10)×Z2 GUT
    Walter Grimus, , Helmut Kühböck
 Show more        doi:10.1016/j.physletb.2006.10.038
    Get rights and conten
}

\bibitem{so10_3.1} Luís Lavoura, Helmut Kühböck, Walter Grimus,
Nuclear Physics B 754 (2006) 1–16.

%--------------------------------------------------------------
\bibitem{so10_4} C.H.~Albright and S.~ Nandi Phys.~Rev.~Lett.~ 73 930 (‎1994)
\comment{
New Approach for the Construction of Fermion Mass ...
link.aps.org/pdf/10.1103/PhysRevLett.73.930
by CH Albright and Satyanarayan Nandi PRL 73 930 (‎1994) - ‎Cited by 28 - ‎Related articles
Regular Article - Theoretical Physics
Journal of High Energy Physics
September 2015, 2015:40
First online: 08 September 2015
Open Access
A realistic pattern of fermion masses from a five-dimensional SO(10) model
    Ferruccio Feruglio, Ketan M. Patel, Denise Vicino 
}		

\bibitem{so10_5}C.H. Albright, S.M. Barr
%Fermion masses in SO(10) with a single adjoint Higgs field
Phys. Rev. D, 58 (1998) 013002
%--------------------------------finish review ---------------------------------------------%

\bibitem{hill1994}
C.~T.~ Hill, Phys.~Lett.~B266 (1991) 491 and {\it ibid} B345 (1995) 483;
Phys.~Rev.~ D87 (2013) 065002.

\bibitem{bhl1990a}
C.~ T.~ Hill, Phys.~ Rev.~ D24, 691 (1990); C.~ T.~ Hill, C.~N.~ Leung, S.~ Rao, Nucl.~
Phys.~ B262, 517, (1985); J. Bagger, S.~ Dimopoulous, E.~ Masso, Phys.~Rev.~ Lett.~55 920 (1985).

\bibitem{bhl1990}
W.~A.~Bardeen, C.~T.~Hill and M.~Lindner, {\sl Phys.~Rev.} {\bf D41} (1990) 1647. 

\bibitem{nambu1989}
Y.~Nambu, %Enrico Fermi Institute Report No.~89-08, 1989 (unpublished); 
in Proceedings of
the 1989 Workshop on Dynamical Symmetry Breaking, edited by T.~Muta and K.~ Yamawaki (Nagoya University, Nagoya, Japan, 1990);\\
%\bibitem{mty1989}
V.A.~Miranski, M.~Tanabashi and K.~Yamawaki, {\sl Mod.~Phys.~Lett.} {\bf A4} (1989) 1043; {\sl Phys.~Lett.} {\bf B221} (1989) 117;\\
H.~Kleinert, the SU(3)-extension of their work in
Chapter 26 of the textbook http://klnrt.de/b6, "Particles an Quantum Field''
World Scientific  Publishing Company, 2016.

\bibitem{Marciano1989}
W.~J.~Marciano, Phys.~Rev.~Lett.~62, (1989) 2793.

\bibitem{DSB_review}
G.~Cvetic, Rev.~Mod.~Phys.~71 (1999) 513-574;\\ 
C.~T.~Hill, E.~H.~Simmons, Phys.~Rept.~381 (2003) 235-402; {\it Erratum-ibid.} 390 (2004) 553-554.

\bibitem{xue2013}
S.-S.~Xue, Phys.~ Lett.~B727 (2013) 308.

\bibitem{xue2014}
S.-S.~Xue, Phys.~ Lett.~B737 (2014) 172.

\bibitem{xue1997} S.-S.~Xue, Phys.~Lett.~B381 (1996) 277, Nucl.~Phys.~B486 (1997) 282, {\it ibid} B580 (2000) 365, Phys.~Rev.~D 61 (2000) 054502, 
J. Phys. G, Nucl. Part. Phys. 29 (2003) 2381.

\bibitem{xueparity} S.-S.~Xue, J. Phys. G, Nucl. Part. Phys. 29 (2003) 2381 and Phys. Lett. B 706 (2011) 213–218.

\bibitem{cky2009} Kingman Cheung, Wai-Yee Keung and Tzu-Chiang Yuan, Phys. Lett. B 682 (2009) 287.

\bibitem{xue2016}
S.-S.~Xue,  Phys. Rev. D93, 073001 (2016), arXiv1506.05994v3.

\bibitem{xue2010}
S.-S.~Xue, Phys.~Rev.~D82, 064039 (2010), 
Phys.~Lett.~B682 (2009) 300 and {\it ibid} B711 (2012) 404.

\bibitem{nn1981} H.B.~Nielson and M.~Ninomiya, Nucl.~ Phys.~ B185 
(1981) 20, {\it ibid} B193 (1981) 173, {\sl Phys.~Lett.} {\bf B105} 
(1981) 219, {\sl Int.~J.~Mod. Phys.} {\bf A6} (1991) 2913.

\bibitem{itzykson}
see for example, C.~Itzykson and J-B.~ Zuber ``{\it Quantum Feild Theory}'' page 161, McGraw-Hill, Inc.~ISBN 0-07-032071-3;\\
H. Kleinert, in Understanding the Fundamental Constituents
of Matter, edited by A. Zichichi (Plenum Press,
New York, 1978), pp. 289–390.

\bibitem{xue2015}
S.-S.~Xue, Phys.~ Lett.~B744 (2015) 88–94.

\bibitem{ATLAS} ATLAS Collaboration, Phys.~Lett.~B 716 (2012) 1, and 
http://atlas.ch/.

\bibitem{CMS} CMS Collaboration, Phys.~Lett.~B 716 (2012) 30-61.

\bibitem{xue2013_1}
S.-S.~Xue, Phys.~ Lett.~B721 (2013) 347.

\bibitem{lq1997}
E. Eichten, J. Preskill, Nucl. Phys. B 268 (1986) 179;\\
M. Creutz, C. Rebbi, M. Tytgat, S.-S. Xue, Phys. Lett. B 402 (1997) 341.

\bibitem{zli}
See for example, ``Gauge theory of elementary particle physics''
by T.~P.~ Cheng and L.~F.~ Li, Oxford University Press Inc. New York, 1984, ISBN 978-019-851961-4 (page 449), the references therein.

\bibitem{mohapatra} Rabindra N.~Mohapatra,Palash B.~Pal, Massive Neutrinos in Physics and Astro-physics,third edition, World Scientific, 2004.

\bibitem{pdg2012} J. Beringer et al. (Particle Data Group Collaboration),
Phys. Rev. D 86, 010001 (2012).

\bibitem{PMNS} %{Gonzalez-Garcia:2014bfa}
M. Gonzalez-Garcia, M. Maltoni, and T. Schwetz, “Updated fit to three neutrino mixing: status
of leptonic CP violation,” arXiv:1409.5439 [hep-ph].

\comment{
\bibitem{giudice}  see for example, 
D.~Buttazzo,  G.~Degrassi,  P.~P.~Giardino,  G.~F.~Giudice, et  al.,
%Investigating the near-criticality of the Higgs boson
JHEP 1312 (2013), 089. 
%1307.3536
\bibitem{DCW2015}
D.~ Bai, J.-W.~ Cui, Y.-L.~ Wu, Phys.~Lett.~B746 (2015) 379–384, showing 
that this transition takes place around $760$ GeV.
}

\bibitem{asCDF}
CDF collaboration, Phys. Rev. Lett. 101, 202001.

\bibitem{asD0}
D0 collaboration, Phys. Rev. Lett. 100, 142002.

\bibitem{kogut}
R.~Fukuda and T.~Kugo, {\sl Nucl.~Phys.} {\bf B117} (1976) 250,\\
W.A.~Bardeen, C.N.~Leung, and S.T.~Love, {\sl Nucl.~Phys.} {\bf B273} (1986)
649; {\it ibid} {\bf B323} (1989) 493.\\
A.~Koci\'c, S.~Hands, B.~Kogut and E.~Dagotto, {\sl Nucl.~Phys.} {\bf B347} 
(1990) 217.\\
G.~Preparata and S.-S.~Xue, Phys.~Lett.~B264, (1991) 35,
{\it ibid} {\bf B302} (1993) 442, {\bf B325} (1994) 161. 

\bibitem{xue2000fi}
S.-S.~Xue, Modern Physics Letters A, Vol.~15 (2000) 1089.

\bibitem{ccon} Planck collaborations, Planck 2015 results. XIII. Cosmological parameters
http://arxiv.org/abs/1502.01589

\bibitem{Aguilar:2001ty}
%A.~Aguilar-Arevalo {\it et al.}  [LSND Collaboration],
%  ``Evidence for neutrino oscillations from the observation of anti-neutrino(electron) appearance in a anti-neutrino(muon) beam,''Phys.\ Rev.\ D {\bf 64} (2001) 112007 
%[hep-ex/0104049];\\
%\bibitem{Aguilar-Arevalo:2013pmq}
A.~A.~Aguilar-Arevalo {\it et al.}  [MiniBooNE Collaboration],
  %``Improved Search for $\bar \nu_\mu \rightarrow \bar \nu_e$ Oscillations in the MiniBooNE Experiment,''
Phys.\ Rev.\ Lett.\  {\bf 110} (2013) 161801.
%[arXiv:1207.4809 [hep-ex], arXiv:1303.2588 [hep-ex]]
%;\\
%\bibitem{Armbruster:2002mp}
%B.~Armbruster {\it et al.}  [KARMEN Collaboration],
%  ``Upper limits for neutrino oscillations muon-anti-neutrino $\rightarrow$ electron-anti-neutrino from muon decay at rest,''Phys.\ Rev.\ D {\bf 65} (2002) 112001.
% [hep-ex/0203021].

\bibitem{giuntibook} C.~Giunti and C.~W.~Kim, Fundamentals of Neutrino Physics and Astrophysics,
Oxford University Press, ISBN 978–0–19–850871–7 (Hbk).


\bibitem{Agashe:2014kda}
Particle Data Group Collaboration, K. Olive et al., “Review of Particle Physics,”
Chin.Phys. C38 (2014) 090001.

\bibitem{xuejpg2003} 
S.-S. Xue, J. Phys. G, Nucl. Part. Phys. 29 (2003) 2381, references therein.

\bibitem{xue2015_1}
S.-S.~Xue, arXiv1601.06845 and the last section in the article arXiv1506.05994v2.


\end{thebibliography}
\end{document}